\documentclass[12pt]{article}
\pdfoutput=1
\usepackage[nosort]{cite}
\usepackage[height=8.85in,width=6.45in]{geometry}
\usepackage{pifont}

\usepackage{times}
\usepackage{slashed}

\usepackage[utf8]{inputenc}
\usepackage{amsmath}
\usepackage{amssymb}
\usepackage{mathtools}
\numberwithin{equation}{section}
\usepackage{slashed}
\usepackage{braket}
\usepackage[svgnames,psnames]{xcolor}
\usepackage[colorlinks,citecolor=DarkGreen,linkcolor=FireBrick,linktocpage,pagebackref]{hyperref}
\usepackage{cite}
\usepackage{graphicx}
\usepackage{tikz}
\usepackage{tikz-cd}

\usepackage{courier}
\usepackage{bm}

\usepackage{dashbox}
\usepackage{subcaption}
\usepackage{enumitem}

\usepackage{upgreek}
\usepackage{mdframed}

\usepackage{blkarray}
\usepackage{arydshln}
\usepackage{dsfont}
\usetikzlibrary{patterns}
\usetikzlibrary{arrows.meta}

\usepackage{calc}

\usepackage{accents}

\newcommand{\ie}{\begin{equation}\begin{aligned}}
\newcommand{\fe}{\end{aligned}\end{equation}}

\renewcommand{\title}[1]{\vbox{\center\LARGE{#1}}\vspace{5mm}}
\renewcommand{\author}[1]{\vbox{\center#1}\vspace{5mm}}
\newcommand{\address}[1]{\vbox{\center\em#1}}

\begin{document}
 
 \begin{titlepage}
  	\hfill MIT-CTP/5504, YITP-SB-2022-39
	 	\\
\title{Non-Invertible Gauss Law and Axions}

\author{Yichul Choi${}^{1,2}$,   Ho Tat Lam${}^3$, and Shu-Heng Shao${}^1$}

		\address{${}^{1}$C.\ N.\ Yang Institute for Theoretical Physics, Stony Brook University\\
        ${}^{2}$Simons Center for Geometry and Physics, Stony Brook University\\
		${}^{3}$Center for Theoretical Physics, Massachusetts Institute of Technology
		}
 
 \abstract{
In axion-Maxwell theory at the minimal axion-photon coupling, we find non-invertible 0- and 1-form global symmetries arising from the naive shift and center symmetries. 
Since the Gauss law is anomalous,  there is no conserved, gauge-invariant, and quantized electric charge. 
Rather, using  half higher gauging, we find a non-invertible Gauss law associated with a non-invertible 1-form global symmetry, which is related to the Page charge.
  These  symmetries act invertibly on the axion field and Wilson line, but  non-invertibly on the monopoles  and axion strings, leading to selection rules related to the Witten effect.  
  We also derive various crossing relations between the defects.  
The non-invertible 0- and 1-form global symmetries mix with other invertible symmetries in a way  reminiscent of  a higher-group symmetry. Using this non-invertible higher symmetry structure, we derive universal inequalities on the energy scales where different infrared symmetries emerge in any renormalization group flow to the axion-Maxwell theory. 
Finally, we discuss implications for the Weak Gravity Conjecture and the Completeness Hypothesis in quantum gravity.
 }

 \end{titlepage}
 
 \tableofcontents
 
 \section{Introduction}

In the past year, a new kind of generalized global symmetries, the \textit{non-invertible symmetry}, has been realized in a variety of quantum systems  in diverse spacetime dimensions. 
See \cite{Koide:2021zxj,Choi:2021kmx,Kaidi:2021xfk,Cordova:2022rer,Benini:2022hzx,Roumpedakis:2022aik,Bhardwaj:2022yxj,Arias-Tamargo:2022nlf,Hayashi:2022fkw,Choi:2022zal,Kaidi:2022uux,Choi:2022jqy,Cordova:2022ieu,Antinucci:2022eat,Bashmakov:2022jtl,Damia:2022rxw,Damia:2022bcd,Moradi:2022lqp,Choi:2022rfe,Bhardwaj:2022lsg,Bartsch:2022mpm,Lin:2022xod,GarciaEtxebarria:2022vzq,Apruzzi:2022rei,Heckman:2022muc,Freed:2022qnc,Niro:2022ctq,Kaidi:2022cpf,Mekareeya:2022spm,Antinucci:2022vyk,Chen:2022cyw,Bashmakov:2022uek,Karasik:2022kkq,Cordova:2022fhg,Decoppet:2022dnz,GarciaEtxebarria:2022jky} for a partial list of references for these recent advances, \cite{Rudelius:2020orz,Heidenreich:2021xpr,Nguyen:2021yld,Kaidi:2021gbs,Wang:2021vki} for earlier discussions of non-invertible symmetries in higher dimensions, and \cite{McGreevy:2022oyu,Cordova:2022ruw} for reviews on generalized global symmetries \cite{Gaiotto:2014kfa}. 
The non-invertible symmetry is implemented by a topological operator without an inverse, which is in particular, not unitary. 
Such examples are ubiquitous  in 1+1d systems \cite{Verlinde:1988sn,Petkova:2000ip,Fuchs:2002cm,Frohlich:2004ef,Frohlich:2006ch,Feiguin:2006ydp,Frohlich:2009gb,Carqueville:2012dk,Aasen:2016dop,Bhardwaj:2017xup,Tachikawa:2017gyf,Chang:2018iay,Ji:2019ugf,Lin:2019hks,Thorngren:2019iar,Gaiotto:2020iye,Komargodski:2020mxz,Aasen:2020jwb,Chang:2020imq,Nguyen:2021naa,Thorngren:2021yso,Sharpe:2021srf,Huang:2021zvu,Huang:2021nvb,Vanhove:2021zop,Burbano:2021loy,Inamura:2022lun,Chang:2022hud,Lin:2022dhv,Robbins:2022wlr}, such as the Kramers-Wannier duality line in the Ising conformal field theory. 

In this paper, we uncover non-invertible global symmetries in the 3+1d axion-Maxwell theory, whose  Lagrangian  in Euclidean signature is
\ie\label{Lagrangian}
{f^2\over 2} d\theta \wedge \star d\theta + {1\over 2e^2} F\wedge \star F - {i K \over 8\pi^2 } \theta F\wedge F\,.
\fe
Here $\theta$ is the dynamical axion scalar field with periodicity $\theta \sim \theta+2\pi $,   $f$ is the axion decay constant, and 
$F=dA$ is the field strength of the dynamical $U(1)$ gauge field $A$.  
Without the axion-photon coupling (i.e., $K=0$), the decoupled theory of the dynamical axion field $\theta$ and photon gauge field $A$ has a shift (0-form) symmetry and a (1-form) electric center symmetry, which act as
\ie\label{K=0}
K=0:~~~~&\theta(x)\to \theta(x)+\alpha\,,\\
&A(x)\to A(x) + \lambda(x)\,,~~~~~d\lambda=0\,,
\fe
respectively. 
For $K>1$, these two symmetries are broken to their $\mathbb{Z}_K$ subgroups, which form  a higher-group symmetry with the other invertible higher-form symmetries \cite{Hidaka:2020iaz,Hidaka:2020izy,Brennan:2020ehu}. 
(See \cite{Nakajima:2022feg} for a higher dimensional generalization.) 
However, at $K=1$, there does not appear to be any invertible symmetries left.

In this paper, we find that  the axion-Maxwell theory at $K=1$ already hosts a rich variety of generalized global symmetries. 
In particular, the shift and center symmetries \eqref{K=0} are resurrected as non-invertible 0- and 1-form global symmetries.\footnote{Recall that a (invertible or non-invertible) $q$-form global symmetry is generated by a codimension-$(q+1)$ topological operator/defect in spacetime. Throughout this paper, we will only work with relativistic quantum field theory in Euclidean signature, in which case the distinction between an ``operator" and a ``defect" is usually not essential, and are sometimes related by a Wick rotation. We will therefore use these two terms interchangeably. However, more generally, defects (topological or not) obey more conditions compared to operators. This is because a well-defined defect should be associated with a Hilbert space when we use it to implement a twist in space. Therefore, a  defect has a preferred normalization, and cannot be arbitrarily rescaled by a $c$-number. 
Furthermore, we can add defects, but we cannot consider general linear combination of defects with complex coefficients.
In contrast, we are allowed to consider arbitrary linear combinations of operators with complex coefficients. 

These non-negative integrality conditions on defects are similar to those for the boundary conditions \cite{Cardy:1989ir}.
 In the special case of  $p$-dimensional topological defects, they can be multiplied by a decoupled $p$-dimensional topological quantum field theory (TQFT). We can also consider linear combinations of topological defects with TQFT coefficients \cite{Roumpedakis:2022aik,Choi:2022zal}. One may view these decoupled TQFTs for $p>1$ as generalization of non-negative integers that can be multiplied to topological line defects. Indeed, for $p=1$, a 1-dimensional (bosonic) topological quantum mechanics is completely characterized by the dimension of its Hilbert space, i.e., a non-negative integer.

By a non-invertible symmetry, we mean that the symmetry generator is not invertible as  a defect, not just as an operator. For example, the Fibonacci line, which obeys the fusion rule $W\times W=1+W$ is invertible as an operator since $W \times (W-1) =1$. But $W$ is not invertible as a defect, since $W-1$, being formally the difference between two defects, is not a well-defined defect associated with a Hilbert space.}

The new non-invertible 1-form symmetry is related to the Page charge \cite{Page:1983mke,Marolf:2000cb}.  
The equations of motion in axion-Maxwell theory (with $K=1$) imply
\ie
- {i\over e^2} d\star F=  {1\over 4\pi^2} d\theta \wedge F\,.
\fe
Since the righthand side is nonzero, the Gauss law is anomalous. 
One can attempt to define a formally conserved  charge  $Q_\text{Page} = \oint_{\Sigma^{(2)}} (- {i\over e^2} \star F -{1\over 4\pi^2} \theta\wedge F)$, known as the Page charge \cite{Page:1983mke,Marolf:2000cb},  but it is not invariant under the periodicity of the axion field $\theta\sim \theta+2\pi$. 
Hence, there is no gauge-invariant, conserved, and quantized electric charge. 
Indeed, the lack of an ordinary electric charge can be understood from the Witten effect \cite{Witten:1979ey}: a magnetic monopole  gains an electric charge by going around an axion string. 

While the operator $``\exp(i \alpha Q_\text{Page})"$ is not gauge-invariant, at any rational angle $\alpha =2\pi p/N$, it has a close cousin that is well-defined.\footnote{Throughout the paper, by a ``rational" angle $\alpha$, we actually mean that $\alpha/2\pi$ is a rational number. We hope this will not cause too much confusions.}
This new surface operator is
\begin{equation}
    \mathcal{D}^{(1)}_{p/N}(\Sigma^{(2)})
    =
    \int [D\phi\,D c]_{\Sigma^{(2)}}
    \exp \left[i
        \oint_{\Sigma^{(2)}} \left(
-{i\over e^2}            \frac{2\pi  p}{N}\star F + \frac{N}{2\pi } \phi dc+ \frac{p}{2\pi } \theta dc + \frac{1}{2\pi} \phi dA
        \right)
    \right] \,,
\end{equation}
where $\phi $ and  $c$ are auxiliary 0- and 1-form fields  living only on $\Sigma^{(2)}$.  
Intuitively, it is a composition of the naive Gauss law operator and a 1+1d $\mathbb{Z}_N$ gauge theory coupled to the bulk fields. 
This construction is similar to how a fractional quantum Hall state cures the ABJ anomaly in massless QED \cite{Choi:2022jqy,Cordova:2022ieu}.\footnote{See \cite{Karasik:2022kkq,GarciaEtxebarria:2022jky} for an alternative non-invertible topological operator labeled by a $U(1)$ angle arising from the ABJ anomaly.}
This new operator is gauge-invariant, topological (and in particular conserved under time evolution), and can be defined on any closed 2-manifold. 
However, it is not invertible, and in particular, is not a unitary operator, i.e., ${\cal D}^{(1)}_{p/N}\times ({\cal D}^{(1)}_{p/N})^\dagger \neq 1$. 
Since ${\cal D}^{(1)}_{p/N}$ is supported on a codimension-2 surface $\Sigma^{(2)}$ in spacetime, we call it a non-invertible 1-form symmetry defect.

\begin{figure}[!t]
    \centering
    \includegraphics[width=0.9\textwidth]{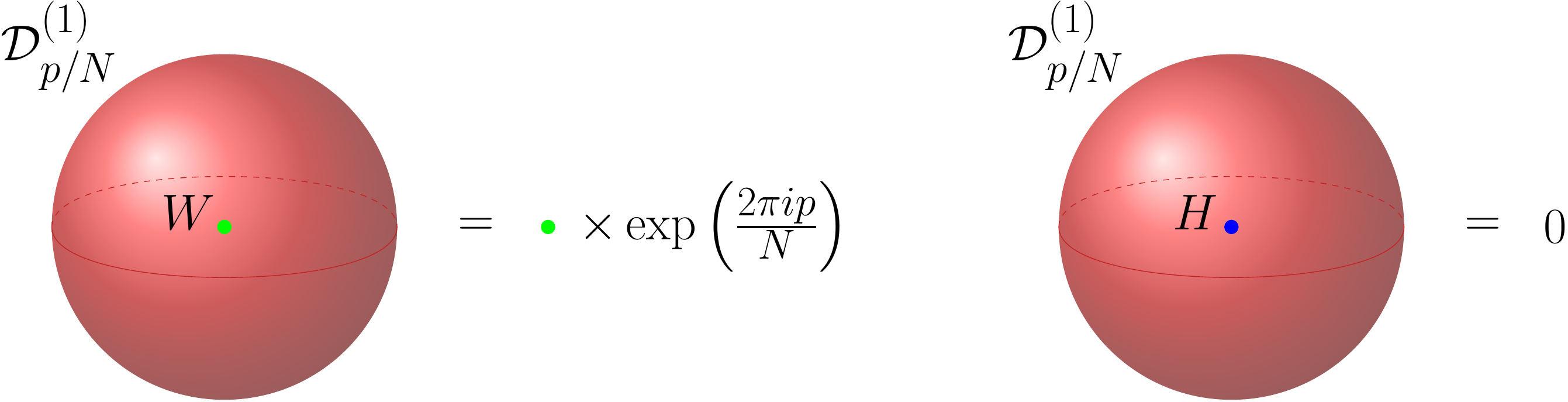}
    \caption{Non-invertible Gauss law implemented by the non-invertible 1-form symmetry surface operator ${\cal D}^{(1)}_{p/N}$ labeled by $p/N\in \mathbb{Q}/\mathbb{Z}$. Here $W$ and $H$ stand for the minimally charged Wilson and 't Hooft lines, located at a point in the 3-dimensional space and extended in the time direction (the time direction  is not shown in the figure). 
    The non-invertible 1-form symmetry  ${\cal D}^{(1)}_{p/N}$ measures the electric charge of the Wilson line invertibly by a phase $e^{2\pi ip/N}$, much as an ordinary Gauss law operator $e^{ i \alpha Q}$ (with $\alpha=2\pi p/N$) does. However, it annihilates the 't Hooft line, and is therefore a non-invertible operator.}
    \label{Fig:gauss}
\end{figure}

To see the non-invertible nature of ${\cal D}^{(1)}_{p/N}$, we can wrap it around an $S^2$ enclosing a heavy electron $W$ of minimal electric charge  (which is represented by a Wilson line localized at a point in space). 
The symmetry operator ${\cal D}^{(1)}_{p/N}$ acts on $W$ invertibly by a phase $\exp(2\pi i p /N)$, measuring its electric charge as in the ordinary Gauss law. 
However,  a heavy  monopole (which is represented by  an 't Hooft line $H$) of minimal magnetic charge is annihilated by  ${\cal D}^{(1)}_{p/N}$. 
This shows that the operator ${\cal D}^{(1)}_{p/N}$ has a kernel and is non-invertible in the presence of a magnetic monopole.\footnote{In the language of \cite{Gorantla:2022eem}, we can view ${\cal D}^{(1)}_{p/N}$ as a time-like global symmetry operator and it acts on the line defects by linking.}
See Figure \ref{Fig:gauss}. 
Just like the ordinary Gauss law  is associated with a $U(1)$ 1-form global symmetry \cite{Gaiotto:2014kfa}, 
here we derive  a \textit{non-invertible Gauss law} associated with a non-invertible 1-form symmetry.

These generalized global symmetries  are typically emergent symmetries in a renormalization group flow to the axion-Maxwell theory. 
Interestingly, some of these symmetries cannot exist without another. 
This leads to universal constraints on the energy scales where these symmetries become emergent. 
Such constraints were known in the context of higher-group symmetries \cite{Cordova:2018cvg,Brennan:2020ehu}, and here we further generalize them to non-invertible symmetries.
Specifically, we show 
\ie
&E_\text{shift}\lesssim E_\text{magnetic},\\
&E_\text{electric} \lesssim \text{min}\{ E_\text{magnetic},E_\text{winding}\}\,.
\fe
Here $E_\text{shift}$ is the scale where the (non-invertible)  shift symmetry is broken, which is related to the scale of the axion potential. 
$E_\text{magnetic}$ is the scale of a dynamical monopole, and $E_\text{electric}$ is the scale where the non-invertible 1-form symmetry is broken, which is related to the scale of an electrically charged particle. Finally, $E_\text{winding}$ is related to the scale of the tension of the axion string. 
It would be interesting to explore phenomenological consequences of these inequalities.

The rest of the paper is organized as follows. 
In Section \ref{sec:noninv} we review the higher-form symmetries in the axion-Maxwell theory and construct new non-invertible 0- and 1-form global  symmetries even in the $K=1$ case. 
In Section \ref{sec:gauging} we provide an alternative construction for the non-invertible 1-form symmetry via half higher gauging. 
In Section \ref{sec:action}, we show that the non-invertible symmetries act invertibly on the axion fields and Wilson lines, but non-invertibly on the 't Hooft lines and axion string worldsheets. 
We discuss the non-invertible Gauss law in Section \ref{sec:gauss}.
 We then discuss various junctions between the symmetry defects and the charged objects in Section \ref{sec:junction}, and derive  crossing  relations and consistency conditions in Section \ref{sec:crossing}. 
We also derive selection rules on correlation functions involving monopoles and axion strings that are related to the Witten effect and charge teleportation (see Section \ref{sec:selection}).  
In Section \ref{sec:higher_structure}, we find the emission of a lower-dimensional defect at the junction between higher-dimensional defects, suggesting a non-invertible generalization of higher-group symmetries. 
Finally, in Section \ref{sec:application}, we derive universal inequalities on the scales where various global symmetries become emergent in any renormalization group flows to the axion-Maxwell theory, and find applications of the non-invertible symmetries to the Weak Gravity Conjecture and the Completeness Hypothesis in quantum gravity. 
Section \ref{sec:summary} summarizes the results of this paper. 
In Appendix \ref{app:higher_group} we review the higher-group symmetry in axion-Maxwell theory with $K>1$ and discuss the junctions of the symmetry defects. Appendix \ref{app:2dZN} discusses the 1+1d $\mathbb{Z}_N$ gauge theory. 

For first-time readers, we recommend Sections \ref{sec:noninv}, \ref{sec:gauss}, \ref{sec:selection}, and \ref{sec:application} for the main results of this paper. 
See also Section \ref{sec:summary} for  a summary.

  \textit{Note added:} After this paper appeared on arXiv, we received  \cite{Yokokura:2022alv} which contains overlapping results.

 \section{Non-Invertible Symmetries of the Axion-Maxwell Theory} \label{sec:noninv}

We first review  the quantization of the axion-photon coupling $K$ in the axion-Maxwell Lagrangian \eqref{Lagrangian}. 
Throughout this paper, we assume every manifold to be spin and in particular oriented. 
On a closed spin four-manifold $X^{(4)}$, ${1\over 8\pi^2} \oint _{X^{(4)}} F\wedge F\in \mathbb{Z}$.  
Therefore, for the $\theta F\wedge F$ term to be compatible with the periodicity of the axion field $\theta\sim \theta+2\pi$, we need $K\in \mathbb{Z}$.

There are four current operators  of interest to us (here the superscripts denote the form degrees of the currents)\footnote{A $q$-form $U(1)^{(q)}$ global symmetry is associated with a conserved $(q+1)$-form current $J^{(q+1)}$ obeying the conservation equation $d\star J^{(q+1)}=0$. Note that the closed current in \cite{Gaiotto:2014kfa} is the Hodge dual of our $J^{(q+1)}$.}:
\ie\label{eq:currents}
&J_\text{shift}^{(1)} = i  f^2 d\theta \,,~~& d\star J^{(1)}_\text{shift} =  {K\over 8\pi^2}F\wedge F\,,\\
& J_\text{winding}^{(3)} = {1\over 2\pi }\star d\theta\,,~~& d\star J^{(3)}_\text{winding} = 0 \,,\\
&J^{(2)}_\text{electric} = -{i\over e^2}  F \,,~~&d\star J^{(2)}_\text{electric} =  {K\over 4\pi^2} d\theta\wedge F\,,\\
&J^{(2)}_\text{magnetic} = {1\over 2\pi }\star F \,,~~& d\star J^{(2)}_\text{magnetic}=0\,.
\fe
Let us explain these currents and their associated higher-form global symmetries in the case when the axion-photon coupling vanishes, i.e., $K=0$. 
(See \cite{Gaiotto:2014kfa} for a more detailed discussion of these symmetries.)

In $K=0$ case, the axion field, which is a free compact scalar field, has a $U(1)^{(0)}_\text{shift}$ shift 0-form global symmetry, $\theta\to \theta+\alpha$ with $\alpha \in [0,2\pi)$, as well as a $U(1)^{(2)}_\text{winding}$ winding 2-form global symmetry which measures the winding number of the axion field.  
The charged object of $U(1)_\text{shift}^{(0)}$ is the axion field $e^{i\theta}$. 
The charged object of $U(1)_\text{winding}^{(2)}$ is the axion string, which sweeps out a 2-dimensional worldsheet in spacetime. 
The defining property of an axion string of winding charge $w$ is 
\ie\label{thetawind}
\oint_\gamma d\theta = 2\pi w\,,
\fe
where $\gamma$ is a loop that links nontrivially with the string worldsheet in spacetime (with linking number 1). 
In other words, $\theta\to \theta+2\pi w$ as the axion field goes around the string.

The free Maxwell gauge theory has an electric center $U(1)^{(1)}_\text{electric}$ 1-form global symmetry that shifts the dynamical 1-form gauge field $A$ by  a flat connection. Dually, there is a magnetic $U(1)^{(1)}_\text{magnetic}$ 1-form global symmetry that shifts the dual 1-form gauge field. 
The charged objects of $U(1)_\text{electric}^{(1)}, U(1)_\text{magnetic}^{(1)}$ are the Wilson line $e^{i\oint A}$, and the monopole worldline (a.k.a., the 't Hooft line).

There is a mixed 't Hooft anomaly between $U(1)^{(0)}_\text{shift}$ and $U(1)^{(2)}_\text{winding}$, and similarly between $U(1)^{(1)}_\text{electric}$ and $U(1)^{(1)}_\text{magnetic}$.

A nonzero axion-photon coupling $K$ violates the conservation equations for the $U(1)^{(0)}_\text{shift}$ symmetry and the $U(1)^{(1)}_\text{electric}$ symmetry. 
Naively, the axion-photon coupling breaks shift and center symmetries to their $\mathbb{Z}_K$ subgroups.
More precisely, these invertible global symmetries combine into a 3-group \cite{Hidaka:2020iaz,Hidaka:2020izy,Brennan:2020ehu}.\footnote{There is also a 3-group symmetry in the axion-Yang-Mills theory \cite{Seiberg:2018ntt,Cordova:2019uob,Brennan:2020ehu}.} 
The $U(1)_\text{magnetic}^{(1)}$ magnetic 1-form symmetry and the $U(1)_\text{winding}^{(2)}$ winding 2-form symmetry are subgroups  and are generated by a topological surface operator/defect
  \ie\label{magnetic_surface}
   \eta^\text{(m)}_\alpha (\Sigma^{(2)}) \equiv \exp \left(i\alpha \oint_{\Sigma^{(2)}} \frac{F}{2\pi}  \right)~.
   \fe
   and  a topological line operator/defect
   \ie\label{winding_line}
   \eta^\text{(w)}_\alpha(\Sigma^{(1)}) \equiv \exp\left( i \alpha \oint_{\Sigma^{(1)}} {d\theta\over2\pi}\right),
   \fe
   respectively.

In this paper, we focus on the case of minimal axion-photon coupling $K=1$, where the $U(1)_\text{shift}^{(0)}$ and $U(1)^{(1)}_\text{electric}$ symmetries appear to be completely broken and the higher group structure trivializes. 
Surprisingly, even at $K=1$, we will show that the $U(1)_\text{shift}^{(0)}$ and $U(1)^{(1)}_\text{electric}$ symmetries actually turn into a non-invertible 0-form and 1-form global symmetries, respectively,   each labeled by the rational numbers $\mathbb{Q}/\mathbb{Z}$.  Furthermore, these non-invertible 0-form and 1-form symmetries mix with the invertible winding 2-form and magnetic 1-form symmetries in a way similar to the mixing in the higher group symmetry.  For example, the intersection of two non-invertible 1-form symmetry operators emits a winding 2-form symmetry operator and so on (see Section \ref{sec:higher_structure}).

For a generic level $K>1$, the non-invertible symmetries combine with the higher group symmetry of \cite{Hidaka:2020iaz,Hidaka:2020izy,Brennan:2020ehu} to form a larger symmetry.
For simplicity, we set $K=1$ from now on and focus on the non-invertible symmetries.

 \subsection{Non-Invertible 0-form Symmetry} \label{sec:noninv0}
 
Here we review the construction in \cite{Choi:2022jqy,Cordova:2022ieu} of non-invertible 0-form symmetry and apply it to the axion-Maxwell theory. 

The  shift  current $J^{(1)}_\text{shift}$ obeys 
\ie\label{Jshift}
 d\star J^{(1)}_\text{shift} =  {1\over 8\pi^2}F\wedge F = {1\over 2} \star J^{(2)}_\text{magnetic} \wedge \star J^{(2)}_\text{magnetic}\,,
 \fe
 which takes the same form as the anomalous conservation equation in the case of  the Adler-Bell-Jackiw (ABJ) anomaly.  
Even though the naive charge operator $\oint \star J^{(1)}_\text{shit}$ is not conserved, it was recently realized that there is a conserved operator implementing  the shift $\theta \to \theta+2\pi/N$  for every positive integer $N$. 
 Following the same construction in \cite{Choi:2022jqy,Cordova:2022ieu} for massless QED, we define this conserved operator  as\footnote{Here we have the option of adding a properly quantized gravitational Chern-Simons term   on $M^{(3)}$. }
 \ie
 {\cal D}^{(0)}_{1/N} (M^{(3)})  = \int [Da]_{M^{(3)}}
 \exp\left[i 
 \oint_{M^{(3)}}
 \left(
 {2\pi \over N} \star  J_\text{shift}^{(1)} +{N\over 4\pi }a\wedge da +{1\over 2\pi} a\wedge dA
 \right)
 \right] \,,
 \fe
 where $a$ is a dynamical 1-form gauge field that only lives on the operator $M^{(3)}$.\footnote{The superscript $(0)$ is to remind us that ${\cal D}^{(0)}_{1/N}$ is a non-invertible 0-form symmetry, i.e., it is supported on a codimension-1 manifold $M^{(3)}$ in spacetime. Similarly, the superscript $(1)$ for ${\cal D}^{(1)}_{p/N}$ defined below means that it is a non-invertible 1-form symmetry supported on a codimension-2 manifold. The subscript $M^{(3)}$ in the path integral of $a$ means that this field only lives on the 3-manifold $M^{(3)}$ where the operator is supported on.} 
${\cal D}^{(0)}_{1/N}$ can be defined on a general closed 3-manifold in 4-dimensional Euclidean spacetime. 
When $M^{(3)}$ is the whole space at a fixed time, ${\cal D}^{(0)}_{1/N}(M^{(3)})$ is an operator acting on the Hilbert space. 
When $M^{(3)}$ extends in the time direction, ${\cal D}^{(0)}_{1/N}(M^{(3)})$  is a defect that modifies the Hamiltonian.

The heuristic way to understand that ${\cal D}^{(0)}_{1/N}$ is a conserved operator is the following. 
Naively, one can integrate out $a$ on $M^{(3)}$, and obtain $a=-A/N$.  
Substituting this back to ${\cal D}^{(0)}_{1/N}(M^{(3)})$ leads to 
$ \exp\left[ i\oint _{M^{(3)}} \left( {2\pi\over N} \star J^{(1)}_\text{shift} - {1\over 4\pi N}A\wedge dA \right)\right]$, which is formally conserved because the integrand is a closed form (see \eqref{Jshift}). 
However, this manipulation is not precise and only serves as a heuristic argument because both $a$ and $A$ are properly normalized gauge fields, and cannot be divided by  a factor of $N$. 
The more rigorous proof of the topological/conserved property of ${\cal D}^{(0)}_{1/N}$ follows from the half gauging construction presented in \cite{Choi:2022jqy}. 
   
   For a more general rational shift, i.e., $\theta\to \theta+2\pi p/N$ with gcd$(p,N)=1$, there is an associated conserved operator:
   \ie \label{eq:0form}
   {\cal D}^{(0)}_{p/N} (M^{(3)}) = \exp\left[ 
   \oint_{M^{(3)}} \left(
   {2\pi i p\over N}\star J^{(1)}_\text{shift} +{\cal A}^{N,p}[dA/N]
   \right)
   \right]\,,
   \fe
   where  ${\cal A}^{N,p}[B]$ is the 2+1d minimal $\mathbb{Z}_N$ TQFT of \cite{Hsin:2018vcg} that couples to a 2-form background gauge field $B$  (see Appendix A of \cite{Choi:2022jqy} for a review) and here we activate the 2-form backgroud $B$ using $dA/N$.  It is the low energy field theory for a $\nu= p/N$ fractional quantum Hall state.  
   (Here we suppress the path integral over the fields for the minimal $\mathbb{Z}_N$ TQFT ${\cal A}^{N,p}$.)
   
   To summarize, the violation of the conservation equation \eqref{Jshift} can be ``cured" by a 2+1d fractional quantum Hall state for every rational shift $\theta\to\theta+ 2\pi p/N$. 
   It leads to an infinite set of gauge-invariant, conserved (and more generally, topological) operators ${\cal D}^{(0)}_{p/N}(M^{(3)})$ labeled by $p/N\in \mathbb{Q}/\mathbb{Z}$. 
   (See \cite{Putrov:2022pua} for an interesting recent discussion on QFT with a $\mathbb{Q}/\mathbb{Z}$ symmetry.) 
   These operators  can be defined on any closed 3-manifold $M^{(3)}$.   
   As demonstrated in \cite{Choi:2022jqy}, the novelty of these new topological operators is that they are non-invertible and do not obey a group multiplication law. In particular, they are not unitary. 
   For example, the product of ${\cal D}^{(0)}_{1/N}$ with its conjugate is  
   \begin{align}\label{DDdagger0}
   \begin{split}
	&   {\cal D}^{(0)}_{1/N} (M^{(3)})\times {\cal D}^{(0)}_{1/N} (M^{(3)})^\dagger\\
	&=  	\int[Da\, D\bar a]_{M^{(3)}}
   \exp\left[ i\oint_{M^{(3)}} \left(
   {N\over 4\pi}a\wedge da -{N\over 4\pi }\bar a\wedge d\bar a 
   +{1\over 2\pi} (a-\bar a) \wedge dA
   \right)
   \right]\neq 1\,,
   \end{split}
   \end{align}
   where the righthand side is a condensation defect from 1-gauging the $\mathbb{Z}_N^{(1)}\subset U(1)^{(1)}_\text{magnetic}$ subgroup of the magnetic 1-form global symmetry \cite{Choi:2022jqy} (see also \cite{Roumpedakis:2022aik,Choi:2022zal}).\footnote{More specifically, the condensation defect on the righthand side of \eqref{DDdagger0} can be written explicitly as a sum over $\eta^\text{(m)}_{2\pi/N} (\Sigma^{(2)})$ in \eqref{magnetic_surface} on $M^{(3)}$:
   \ie\label{condensation1form}
   \frac{1}{|H^0(M^{(3)};\mathbb{Z}_N)|}\sum_{\Sigma^{(2)}\in H_2(M^{(3)};\mathbb{Z}_N)}e^{i\pi Q\left(\text{PD}_{M^{(3)}}(\Sigma^{(2)})\right)}\eta^\text{(m)}_{2\pi /N}(\Sigma^{(2)})\,,
   \fe
   where $\text{PD}_{M^{(3)}}(\Sigma^{(2)})$ is the Poincar\'e dual of $\Sigma^{(2)}$ on $M^{(3)}$ and
   \ie
   Q(a)=\begin{dcases} 
   	a\cup\beta(a)~,\quad &\text{even } N
   	\\
   	0~,\quad &\text{odd } N
   	\end{dcases}~.
   \fe
   Here, $\beta(a)$ is the Bockstein homomorphism associated to the short exact sequence $1\rightarrow \mathbb{Z}_N\rightarrow \mathbb{Z}_{N^2}\rightarrow \mathbb{Z}_N\rightarrow 1$.}

\subsection{Non-Invertible 1-form  Symmetry} \label{sec:noninv1}

Having resurrected the shift 0-form symmetry as a non-invertible symmetry ${\cal D}^{(0)}_{p/N}$ labeled by the rational numbers, we next proceed to study the fate of the electric 1-form symmetry. 
The 2-form current for the electric 1-form symmetry obeys an anomalous conservation equation:
\ie\label{Je}
d\star J^{(2)} _\text{electric} =  {1\over 4\pi^2} d\theta\wedge F =  \star J^{(3)}_\text{winding} \wedge \star J^{(2)}_\text{magnetic}\,,
\fe
where $J^{(2)}_\text{electric} = -{i\over e^2} F$. 
It appears that the $U(1)^{(1)}_\text{electric}$ symmetry in the absence of the axion-photon coupling is broken, and there is no way to define a topological operator of codimension-2.  
In other words, the Gauss law is anomalous \cite{Fischler:1983sc}. 
Interestingly, as we will see, the electric 1-form symmetry survives as a non-invertible symmetry  labeled by elements in $\mathbb{Q}/\mathbb{Z}$.

Naively,   we might attempt to define the electric 1-form symmetry operator as 
\begin{equation} \label{eq:electric_U}
    {U}_{\alpha}(\Sigma^{(2)}) = \exp \left[
       i\alpha   \oint_{\Sigma^{(2)}}
       \star J^{(2)}_{\text{electric}} 
    \right] \,.
\end{equation}
While this is gauge invariant, it is not conserved. 
Next, we might attempt to define  \cite{Heidenreich:2021xpr}
\begin{equation} \label{eq:electric_U_hat}
    \hat{U}_{\alpha}(\Sigma^{(2)}) = \exp(i\alpha Q_\text{Page}) \equiv \exp \left[
       i\alpha   \oint_{\Sigma^{(2)}} \left(
       \star J^{(2)}_{\text{electric}} - \frac{1}{4\pi^2} \theta dA
        \right)
    \right] \,,
\end{equation}
which is formally conserved according to \eqref{Je}. 
However, the $\theta dA$ term does not respect the $2\pi $ periodicity of the axion field, and  thus $\hat U_\alpha$ and $Q_\text{Page}$ are not valid operators.\footnote{One might try to preserve the $2\pi$ periodicity of the axion by replacing the $\theta dA$ term with $A d\theta$.  The integral of the latter is then not gauge invariant under the gauge symmetry of $A$ because  the coefficient $\frac{\alpha}{4\pi^2}$ is not properly quantized. 
We can shuffle the gauge non-invariance around, but we cannot completely get rid of it.} 
In the context of supergravity, the charge $Q_\text{Page}$, which is not gauge invariant but formally topological (and in particular conserved),  is known as the Page charge \cite{Page:1983mke,Marolf:2000cb} (see also \cite{Diaconescu:2003bm,Moore:2004jv}).

Let us be less ambitious and try to construct  a topological operator when $\alpha=2\pi p/N$ with gcd$(p,N)=1$. 
Then there is a close cousin  of $\hat U_{2\pi p/N}$ that is gauge invariant and topological (and in particular conserved under time evolution). 
This new operator  is defined as
\begin{equation} \label{eq:electric_worldvolume}
    \mathcal{D}^{(1)}_{p/N}(\Sigma^{(2)})
    =
    \int [D\phi\,D c]_{\Sigma^{(2)}}
    \exp \left[
        \oint_{\Sigma^{(2)}} \left(
            \frac{2\pi i p}{N} \star J^{(2)}_{\text{electric}} + \frac{iN}{2\pi } \phi dc+ \frac{ip}{2\pi } \theta dc + \frac{i}{2\pi} \phi dA
        \right)
    \right] \,.
\end{equation}
Here $\phi \sim \phi+ 2\pi$ is a compact scalar field and $c$ is a $U(1)$ 1-form gauge field, both living only on $\Sigma^{(2)}$. 
Since all the coefficients are properly quantized, it is clear that ${\cal D}^{(1)}_{p/N}$ is gauge-invariant. 
The last three terms from \eqref{eq:electric_worldvolume} define a 1+1d $\mathbb{Z}_N$ gauge theory living on $\Sigma^{(2)}$, where the background gauge fields for its $\mathbb{Z}_N^{(0)} \times \mathbb{Z}_N^{(1)}$ symmetry are activated by $\frac{p}{N}d\theta $ and $\frac{1}{N}dA$, respectively.

We now discuss the relation between ${\cal D}^{(1)}_{p/N}$ and $\hat U_{2\pi p/N}$. 
If we integrate out $\phi$ in  \eqref{eq:electric_worldvolume}, we get $N dc = - dA$.
If we further substitute   $c = -A/N$ into \eqref{eq:electric_worldvolume}, then we retrieve $\hat U_{2\pi p/N}$ \eqref{eq:electric_U_hat}. 
However, this substitution is not allowed because both $\phi$ and $c$ are compact fields and cannot be divided by $N$. 
Nonetheless, this gives a heuristic argument why  the operator \eqref{eq:electric_worldvolume} should be topological.
In this sense ${\cal D}^{(0)}_{p/N}$ is a close cousin of $\hat U_{2\pi p/N}$, but they are different.

More rigorously, we can verify the topological nature of the operator ${\cal D}^{(1)}_{p/N}$ by introducing the notion of  half higher gauging. This will be discussed in Section \ref{sec:gauging}.

Since ${\cal D}^{(1)}_{p/N}$ involves a non-invertible topological phase, i.e., the 1+1d $\mathbb{Z}_N$ gauge theory ${iN\over 2\pi} \phi dc$, on its worldsheet $\Sigma^{(2)}$, it is not an invertible operator. 
In particular, it is not unitary. 
Let us demonstrate this by explicitly computing the product:
\begin{align}\label{DDdagger1}
\begin{split}
{\cal D}^{(1)}_{p/N} (&\Sigma^{(2)})\times {\cal D}^{(1)}_{p/N} (\Sigma^{(2)})^\dagger=\int [D\phi\,D\bar\phi\,D c\,D\bar c]_{\Sigma^{(2)}}
 \\& \exp \left[
 i \oint_{\Sigma^{(2)}}\left(
 {N\over 2\pi} \phi d c 
 - {N\over 2\pi} \bar\phi d \bar c 
 +{p\over 2\pi }\theta d(c-\bar c)
 +{1\over 2\pi } (\phi-\bar \phi)dA
\right)
 \right] \neq 1\,.
\end{split}
\end{align}
The righthand side is a condensation defect from 2-gauging a $\mathbb{Z}_N^{(1)}\times \mathbb{Z}_N^{(2)}\subset U(1)^{(1)}_\text{magnetic}\times U(1)^{(2)}_\text{winding}$ symmetry along $\Sigma^{(2)}$, which is not a trivial operator. 
We can further simplify this condensation defect by defining $ c' = c-\bar c$ and $\phi' = \phi -\bar \phi$ and rewriting it as
\begin{align}\label{condensation1}
\begin{split}
 & 
\int [D\phi\,D c']_{\Sigma^{(2)}} \exp \left[
i \oint_{\Sigma^{(2)}}
\left(
 {N\over 2\pi} \phi d  c' 
  +{p\over 2\pi }\theta d c'\right)\right]\\
\times &
    \int [D\phi'\,D\bar c]_{\Sigma^{(2)}}
    \exp\left[ i\oint_{\Sigma^{(2)} }\left(
 {N\over 2\pi} \phi' d \bar c 
 +{1\over 2\pi }  \phi' dA
\right)
 \right]   \,,
\end{split}
\end{align}
where the first line is the condensation defect from 2-gauging the $\mathbb{Z}_N^{(2)}\subset U(1)^{(2)}_\text{winding}$ symmetry, while the second line is a sum of the $\mathbb{Z}_N^{(1)}\subset U(1)^{(1)}_\text{magnetic}$ magnetic 1-form symmetry operators.\footnote{Explicitly, the first line is a sum over winding $\mathbb{Z}_N^{(2)}$ 2-form symmetry defect  \eqref{winding_line} around the 1-cycles on $\Sigma^{(2)}$
\ie\label{condensation2form}
\frac{1}{|H^0(\Sigma^{(2)};\mathbb{Z}_N)|}\sum_{\Sigma^{(1)}\in H_1(\Sigma^{(2)};\mathbb{Z}_N)}\eta^\text{(w)}_{2\pi p/N}(\Sigma^{(1)})~,
\fe
while the second line is a sum over magnetic $\mathbb{Z}_N^{(1)}$ 1-form symmetry defect \eqref{magnetic_surface}, $\sum_{n=1}^N\left[\eta^\text{(m)}_{2\pi /N}(\Sigma^{(2)})\right]^n$.}

To summarize, we have constructed a gauge-invariant and topological operator ${\cal D}^{(1)}_{p/N}$ that can be defined on any closed 2-manifold $\Sigma^{(2)}$. 
The price we pay is that it is a non-invertible operator generating a non-invertible 1-form symmetry. 
Intuitively, the 1+1d $\mathbb{Z}_N$ gauge theory on ${\cal D}^{(1)}_{p/N}$ ``cures" the anomalous conservation equation \eqref{Je}. 
The non-invertible symmetry ${\cal D}^{(1)}_{p/N}$ is a gauge-invariant cousin  of  $\hat U_{2\pi p/N} = \exp(2\pi i p Q_\text{Page}/N)$, but they are not the same. 
 Our construction is similar to that in \cite{Apruzzi:2022rei} for IIB supergravity.

Finally, we comment on the possible topological counterterms one can add to the non-invertible surface defect ${\cal D}^{(1)}_{p/N}$. 
 Given a topological surface defect supported on $\Sigma^{(2)}$, one can always dress it with an Euler counterterm $\exp(\lambda \oint_{\Sigma^{(2)}} R)$ with $\lambda\in \mathbb{R}$. Therefore, the ``quantum dimension" of a surface defect on a general 2-manifold is subject to the ambiguity from this counterterm. 
In particular, we can always choose a counterterm such that the expectation value of ${\cal D}^{(1)}_{p/N}$ on $S^2$ (with no other operator insertions) is 1, i.e., 
\ie\label{S2qdim}
\langle {\cal D}_{p/N}^{(1)}\rangle_{S^2}=1\,.
\fe
See \cite{Roumpedakis:2022aik} for related discussions on this counterterm.

\section{Half Higher Gauging}\label{sec:gauging}

The non-invertible 0-form symmetry defect $\mathcal{D}^{(0)}_{p/N}$ in the axion-Maxwell theory can also be obtained by gauging a discrete subgroup of the magnetic 1-form symmetry in  half of the spacetime while imposing the Dirichlet boundary condition for the corresponding discrete gauge field \cite{Choi:2021kmx,Kaidi:2021xfk,Choi:2022zal}, a procedure known as half gauging. 
We refer the readers to \cite{Choi:2022jqy} for more details.

In this section, we will generalize this procedure to construct topological defects of codimension greater than 1. 
We will first explain the procedure of \emph{half higher gauging} \cite{Damia:2022bcd, Kaidi:2022cpf} in  general, and then apply it to the axion-Maxwell theory to rederive the non-invertible 1-form symmetry  $\mathcal{D}^{(1)}_{p/N}$.
This alternative construction of $\mathcal{D}^{(1)}_{p/N}$  provides a rigorous proof of its topological nature as well as a way of determining its action on other operators/defects.
The half higher gauging has been previously used in \cite{Damia:2022bcd} to produce a non-invertible 1-form symmetry defect in the 5-dimensional Maxwell-Chern-Simons theory.

\subsection{Self-Duality under Higher Gauging}

In general, half gauging of a discrete symmetry produces a topological interface between two different QFTs.
However, when a given QFT is self-dual under gauging the discrete symmetry, the half gauging generates a codimension-1 topological defect in a single theory, which implements a non-invertible 0-form symmetry \cite{Choi:2021kmx,Kaidi:2021xfk,Choi:2022zal}.
We now generalize this construction to produce topological defects of codimension greater than 1.
To this end, we first define the notion of \emph{self-duality under higher gauging}, or \emph{higher self-duality} for short.

Higher gauging \cite{Roumpedakis:2022aik} of a discrete higher-form symmetry generates a topological defect, known as the condensation defect \cite{Kong:2013aya,Kong:2014qka,Else:2017yqj,Gaiotto:2019xmp,Roumpedakis:2022aik,Choi:2022zal}. 
More specifically, $p$-gauging of a $q$-form global symmetry is defined by inserting a network of the $q$-form symmetry defects along a codimension-$p$ manifold in spacetime.\footnote{In 2+1d TQFT, 0-gauging a 1-form symmetry is known as ``condensation," while 1-gauging a 1-form symmetry is known as ``condensation confined to a line" \cite{Kong:2013aya}.}

We will be interested in the case where the higher gauging leads to a ``trivial" condensation defect. 
More precisely, by a trivial condensation defect, we mean that its insertion on any closed manifold can be removed by topological local counterterms on the worldvolume of the  defect.\footnote{There can still be nontrivial contact terms when a trivial condensation defect intersects with other operators.
Determining such contact terms requires more data such as a choice of a symmetry fractionalization class. See, for instance, \cite{Chen:2014wse,Barkeshli:2014cna,Benini:2018reh}, and also \cite{Delmastro:2022pfo,Brennan:2022tyl} for recent discussions.  There are also other notions of trivial condensation defects that we do not explore here.}   
In particular, this implies that the quantum dimension of a trivial condensation defect can also be chosen to be 1 by an appropriate topological local counterterm. 
In this case, we will say that the theory is self-dual under the corresponding higher gauging. 
This generalizes the self-duality of a QFT under the ordinary gauging, in which case we mean that the QFTs before and after gauging are isomorphic up to a classical counterterm \cite{Choi:2021kmx}.

A simple example is the 2+1d Ising TQFT, which has a fermion line $\psi$ generating a $\mathbb{Z}_2^{(1)}$ 1-form symmetry and a non-invertible line $\sigma$. They obey the fusion rule
\ie
\psi\times \psi =1\,,~~~\sigma\times \sigma = 1+\psi \,,~~~ \psi \times \sigma = \sigma\times \psi = \sigma\,.
\fe
The condensation defect from 1-gauging the $\mathbb{Z}_2^{(1)}$ symmetry generated by the fermion line is trivial \cite{Roumpedakis:2022aik}. Thus, the 2+1d Ising TQFT is self-dual under the 1-gauging of the $\mathbb{Z}_2^{(1)}$ 1-form symmetry.

Next, we discuss the quantum symmetry under gauging. 
Recall that (0-)gauging a discrete $q$-form symmetry in a $d$-dimensional QFT leads to a dual $(d-q-2)$-form symmetry, sometimes referred to as a quantum symmetry \cite{Vafa:1989ih,Gaiotto:2014kfa,Tachikawa:2017gyf}.\footnote{For a more complete description of quantum symmetries, see \cite{Bhardwaj:2022lsg,Bartsch:2022mpm}.}
When a QFT is self-dual under gauging a discrete $q$-form symmetry, the $(d-q-2)$-form quantum symmetry in the gauged theory and the $q$-form symmetry in the original theory are identified \cite{Choi:2021kmx,Choi:2022zal}.
This is possible only if $q = d-q-2 \Leftrightarrow q = (d-2)/2$, which in particular requires $d$ to be even.
Alternatively, regardless of whether $d$ is even or odd, it is possible to have a theory which is self-dual under gauging a discrete $q\text{-form} \times (d-q-2)\text{-form}$ symmetry for any integer $0\leq q \leq d-2$.
For instance, see \cite[Appendix~D]{Kaidi:2021xfk} for examples of such QFTs with $d=3$ and $q=0$.\footnote{There is a cheap way to obtain rather trivial examples of QFTs which are self-dual under this kind of gauging.
Consider a  QFT $\mathcal{Q}$ with a non-anomalous, finite, abelian, $q$-form symmetry $G^{(q)}$ with $0 \leq q \leq d-2$.
The theory $\mathcal{Q}/G^{(q)}$ obtained by gauging $G^{(q)}$  then has a quantum $(d-q-2)$-form symmetry described by the Pontryagin dual group $\hat{G}^{(d-q-2)}$.
Now, consider taking the product of these two theories, $\mathcal{Q}^\prime \equiv \mathcal{Q} \times \mathcal{Q}/G^{(q)}$.
The theory $\mathcal{Q}^\prime$ is then self-dual under gauging the $G^{(q)} \times \hat{G}^{(d-q-2)}$ symmetry, since such a gauging simply swaps the two factors $\mathcal{Q}$ and $\mathcal{Q}/G^{(q)}$.
When $q=0$, the duality defect in $\mathcal{Q}^\prime$ obtained from half gauging the $G^{(q)} \times \hat{G}^{(d-q-2)}$ symmetry gives a simple example of a non-invertible symmetry defect. 
The latter acts non-invertibly on local operators, and maps the charged local operators from  $\mathcal{Q}$  to non-local operators  attached to the $G^{(q=0)}$ Wilson lines  in  $\mathcal{Q}/G^{(q=0)}$.\label{fn:product}}

Similarly, upon higher gauging a discrete higher-form symmetry, one obtains higher quantum symmetry defects which are topological defects supported on the condensation defect \cite{Roumpedakis:2022aik}.
For instance, if we $p$-gauge a discrete $q$-form symmetry in a $d$-dimensional QFT, then the corresponding higher quantum symmetry defects are of dimension $q+1-p$.\footnote{We assume $p\leq q+1$, otherwise  higher gauging is not well-defined.}
The symmetry defect for the original $q$-form symmetry is of dimension $d-q-1$.
If the theory is self-dual under such a higher gauging, the higher quantum symmetry has to be identified with the original $q$-form symmetry, which is possible only if $d-q-1 = q+1-p$, that is,
\begin{equation}\label{eq:higher_self_dual}
q = (d+p-2)/2   \,.
\end{equation} 
In particular, this is possible only if $d+p$ is even.
In the example of the 2+1d Ising TQFT which is self-dual under 1-gauging the $\mathbb{Z}_2^{(1)}$ 1-form symmetry, we have $d=3$, $p=1$ and $q=1$, which indeed satisfies the condition \eqref{eq:higher_self_dual}.

Alternatively, similar to the ordinary 0-gauging case, it is possible to have a $d$-dimensional theory which is self-dual under $p$-gauging a discrete $q\text{-form} \times (d+p-q-2)\text{-form}$ symmetry for any integer $0\leq q \leq d-1$.\footnote{For $p=0$, we require $0 \leq q \leq d-2$ to avoid $(-1)$-form symmetries, although one may still draw the same conclusion even when $p=0$ and $q= d-1$ by defining the ``gauging'' of a discrete $(-1)$-form symmetry as taking the direct sum of copies of the QFT with different values of discsrete parameters that are responsible for the $(-1)$-form symmetry.}
As we will see, the axion-Maxwell theory provides such an example with $d=4$, $p=1$ and $q=1$.

\subsection{Topological Defects from Half Higher Gauging}

Having introduced the higher gauging and higher self-duality, we next use it to generate topological defects. 

Condensation defects always admit a topological boundary condition, because one can impose the topological Dirichlet boundary condition for the discrete gauge field living on the defect  \cite{Kaidi:2022cpf}.
If one views the higher gauging as a summation over insertions of symmetry defects along a submanifold, the Dirichlet boundary condition means that the symmetry defects cannot end on the boundary of the submanifold. 
This leads to  a boundary condition that is manifestly topological.
We refer to the procedure of higher gauging with the Dirichlet boundary condition as  \textit{half higher gauging}.

Half higher gauging generates a topological defect living at the boundary of a higher dimensional condensation defect. 
As common in the literature, we refer to such an $n$-dimensional defect (which is generally not necessarily topological) living at the boundary of an $n+1$-dimensional topological defect as a twist defect, or a monodromy defect. 
A twist defect is  to be contrasted with a genuinely $n$-dimensional defect that is not attached to anything else.

\begin{figure}[!h]
    \centering
    \includegraphics[width=\textwidth]{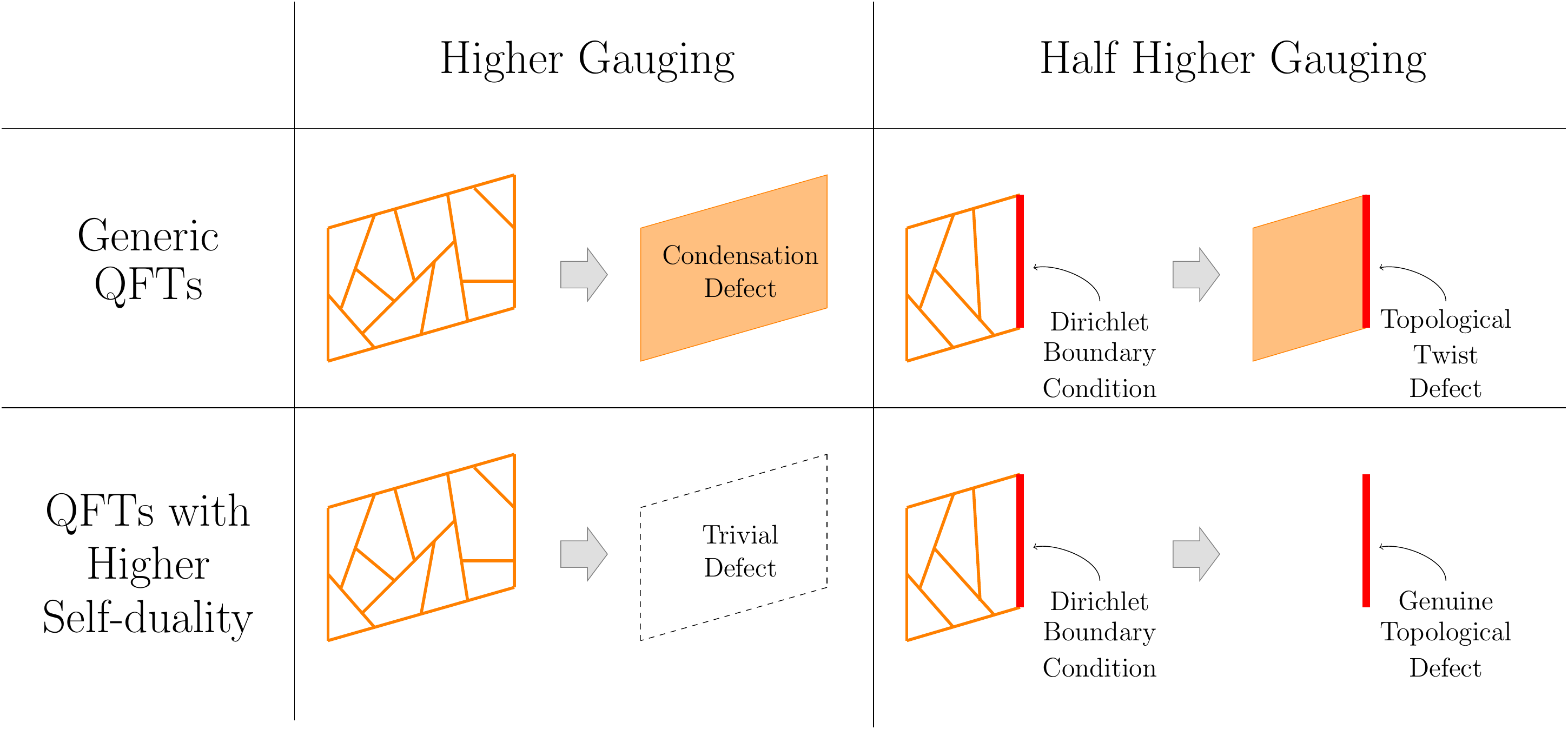}
    \caption{In a general QFT, higher gauging of a higher-form symmetry on a closed submanifold generates a nontrivial condensation defect. Half higher gauging corresponds to higher gauging the  symmetry on a submanifold with a topological  Dirichlet boundary condition imposed on the boundary. This generates a topological twist defect which lives on the boundary of the condensation defect.
    If the theory is self-dual under the higher gauging, then the resulting condensation defect is  trivial, and the half higher gauging generates a genuine topological defect not attached to anything.}
    \label{Fig:halfhigher}
\end{figure}

If the theory is self-dual under a $p$-gauging, i.e., the corresponding codimension-$p$ condensation defect is trivial, then the half $p$-gauging generates a genuine codimension-$(p+1)$ topological defect.\footnote{This is to be contrasted with the discussions in \cite{Kaidi:2022cpf}, where the twist defect becomes a genuine topological operator upon gauging the corresponding condensation defect.
Here, we do not need to gauge the condensation defect, since we assume the theory is self-dual under the corresponding higher gauging and the condensation defect of interest is already trivial.}
When $p=0$, this is the construction of codimension-1 topological defect from half gauging \cite{Choi:2021kmx,Kaidi:2021xfk,Choi:2022zal}.
Some $p=1$ examples were discussed in \cite[Appendix~D]{Kaidi:2021xfk}. 
Later, we will see that the non-invertible 1-form symmetry defect $\mathcal{D}^{(1)}_{p/N}$ in the axion-Maxwell theory also arises from  half  1-gauging.
See Figure \ref{Fig:halfhigher} for a summary of this discussion.

We end this subsection with a simple example of a topological defect obtained from half higher gauging. 
Since the 2+1d Ising TQFT is self-dual under 1-gauging the $\mathbb{Z}_2^{(1)}$ 1-form symmetry, half 1-gauging it generates a genuine line defect. 
The latter is nothing but the non-invertible $\sigma$ line.

\subsection{Higher Self-Duality of Axion-Maxwell Theory}

We now return to the axion-Maxwell theory which is of our main interest.
We claim that the axion-Maxwell theory is self-dual under  1-gauging of a discrete subgroup of the $U(1)_\text{magnetic}^{(1)} \times U(1)_\text{winding}^{(2)}$ symmetry with an appropriate choice of the discrete torsion.
Specifically, for any positive integer $N$, consider a $\mathbb{Z}_N^{(1)} \times \mathbb{Z}_N^{(2)} \subset U(1)_\text{magnetic}^{(1)} \times U(1)_\text{winding}^{(2)}$ subgroup generated by the symmetry defects\footnote{The minus sign in $\eta^\text{(w)}_{-2\pi p/N}$ is chosen such that the non-invertible 1-form symmetry ${\cal D}^{(1)}_{p/N}$ acts on a Wilson line of electric charge $q$ by a phase $e^{2\pi i pq/N}$ (rather than $e^{-2\pi i pq/N}$). }
\begin{equation} \label{eq:invdefects}
    \eta^{\text{(m)}}_{2\pi/N} (\Sigma^{(2)}) = \exp \left(
        \frac{2\pi i}{N}\oint_{\Sigma^{(2)}} \frac{F}{2\pi} 
    \right) \,, \quad
    \eta^{\text{(w)}}_{-2\pi p/N} (\Sigma^{(1)}) = \exp \left(-
        \frac{2\pi ip}{N} \oint_{\Sigma^{(1)}} \frac{d\theta}{2\pi}
    \right) \,.
\end{equation}
Here, $\Sigma^{(n)}$ is a closed $n$-dimensional submanifold   in spacetime. 
$p$ is an integer coprime to $N$, and we have $(\eta^{\text{(m)}}_{2\pi/N})^N = (\eta^{\text{(w)}}_{-2\pi p/N})^N = 1$.

Next, we construct the condensation defect obtained by 1-gauging the $\mathbb{Z}_N^{(1)} \times \mathbb{Z}_N^{(2)}$ symmetry with a particular choice of a discrete torsion, supported on a codimension-1 closed submanifold $\Sigma^{(3)}$:
\begin{align} \label{eq:cond1}
	\begin{split}
		\mathcal{C}(\Sigma^{(3)})
		&= \frac{1}{|H^0|}\frac{|H^0|}{|H^1|}  
		\sum_{\substack{\Sigma^{(2)}\in H_2 \\  \Sigma^{(1)}\in H_1}} 
		\exp \left(\frac{2\pi i}{N} \#(\Sigma^{(2)},\Sigma^{(1)}) \right) \eta^\text{(m)}_{2\pi /N} (\Sigma^{(2)}) \eta^\text{(w)}_{-2\pi p/N} (\Sigma^{(1)}) \\
		&= \frac{1}{|H^1|}
		\sum_{\substack{b^{(1)}\in H^1 \\  b^{(2)}\in H^2}} \exp\left[ 
		\frac{2\pi i}{N}\oint_{\Sigma^{(3)}} \left( b^{(1)} \cup b^{(2)} + b^{(1)} \cup \frac{F}{2\pi} - p \, b^{(2)} \cup \frac{d\theta}{2\pi}
		\right)
		\right] \,.
	\end{split}
\end{align}
Here, $H^i$, $H_i$ are abbreviations for $H^i(\Sigma^{(3)};\mathbb{Z}_N)$ and $H_i(\Sigma^{(3)};\mathbb{Z}_N)$, respectively. We will use this notation throughout this section.
$\#(\Sigma^{(2)},\Sigma^{(1)})$ denotes the intersection number of $\Sigma^{(2)}$ and $\Sigma^{(1)}$ mod $N$ inside $\Sigma^{(3)}$.
We assume that $\Sigma^{(3)}$ is oriented, and $b^{(1)}$ and $b^{(2)}$ are discrete $\mathbb{Z}_N$ 1-form and 2-form gauge fields living on the condensation defect, which are Poincar\'e dual to $\Sigma^{(2)}$ and $\Sigma^{(1)}$ inside $\Sigma^{(3)}$, respectively.
When we take a cup product between either $F/2\pi$ or $d\theta/2\pi$ with a discrete gauge field, we are effectively treating the former as $\mathbb{Z}_N$ cocycles by taking their values modulo $N$.

We now show that this is actually a trivial condensation defect on any closed 3-manifold. 
By performing a field redefinition, $\tilde{b}^{(1)} \equiv b^{(1)} - pd\theta/2\pi$, $\tilde{b}^{(2)} \equiv b^{(2)} + F/2\pi$, we rewrite \eqref{eq:cond1} as
\begin{align}
	\begin{split}
		\mathcal{C}(\Sigma^{(3)}) = \left(
		\frac{1}{|H^1|}
		\sum_{\substack{\tilde{b}^{(1)}\in H^1 \\  \tilde{b}^{(2)}\in H^2}} \exp \left(
		\frac{2\pi i}{N} \oint_{\Sigma^{(3)}} \tilde{b}^{(1)} \cup \tilde{b}^{(2)}
		\right)
		\right) \times
		\exp \left(
		\frac{2\pi i p}{N} \oint_{\Sigma^{(3)}} \frac{d\theta}{2\pi} \wedge \frac{F}{2\pi}
		\right) \,.
	\end{split}
\end{align}
The factor in the first pair of parentheses is the partition function of a decoupled 2+1d invertible field theory, whose value evaluates to 1 on every closed 3-manifold $\Sigma^{(3)}$. 
Next, using   \eqref{Je}, we can rewrite the condensation defect as
\ie\label{trivialc}
\mathcal{C}(\Sigma^{(3)})=\exp \left(
        \frac{2\pi i p}{N} \oint_{\Sigma^{(3)}} \frac{d\theta}{2\pi} \wedge \frac{F}{2\pi}
    \right)=\exp\left(
\frac{2\pi i p}{N} \oint_{\Sigma^{(3)}}d\star J^{(2)}_{\text{electric}}\right)=1\,,
\fe
where in the last equality we have used   the fact that $\Sigma^{(3)}$ is closed and $J^{(2)}_\text{electric}$ is a well-defined operator.
We conclude that the axion-Maxwell theory is self-dual under this particular 1-gauging.

Recall that if a QFT is self-dual under a higher gauging, then we expect the higher quantum symmetry defects living on the trivial condensation defect to be identified with the original symmetry defects from the bulk. 
We now verify this for the axion-Maxwell theory. 
The higher quantum symmetry defects in this case are the Wilson lines/surfaces for the discrete gauge fields $b^{(1)}$ and $b^{(2)}$ living on $\mathcal{C}(\Sigma^{3})$. Using the equations of motion of $b^{(2)}$ and $b^{(1)}$ on the trivial condensation defect \eqref{eq:cond1}, we  identify the insertion of the Wilson line $\exp(i\oint_{\Sigma^{(1)}} b^{(1)})$ with the $\mathbb{Z}_N^{{(2)}}$ 2-form symmetry operator $\eta^\text{(w)}_{2\pi p/N} (\Sigma^{(1)})$, and the insertion of the Wilson surface $\exp(i\oint_{\Sigma^{(2)}} b^{(2)})$ with the $\mathbb{Z}_N^{{(1)}}$ 1-form symmetry operator $\eta^\text{(m)}_{-2\pi/N} (\Sigma^{(2)})$. This is consistent with the expectation.

\subsection{Half Higher Gauging in Axion-Maxwell Theory}

As the axion-Maxwell theory is self-dual under the 1-gauging of the $\mathbb{Z}_N^{(1)} \times \mathbb{Z}_N^{(2)} $ symmetry in \eqref{eq:invdefects}, we can proceed to construct the non-invertible 1-form symmetry $\mathcal{D}^{(1)}_{p/N}$ from half higher gauging. 
Since half higher gauging is a topological manipulation, this alternative construction rigorously proves that $\mathcal{D}^{(1)}_{p/N}$ is topological.

For this purpose, we place the condensation defect  on a 3-manifold $\Sigma^{(3)}$ with a boundary, $\partial \Sigma^{(3)} = \Sigma^{(2)} $.
We impose the Dirichlet boundary conditions for the discrete $\mathbb{Z}_N$ gauge fields $b^{(1)}$ and $b^{(2)}$ living on the defect.
The condensation defect is now
\begin{align} \label{eq:halfcond}
	\begin{split}
		\mathcal{C}(\Sigma^{(3)},\partial \Sigma^{(3)})\equiv \frac{1}{|H^1_\partial|} 
		\sum_{\substack{b^{(1)}\in H^1_\partial \\  b^{(2)}\in H^2_\partial}} \exp\left[ 
		\frac{2\pi i}{N}\int_{\Sigma^{(3)}} \left( b^{(1)} \cup b^{(2)} + b^{(1)} \cup \frac{F}{2\pi} - p \, b^{(2)} \cup \frac{d\theta}{2\pi}
		\right)
		\right] \,,
	\end{split}
\end{align}
where $H^i_\partial$ is an abbreviation for the relative cohomology group $H^i (\Sigma^{(3)},\partial \Sigma^{(3)};\mathbb{Z}_N)$ of the pair $(\Sigma^{(3)},\partial \Sigma^{(3)})$. 
The elements of this cohomology group are gauge inequivalent classes of $\mathbb{Z}_N$ $i$-cocycles that vanish on the boundary $\partial \Sigma^{(3)}$, i.e.,  discrete gauge field configurations satisfying the Dirichlet boundary condition.
The expression \eqref{eq:halfcond} is manifestly topological.

Even though ${\cal C}$ is trivial on a closed 3-manifold, we will see that it is nontrivial on a 3-manifold with boundary. 
Similar to before, we rewrite \eqref{eq:halfcond} as
\begin{align} \label{eq:halfcond2}
	\begin{split}
		\mathcal{C}(\Sigma^{(3)},\partial \Sigma^{(3)}) = &\,\exp \left(
		\frac{2\pi i p}{N} \int_{\Sigma^{(3)}} \frac{d\theta}{2\pi} \wedge \frac{F}{2\pi}
		\right)\times 
		\\ &\,
		\frac{1}{|H^1_\partial|}
		\sum_{\substack{b^{(1)}\in H^1_\partial \\  b^{(2)}\in H^2_\partial}} \exp \left[
		\frac{2\pi i}{N} \int_{\Sigma^{(3)}} \left(b^{(1)} - p\frac{d\theta}{2\pi}\right) \cup \left(b^{(2)}+\frac{F}{2\pi}\right)\right] \,.
	\end{split}
\end{align}
Using \eqref{Je} 
and  Stokes' theorem, the first line in \eqref{eq:halfcond2} becomes 
\begin{equation} \label{eq:stokes}
    \exp \left(
            \frac{2\pi i p}{N} \int_{\Sigma^{(3)}} \frac{d\theta}{2\pi} \wedge \frac{F}{2\pi}
        \right) 
        = \exp \left( 
          \frac{2\pi ip}{N} \oint_{\Sigma^{(2)}} \star J_{\text{electric}}^{(2)}
        \right) \,.
\end{equation}

Next, as discussed in the previous subsection,  the second line in \eqref{eq:halfcond2} is the partition function of a 2+1d invertible theory, which  evaluates to 1 on any closed 3-manifold.
However, when $\partial \Sigma^{(3)} \neq \emptyset$, this reduces to (the partition function of) a 1+1d QFT living on $\Sigma^{(2)} = \partial \Sigma^{(3)}$, coupled to the bulk axion and $U(1)$ gauge fields.
In Appendix \ref{app:2dZN}, we show that with the Dirichlet boundary conditions imposed on $b^{(1)}$ and $b^{(2)}$, the invertible field theory on $\Sigma^{(3)}$ reduces to a 1+1d $\mathbb{Z}_N$ gauge theory on the boundary $\Sigma^{(2)}$.
In particular, there we derive
\begin{align} \label{eq:ZN}
	\begin{split}
		{\frac{1}{|H^1_\partial|}}\sum_{\substack{b^{(1)}\in H^1_\partial \\  b^{(2)}\in H^2_\partial}} &\,\exp \left[
		\frac{2\pi i}{N} \int_{\Sigma^{(3)}} \left(b^{(1)} - p\frac{d\theta}{2\pi}\right) \cup \left(b^{(2)}+\frac{F}{2\pi}\right)
		\right] \\
		= 
		\int [D\phi\,D c]_{\Sigma^{(2)}}
		&\,\exp \left[ \oint_{\Sigma^{(2)}}\left(
		\frac{iN}{2\pi} \phi dc + \frac{ip}{2\pi} \theta dc
		+ \frac{i}{2\pi} \phi F
		\right)
		\right]
	\end{split}
\end{align}
where $\phi \sim \phi + 2\pi$ is a periodic scalar and $c$ is a $U(1)$ 1-form gauge field, both living only on $\Sigma^{(2)} = \partial \Sigma^{(3)}$.

Combining \eqref{eq:stokes} and \eqref{eq:ZN}, we obtain
\begin{align}
\begin{split}
    &\mathcal{C}(\Sigma^{(3)},\partial \Sigma^{(3)} = \Sigma^{(2)}) \\
    &= 
    \int [D\phi\,D c]_{\Sigma^{(2)}}
    \exp \left[
        i\oint_{\Sigma^{(2)}} \left(
            \frac{2\pi p}{N} \star J^{(2)}_{\text{electric}} + \frac{N}{2\pi} \phi dc + \frac{p}{2\pi} \theta dc + \frac{1}{2\pi} \phi F
        \right)
    \right] \\
    &= \mathcal{D}^{(1)}_{p/N}(\Sigma^{(2)}) \,.
\end{split}
\end{align}
The half 1-gauging on $\Sigma^{(3)}$ precisely reproduces the non-invertible 1-form symmetry defect $\mathcal{D}^{(1)}_{p/N}$ \eqref{eq:electric_worldvolume} on the boundary $\Sigma^{(2)} =\partial \Sigma^{(3)}$ of the trivial condensation defect as claimed.

We conclude that the non-invertible 1-form symmetry  $\mathcal{D}^{(1)}_{p/N}$ in the axion-Maxwell theory can be constructed from  half higher gauging of the $\mathbb{Z}_N^{(1)}\times\mathbb{Z}_N^{(2)}$ global symmetry.
Since half higher gauging with the Dirichlet boundary condition always produces a manifestly topological defect, this proves that $\mathcal{D}^{(1)}_{p/N}$  is  topological.

\section{Action of Non-Invertible Symmetries} \label{sec:action}

In this section, we analyze the action of non-invertible symmetries in the axion-Maxwell theory on other operators and defects.
In particular, we will see that the non-invertible 1-form symmetry $\mathcal{D}^{(1)}_{p/N}$ acts non-invertibly on the worldline of monopoles (a.k.a., the 't Hooft line) and the worldsheet of axion strings.

Let us introduce some notations. For the Wilson line, we define  $W^q \equiv \exp \left(iq \oint A \right)$. 
Next,  we use   $H_m$ to stand for \textit{any} 't Hooft line with magnetic charge $m$ under $U(1)^{(1)}_\text{magnetic}$. 
Note that the choice of $H_m$ is far from unique, but most of our conclusions below hold universally true.  
(For example, one can always stack  a decoupled quantum mechanics with the worldline of $H_m$ to produce another line defect with the same quantum number.) 
Similarly, we use    $S_w$ to stand for \textit{any} axion string worldsheet with winding charge $w$ under $U(1)^{(2)}_\text{winding}$.

We  denote the charge 1 Wilson line, charge 1 't Hooft line, and charge 1 axion string worldsheet by $W \equiv \exp (i\oint A), H \equiv H_1$, and $S \equiv S_1$, respectively.  
We will sometimes refer to these defects $W,H,S$ as the minimal Wilson, 't Hooft, and axion string defects, respectively. 
For  $H$ and $S$, we emphasize again that the choice is not unique, and by ``minimal" we only mean that their quantum numbers take  the smallest possible values.

The quantum numbers  $q$, $m$ and $w$ are all integers. 
While $m$ of the 't Hooft line and $w$ of the axion string are respectively the charges of the $U(1)^{(1)}_\text{magnetic}$ and $U(1)^{(2)}_\text{winding}$ global symmetries, the role of the ``electric charge" $q$ for the Wilson line is more obscure. 
We will see below that $q$ is related to the eigenvalue of the non-invertible 1-form global symmetry ${\cal D}^{(1)}_{p/N}$.

 Generally for a non-invertible symmetry, there are two related but different ways to define its action on other operators and defects \cite{Chang:2018iay}. 
For the non-invertible 1-form symmetry, they are discussed respectively in Sections \ref{sec:D1action} and \ref{sec:gauss}, with the latter leading to a non-invertible Gauss law.

\subsection{Action of Non-Invertible 0-form Symmetry}

The non-invertible 0-form symmetry defect $\mathcal{D}^{(0)}_{p/N}$ acts on the operator $e^{i\theta}$ as well as on the 't Hooft line, which can be understood from the half gauging construction.
This was discussed in \cite{Choi:2022jqy}, and we briefly review it here.

The axion field $\theta$ is shifted by $2\pi p/N$ under the action of $\mathcal{D}^{(0)}_{p/N}$ due to the $\star J^{(1)}_{\text{shift}}$ term in \eqref{eq:0form}.
That is, as we sweep  $\mathcal{D}^{(0)}_{p/N}$ past  $e^{i\theta}$, we have
\begin{equation}
    \mathcal{D}^{(0)}_{p/N}\,:\quad
    \exp\left(i\theta \right) \mapsto \exp\left(
        \frac{2\pi ip}{N}
    \right)
    \exp\left(i\theta \right) \,.
\end{equation}
So $\mathcal{D}^{(0)}_{p/N}$ acts invertibly on $e^{i\theta}$ as a $\mathbb{Z}_N^{(0)}$ symmetry.

In contrast, $\mathcal{D}^{(0)}_{p/N}$ acts non-invertibly on 't Hooft lines.
Let $H(\gamma)$ be the minimal 't Hooft line on a loop $\gamma$.
For simplicity, we assume $\gamma$ to be contractible, and $\gamma = \partial \Sigma^{(2)}$ for some 2-dimensional surface $\Sigma^{(2)}$.
Then, as we sweep $\mathcal{D}^{(0)}_{p/N}$ past the 't Hooft line $H(\gamma)$, we have
\begin{equation}
    \mathcal{D}^{(0)}_{p/N}\,:\quad
    H(\gamma) \mapsto H(\gamma)\exp \left(
        \frac{2\pi i p}{N}\int_{\Sigma^{(2)}} \frac{F}{2\pi}
    \right)\,.
\end{equation}
For the derivation of this result and for the case of non-contractible $\gamma$, see \cite{Choi:2022jqy,Cordova:2022ieu}.
In particular, the minimal 't Hooft line is annihilated when surrounded by the $\mathcal{D}^{(0)}_{p/N}$ defect,  demonstrating the non-invertible nature of the latter. 
More generally, when $\mathcal{D}^{(0)}_{p/N}$ surrounds a non-minimal 't Hooft line $H_m$, the line is annihilated if $m\neq 0\text{ mod }N$, and turns into $H_m W^{pm/N}$ if $m=0\text{ mod } N$.

\begin{figure}[!t]
    \centering
    \includegraphics[width=0.8\textwidth]{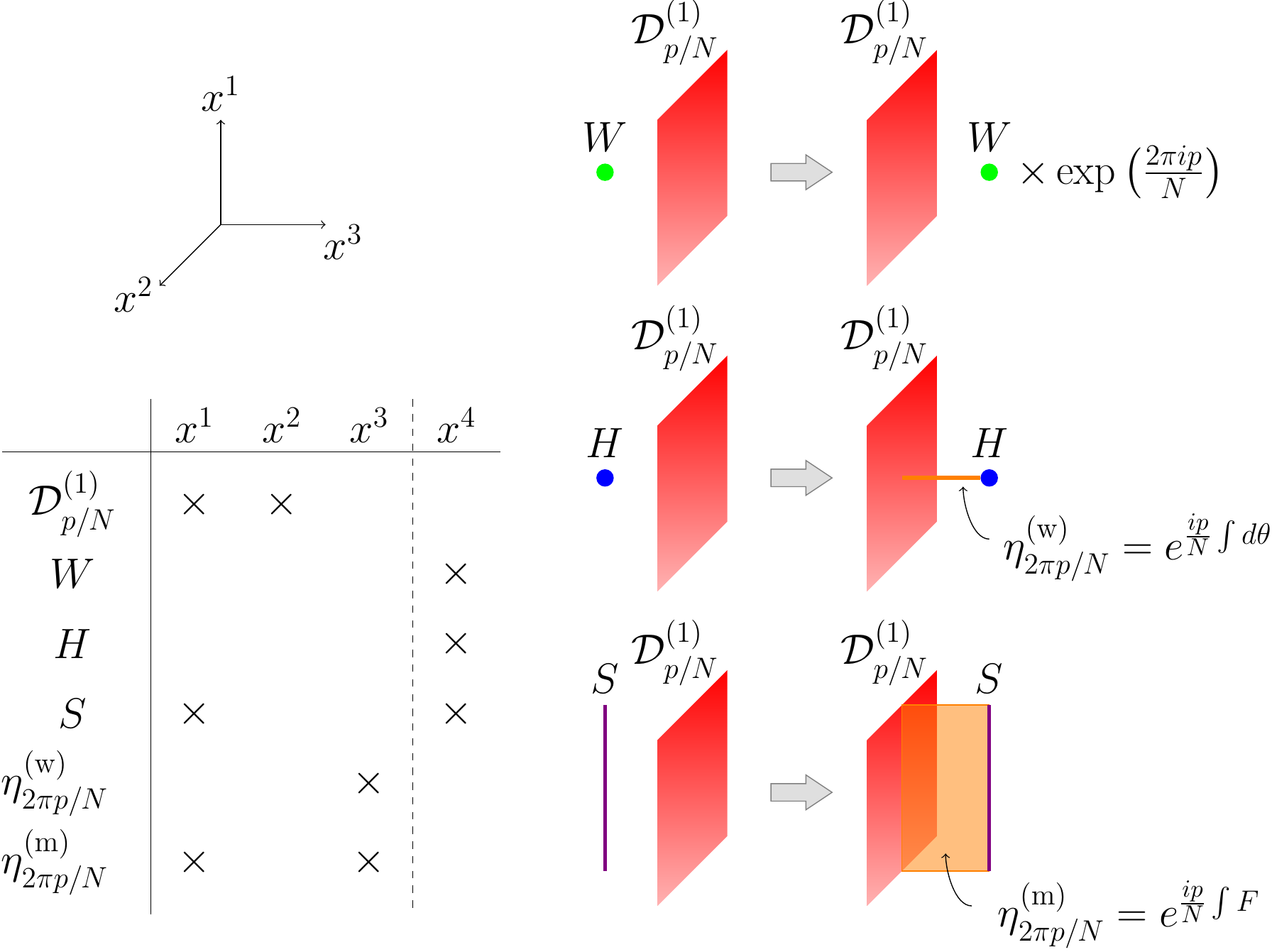}
    \caption{The action of the non-invertible 1-form symmetry defect $\mathcal{D}^{(1)}_{p/N}$ on extended operators/defects. 
    $W$, $H$, $S$ are the minimal Wilson line, 't Hooft line, and axion string worldsheet, respectively.
    For simplicity, we assume that the spacetime manifold is locally  $\mathbb{R}^4$. The directions along which various operators extend are summarized in the table, and the $x^4$-direction is suppressed in the drawing. The non-invertible 1-form symmetry $\mathcal{D}^{(1)}_{p/N}$ acts on the Wilson line $W$ invertibly by a phase, but it acts on the 't Hooft line $H$ and the axion string worldsheet $S$ non-invertibly.}
    \label{Fig:action}
\end{figure}

\subsection{Action of Non-Invertible 1-form Symmetry}\label{sec:D1action}

We move on to discuss the action of the non-invertible 1-form symmetry $\mathcal{D}^{(1)}_{p/N}$.
Since $\mathcal{D}^{(1)}_{p/N}$ is a codimension-2 topological operator, it can  act on extended operators of dimension 1 or higher.
On the other hand, it necessarily acts trivially on all local operators  as we can always shrink  $\mathcal{D}^{(1)}_{p/N}$ without crossing the local operator insertions.
Indeed, below we will show that $\mathcal{D}^{(1)}_{p/N}$ acts invertibly on the Wilson lines, but non-invertibly on the 't Hooft lines  and axion string worldsheets.
The results are summarized in Figure \ref{Fig:action}.

To begin with, the non-invertible 1-form symmetry  acts on Wilson lines invertibly, just like an ordinary electric 1-form symmetry. 
As we move $\mathcal{D}^{(1)}_{p/N}$  past the Wilson line as in Figure \ref{Fig:action}, we get a phase  $\exp \left( \frac{2\pi i pq}{N} \right)$:
\begin{equation}\label{eq:action_Wilson}
    \mathcal{D}^{(1)}_{p/N} \,:\quad 
    W^q \mapsto W^q \times \exp\left( \frac{2\pi i p q}{N}\right) \,.
\end{equation}
This follows from the first term $\star J^{(2)}_\text{electric}$ in \eqref{eq:electric_worldvolume}, which implements an invertible 1-form symmetry action.

The non-invertible 1-form symmetry $\mathcal{D}^{(1)}_{p/N}$ acts on the 't Hooft lines and the axion string worldsheets as well.
The action is more intricate in these cases, and in particular, it will be non-invertible.
To determine the action of $\mathcal{D}^{(1)}_{p/N}$ on the 't Hooft lines and the axion string worldsheets, it is convenient to utilize the half higher gauging construction in Section \ref{sec:gauging}.

The action can be determined by considering an effective theory on the 2+1d worldvolume $\Sigma^{(3)}$ of the trivial condensation defect ${\cal C}(\Sigma^{(3)})$. 
The non-invertible symmetry defect $\mathcal{D}^{(1)}_{p/N}$ is realized at the boundary of  $\mathcal{C}(\Sigma^{(3)})$ supported on $\Sigma^{(3)}$. 
From the $\Sigma^{(3)}$ point of view, ${\cal D}^{(1)}_{p/N}$ is a non-invertible 0-form symmetry from half gauging a $\mathbb{Z}_N^{(0)}\times \mathbb{Z}_N^{(1)}$ symmetry.\footnote{Recall that the quantum symmetry of a discrete 0-form symmetry is a 1-form symmetry in 2+1d, and vice versa \cite{Gaiotto:2014kfa,Tachikawa:2017gyf}. Therefore, there are 2+1d QFTs invariant under gauging the product of a 0-form symmetry and a 1-form symmetry. The simplest example is to take any QFT $\cal Q$ with a $\mathbb{Z}_N^{(0)}$ global symmetry, and consider the product QFT ${\cal Q} \times {\cal Q}/\mathbb{Z}_N^{(0)}$. This product QFT is invariant under gauging $\mathbb{Z}_N^{(0)}\times \mathbb{Z}_N^{(1)}$.  Examples of non-invertible duality defects from the invariance of gauging a product of a 0-form and a 1-form symmetries have been discussed in \cite{Kaidi:2021xfk}. See footnote \ref{fn:product}.} The 't Hooft line $H$ and axion string worldsheet $S$ are the charged objects under this effective  $\mathbb{Z}_N^{(0)}\times \mathbb{Z}_N^{(1)}$ symmetry. 
They  are 0d and 1d objects from the $\Sigma^{(3)}$ point of view. 
See  Figure \ref{Fig:action} for the configuration of these operators and defects, with $\Sigma^{(3)}$ defined by $x^3\ge 0$ and $x^4=0$.

Therefore, as we bring ${\cal D}^{(1)}_{p/N}$ past the 't Hooft line $H_m$, the latter becomes not gauge invariant and is attached to a Wilson line associated with the effective $\mathbb{Z}_N^{(0)}$ gauge symmetry   from the $\Sigma^{(3)}$ point of view. 
As discussed in the paragraph below \eqref{trivialc}, this Wilson line $\exp \left( i m \int b^{(1)} \right)$ is identified with $\eta^\text{(w)}_{2\pi pm /N}$ in the 3+1d bulk. 
We thus obtain the following action on the 't Hooft line $H_m$:
\begin{equation} \label{eq:action_tHooftm}
    \mathcal{D}^{(1)}_{p/N} \,:\quad 
    H_m \mapsto H_m \times \exp\left( \frac{ i p m}{N} \int d\theta \right) \,,
\end{equation}
where the $\exp\left( \frac{i pm}{N} \int d\theta \right)$ line is stretched between the 't Hooft line and $\mathcal{D}^{(1)}_{p/N}$. Any potential phase on the righthand side of \eqref{eq:action_tHooftm} can be absorbed by redefining the 0-dimensional junction between the winding symmetry line and the 't Hooft line.  
The action is non-invertible if  $m \neq 0$ mod $N$ in the sense the action creates a topological line defect $ \exp\left( \frac{ i p m}{N} \int d\theta \right)$. 
In Section \ref{sec:gauss} we give a complimentary interpretation of the non-invertible action and discuss the $m=0$ mod $N$ case there.

\begin{figure}[!t]
    \centering
    \includegraphics[width=\textwidth]{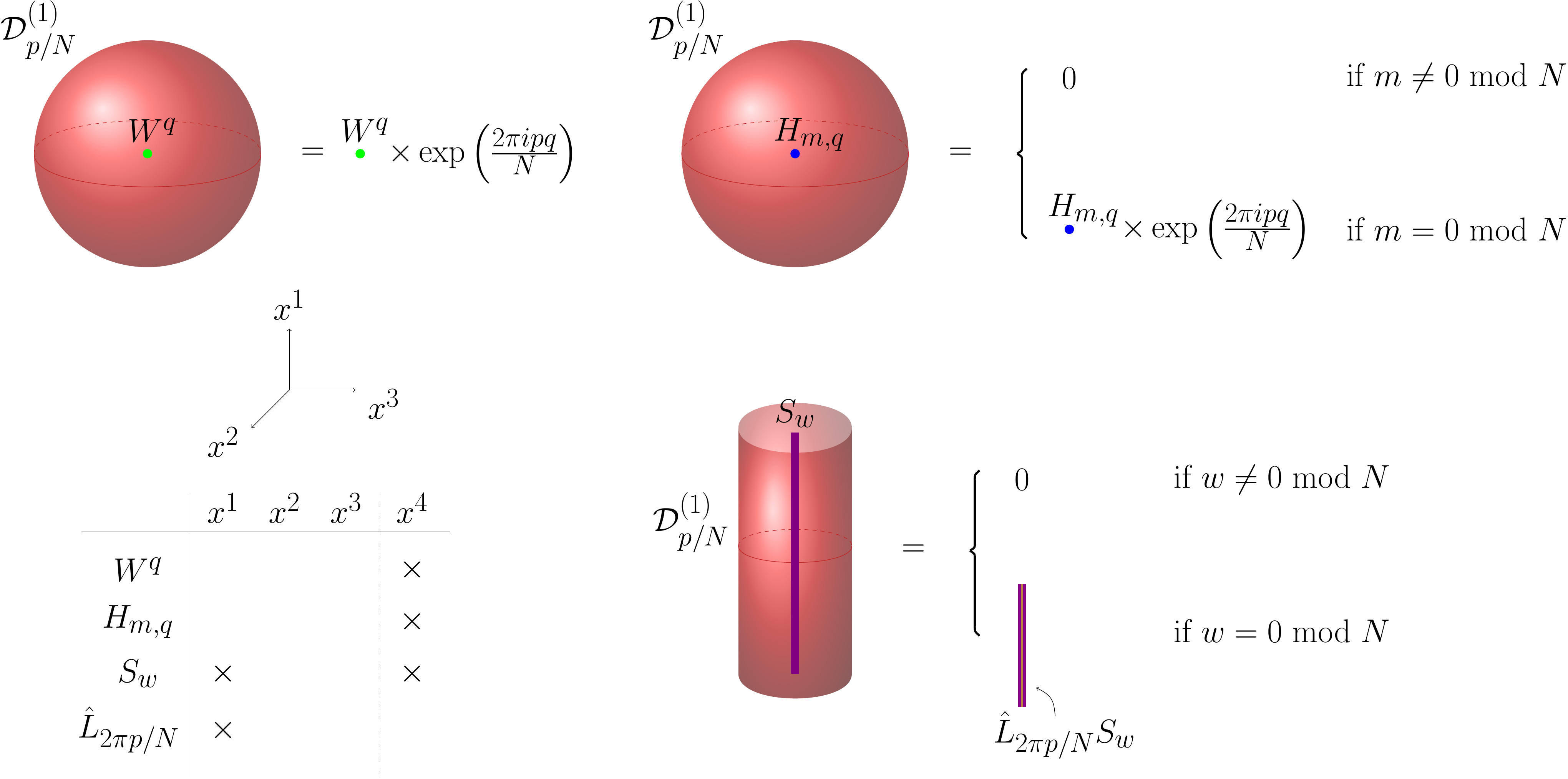}
    \caption{The non-invertible Gauss law. The non-invertible operator ${\cal D}^{(1)}_{p/N}$ measures the electric charge invertibly on the Wilson line $W^q$. On the other hand, the dyonic line $H_{m,q}$ is annihilated when measured by  $\mathcal{D}^{(1)}_{p/N}$ if $m\neq 0$ mod $N$. Physically, it means that the electric charge $q$ of a dyon $H_{m,q}$ with magnetic charge $m$ is only well-defined modulo $m$, i.e., $q\sim q+m$. 
    Similarly,  the  axion string worldsheet $S_w$ is annihilated by ${\cal D}^{(1)}_{p/N}$ if $w\neq 0$ mod $N$.}
    \label{Fig:action2}
\end{figure}

Similarly, as we bring ${\cal D}^{(1)}_{p/N}$ past the axion string $S_w$, the latter is no longer gauge invariant because it carries charge $-pw$ under the $\mathbb{Z}_N^{(2)}$ symmetry in \eqref{eq:invdefects}. 
From the $\Sigma^{(3)}$ point of view, the axion string  is attached to a charge $-pw$ Wilson surface associated with the effective $\mathbb{Z}_N^{(1)}$ gauge symmetry. 
As discussed in  the paragraph below \eqref{trivialc}, this Wilson surface $\exp\left( - i p w\int b^{(2)}\right)$ is identified with  $\eta^\text{(m)}_{2\pi pw/N}$ in the 3+1d bulk. 
We thus obtain the following action on the axion string worldsheet $S_w$:
\begin{equation} \label{eq:action_stringw}
    \mathcal{D}^{(1)}_{p/N} \,:\quad 
    S_w \mapsto S_w \times \exp\left( \frac{ i p w}{N} \int F \right) \,,
\end{equation}
where the surface $\exp\left( \frac{ i pw}{N} \int F \right)$ is stretched between the axion string worldsheet and the defect $\mathcal{D}^{(1)}_{p/N}$.
The action is non-invertible if $w \neq 0$ mod $N$, in the sense that the action creates a topological surface defect $\exp\left( \frac{ i p w}{N} \int F \right)$.

\subsection{Non-Invertible Gauss Law}\label{sec:gauss}

In ordinary electromagnetism, Gauss law states that the total electric charge  can be measured by a closed surface integral $Q_\text{Maxwell}  =\oint_{\Sigma^{(2)}} \star F$.  
The electric charge is quantized and topological -- Gauss law $d\star F=0$  implies that $Q_\text{Maxwell}$  depends topologically on the choice of the surface $\Sigma^{(2)}$.

In axion-Maxwell theory, the Gauss law is anomalous \eqref{Je} \cite{Fischler:1983sc}, and hence $Q_\text{Maxwell}$ is not topological.  
That is, there is no conserved, gauge-invariant, and quantized electric charge. 
This point was emphasized, for example, in \cite{Marolf:2000cb} (see also \cite{Bachas:2000ik,Taylor:2000za}).  
Instead, what we have is the non-invertible 1-form symmetry ${\cal D}^{(1)}_{p/N}(\Sigma^{(2)})$, which is both topological (and in particular conserved) and gauge-invariant.\footnote{The ``Maxwell charge" discussed in \cite{Marolf:2000cb} is conserved and gauge-invariant, but it is only defined at infinity with a specific fall-off condition on the fields,  and is therefore not topological. In contrast, our ${\cal D}^{(1)}_{p/N}$ is gauge-invariant and topological. It can be defined on any closed 2-manifold.}

We can surround ${\cal D}^{(1)}_{p/N}(S^2)$ around a Wilson line, which is a point in space. 
See Figure \ref{Fig:action2} for this configuration, which can be deformed from Figure \ref{Fig:action}. 
Denote this action by $\cdot$, we have\footnote{Here we choose an Euler counterterm such that the expectation value of ${\cal D}^{(1)}_{p/N}$ on $S^2$ with no other operator insertions is 1. See \eqref{S2qdim}.} 
\ie\label{dotactionW}
{\cal D}^{(1)}_{p/N} \cdot W^q  = e^{2\pi i pq /N} W^q
\fe  
In this sense, ${\cal D}^{(1)}_\alpha$ behaves as $``\exp (i\alpha Q_\text{Page}  )"$ on the Wilson lines (see \eqref{eq:electric_U_hat}). 
However, $Q_\text{Page} = \oint_{\Sigma^{(2)}}(\star J_\text{electric}^{(2)} -{1\over 4\pi^2} \theta dA)$ is not a well-defined operator, but the non-invertible 1-form symmetry is.

\begin{figure}[!t]
    \centering
    \includegraphics[width=0.9\textwidth]{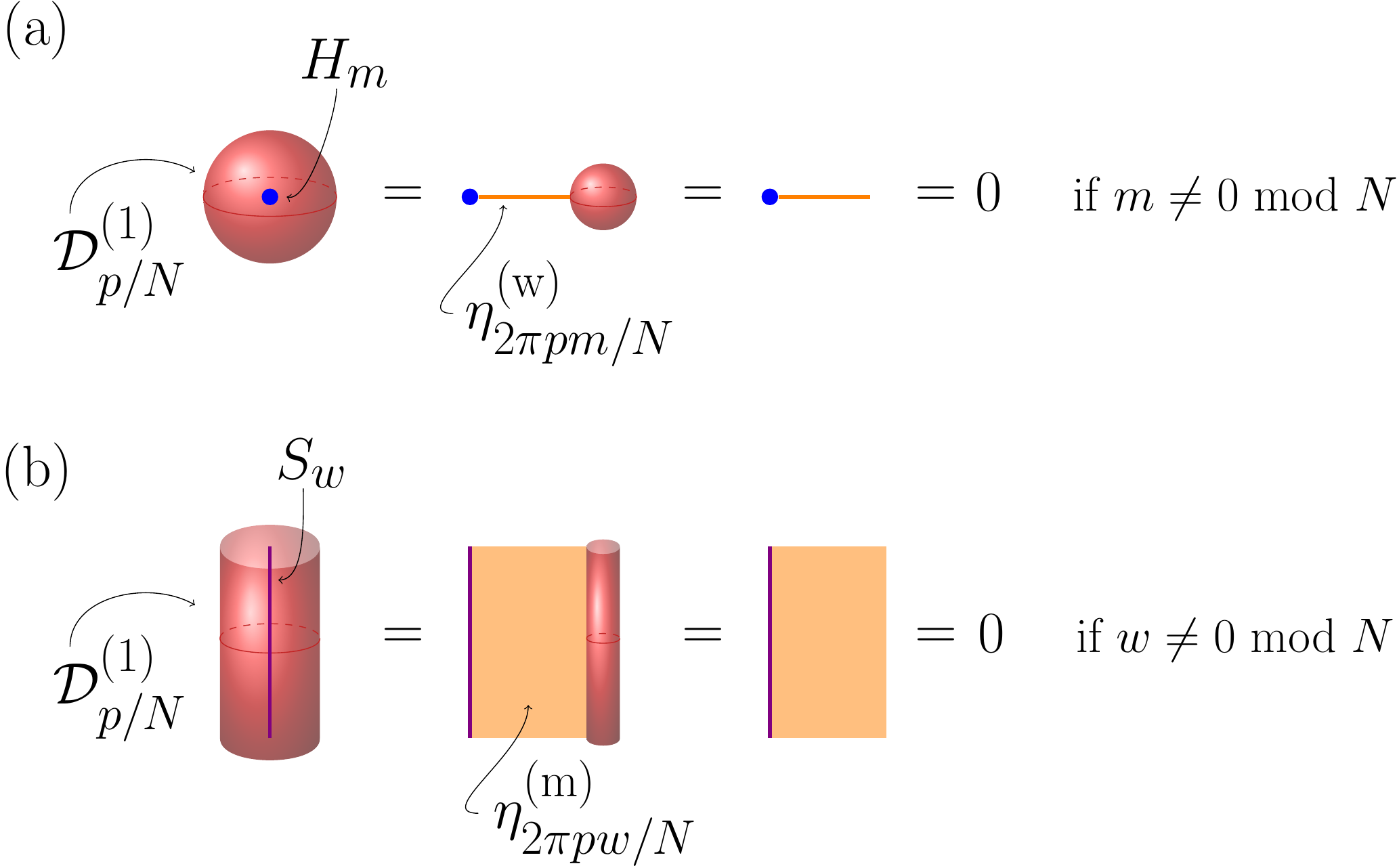}
    \caption{(a) Non-invertible action of $\mathcal{D}^{(1)}_{p/N}$ on the 't Hooft line $H_m$ for $m\neq 0$ mod $N$. We first pull $\mathcal{D}^{(1)}_{p/N}$ past $H_m$, and then shrink the former to a point. This  generates a putative topological endpoint for the topological line $\eta^{\text{(w)}}_{2\pi pm/N}$. However, the line $\eta^{\text{(w)}}_{2\pi pm/N}$ does not admit any topological endpoint, since it acts faithfully on the other operators \cite{Chang:2018iay}. Therefore, such a configuration results in a vanishing correlation function. (b) Non-invertible action of $\mathcal{D}^{(1)}_{p/N}$ on the axion string worldsheet $S_w$ for $w\neq 0$ mod $N$. Similar to before, we first pull  $\mathcal{D}^{(1)}_{p/N}$ past $S_w$, and then shrink the former to a line. This generates a putative topological boundary line for the topological surface $\eta^{\text{(m)}}_{2\pi pw/N}$, which does not exist. Therefore, such a configuration leads to a vanishing correlation function. Both the 't Hooft line and the axion string worldsheet are extended in time, and they are point and line in space as shown in the figure, respectively.}
    \label{Fig:tadpole}
\end{figure}

Next, from \eqref{eq:action_tHooftm}, it follows that  ${\cal D}^{(1)}_{p/N}(S^2)$ annihilates the 't Hooft line $H_m$ with $m\neq 0$ mod $N$.
This can be derived by closing the non-invertible 1-form symmetry defect to the left in Figure \ref{Fig:action} on the righthand side, creating an empty bubble at which the winding symmetry line $\eta^\text{(w)}_{2\pi p/N}$ terminates. 
However, by shrinking the bubble, it gives  a topological endpoint of the winding symmetry line, which cannot exist because the the latter acts faithfully.
It means that any such configuration gives zero correlation function, which is analogous to the ``vanishing tadpole'' condition discussed in \cite{Chang:2018iay}.
This is  why equating ${\cal D}^{(1)}_\alpha$ with $``\exp (i \alpha Q_\text{Page}  )"$ (with $\alpha=2\pi p/N$) is only true for Wilson lines, but not for 't Hooft lines. 
${\cal D}^{(1)}_{p/N}(S^2)$ is a non-invertible operator since it has a kernel.
See Figure \ref{Fig:tadpole}(a) for the illustration.

If $m = 0$ mod $N$, the line $\exp({ipm\over N} \int d\theta)$ becomes trivial, leaving behind a topological point operator on $H_m$. 
If we assume $H_m$ to be a simple line, i.e., it cannot be decomposed into a direct sum of other line defects, then there is a unique topological point operator on $H_m$ (see, for example, \cite{Chang:2018iay}). 
Then $\mathcal{D}^{(1)}_{p/N}$ acts invertibly on $H_m$, with a possible phase factor $e^{ 2\pi i pq/N}$ reflecting  a nontrivial electric charge $q$. 
To keep track of this quantum number, we use a more refined notation and denote such a  simple, dyonic line of magnetic charge $m$ and electric charge $q$ by $H_{m,q}$. 
From this point on, we define $H_m = H_{m,0}$.  
Importantly, the electric charge $q$ is only defined modulo $m$, i.e., 
\ie
q\sim q+m\,.
\fe 
To conclude, the non-invertible 1-form symmetry acts on the dyonic line $H_{m,q}$ as
\ie\label{dotactionH}
{\cal D}^{(1)}_{p/N}  \cdot H_{m,q}
=\begin{cases}
0~~~~~~~~&\text{if}~~m\neq 0 ~\text{mod}~N\\
e^{2\pi i pq/N}H_{m,q}~~~~&\text{if}~~m=0 ~\text{mod}~N
\end{cases}
\fe
We dub \eqref{dotactionW} and \eqref{dotactionH} as  the \textit{non-invertible Gauss law}.

There is a simple physical interpretation of the non-invertible Gauss law. 
A monopole with magnetic charge $m$ can gain $m$ units of electric charge by going around an axion string, which implements the Witten effect. Therefore,  the electric charge is conserved only modulo $m$ in the presence of a charge $m$ 't Hooft line.  
This conservation law is exactly captured by the non-invertible Gauss law measured by the non-invertible 1-form symmetry.
(See \cite{Chen:2022cyw} for an analogous conservation law in a different context.)

Consider a configuration of multiple dyonic lines $H_{m_i,q_i}$ in a region in space and extended in time.  Unlike an ordinary invertible symmetry, the action of the non-invertible 1-form symmetry on this region does not generally reduce to the product of the individual actions, i.e., ${\cal D}^{(1)}_{p/N}\cdot \left( \prod_i H_{m_i ,q_i} \right)\neq \prod_i {\cal D}^{(1)}_{p/N}\cdot H_{m_i,q_i}$.
This is because as we break ${\cal D}^{(1)}_{p/N}$ into smaller spheres encircling individual dyons, we need to perform a nontrivial crossing move. 
The crossing relation between the non-invertible 1-form symmetry defects generally produces other defects connecting them. 
This is analogous to the move in Figure 20 of \cite{Koide:2021zxj} in 3+1d and to that in Figure 30 of \cite{Chang:2018iay} in the Ising CFT. 
We leave the study of the crossing between ${\cal D}^{(1)}_{p/N}$ for the future.

Similarly, we can define an action of the non-invertible 1-form symmetry on the axion string worldsheet as in Figure \ref{Fig:action2}. 
This action is non-invertible
\ie \label{eq:inv_action_string}
{\cal D}^{(1)}_{p/N}\cdot S_w 
=\begin{cases}
0~~~~~~~~~~~~~~~~~~~~~~~~~~~~~~~~~~~~~~~&\text{if}~~w\neq 0 ~\text{mod}~N\\
\hat{L}_{2\pi p / N}(\gamma)\, S_w~~\,~~~&\text{if}~~w=0 ~\text{mod}~N\,,
\end{cases}
\fe
where $\gamma$ is the 1d curve that ${\cal D}^{(1)}_{p/N}$ shrinks towards on $S_w$. 
Here $\hat{L}_{2\pi p/N}$ is a topological line that only lives on the axion string worldsheet, which we will discuss in Section \ref{sec:junction}. 
The case of $w\neq 0 $ mod $N$ can be derived by closing the  $\mathcal{D}^{(1)}_{p/N}$ defect to the left in Figure \ref{Fig:action} on the righthand side and using the fact that the magnetic symmetry surface $\eta^\text{(m)}_\alpha$ does not admit a topological boundary condition.
See Figure \ref{Fig:tadpole}(b) for the illustration.

\section{Selection Rules of Monopoles and  Strings} \label{sec:selection_rules}

The action of $\mathcal{D}^{(1)}_{p/N}$ discussed in Section \ref{sec:action} implies the existence of various topological junctions between the defects.
For instance, the action on the 't Hooft lines implies that the winding symmetry defect $\eta^\text{(w)}_\alpha$ can end topologically on an 't Hooft line.
Similarly, the action on the axion string worldsheets implies that the magnetic symmetry defect $\eta^\text{(m)}_\alpha$ can end topologically on an axion string worldsheet.

In this section, we will explain why such topological junctions exist, and also discuss the existence of non-topological junction configurations where a Wilson line end on either an 't Hooft line or an axion string worldsheet.
This will be based on the anomaly inflow arguments on the 't Hooft lines and the axion string worldsheets.
The fact that a Wilson line is endable \cite{Rudelius:2020orz,Heidenreich:2021xpr} on other extended operators is another manifestation of the absence of an ordinary, invertible electric 1-form symmetry in the axion-Maxwell theory.

We will further derive crossing relations between various defects. 
These crossing relations obey certain consistency conditions, reminiscent of the pentagon identity in fusion category. 
(However, the participating defects are not all topological.)

Finally, we will  derive  selection rules for correlation functions involving monopoles and axion strings.
In particular, these selection rules give a global symmetry interpretation of the Witten effect \cite{Witten:1979ey} in the presence of dynamical axions.

\subsection{Worldvolume Actions and Anomaly Inflow}

We first review the electrically charged degrees of freedom on the monopoles and axion strings via anomaly inflow. 
This will be crucial later in understanding the allowed junction configurations.
The materials  in this subsection are standard and can be found in, for example, \cite{Callan:1984sa,Naculich:1987ci,Fukuda:2020imw,Fan:2021ntg,Heidenreich:2021yda}.

\subsubsection*{Monopole Worldline}

The Witten effect states that as $\theta \rightarrow \theta+2\pi$, the 't Hooft line acquires an electric charge. 
 This can be understood as the inflow of a mixed anomaly between the $S^1$ space parameterized by $\theta$ and the $U(1)$ gauge group \cite{Cordova:2019jnf,Cordova:2019uob,Fukuda:2020imw}.
For an 't Hooft line $H_m$ of magnetic charge $m$, the anomaly polynomial 3-form for this worldline anomaly is
\begin{equation} \label{eq:inflow_monopole}
    \mathcal{I}^{\text{monopole}}_3 = -\frac{m}{(2\pi)^2} d\theta \wedge F \,.
\end{equation}
Since the axion field $\theta$ is dynamical (as well as the photon field $A$), the worldline anomaly \eqref{eq:inflow_monopole}  must be canceled by additional degrees of freedom living on the monopole worldline  \cite{Fukuda:2020imw,Fan:2021ntg}.

A natural choice of a quantum mechanical system which carries an anomaly that is opposite to \eqref{eq:inflow_monopole} is that of the particle on a circle quantum mechanics \cite{Cordova:2019jnf} (see also \cite{Gaiotto:2017yup,Kikuchi:2018gfo}).
Let $\gamma$ be a closed loop where the 't Hooft line $H_m(\gamma)$ is supported on.
The worldline action of $H_m$ contains the quantum mechanical degrees of freedom of a dynamical compact scalar field $\sigma \sim \sigma + 2\pi$ \cite{Jackiw:1975ep},
\begin{equation} \label{eq:particle_on_circle}
    \int [D\sigma]_\gamma \exp \left[
        -\oint_\gamma d\tau \left(
            \frac{l_\sigma}{2} (\dot{\sigma} - m A_\tau)^2 - \frac{i}{2\pi} \theta (\dot{\sigma}-m A_\tau)
        \right)
    \right] \,,
\end{equation}
where $\tau$ is the coordinate along $\gamma$, $\dot{\sigma} \equiv d\sigma/d\tau$, and $l_\sigma$ is a parameter with the dimension of  length. 
Under the bulk gauge transformation $A \rightarrow A + d\lambda$, the $\sigma$ field transforms as $\sigma \rightarrow \sigma + m\lambda$.
Thus, \eqref{eq:particle_on_circle} is invariant under the bulk gauge transformation.
On the other hand, under $\theta \rightarrow \theta + 2\pi$, \eqref{eq:particle_on_circle} acquires an anomalous phase 
\begin{equation}
    \exp \left( im \oint_\gamma A \right) \,,
\end{equation}
which precisely cancels the anomaly inflow \eqref{eq:inflow_monopole}.
Thus, $H_m$ is a well-defined, gauge-invariant line defect.

We emphasize that the choice of the quantum mechanical degrees of freedom on the worldline of a monopole is far from unique, and the choice in \eqref{eq:particle_on_circle} is only one example.\footnote{For instance, we can define another 't Hooft line $H_m'$ by  replacing \eqref{eq:particle_on_circle} with
\ie
    \int [D\sigma]_\gamma \exp \left[
        -\oint_\gamma d\tau \left(
            \frac{l_\sigma}{2} (\dot{\sigma} -  A_\tau)^2 - \frac{im}{2\pi} \theta (\dot{\sigma}-A_\tau)
        \right)
    \right]
\fe 
with gauge transformation $\sigma \rightarrow \sigma+\lambda, A\rightarrow A+d\lambda$. 
This alternative quantum mechanics carries the same anomaly as \eqref{eq:particle_on_circle}. Let $Q= i \ell_\sigma \dot\sigma$ be the conserved charge of the $U(1)$ symmetry of the worldline quantum mechanics on $H_m'$. In $H_m'$, we have $\partial_\tau Q = -{m\over 2\pi }\partial_\tau\theta$, leading to $m$ topological point operators on the worldline:
\ie
\exp\left( {2\pi i k \over m}Q+i k \theta\right)\,,~~~k=0,1,\cdots, m-1\,.
\fe
Therefore,  $H_m'$ is not simple in the sense that it can be decomposed into sums of other lines. In contrast, the line $H_m$ using  \eqref{eq:particle_on_circle} is a simple line.}
The only requirement is that the added degrees of freedom should properly cancel the worldline anomaly \eqref{eq:inflow_monopole}.
Independent of a particular choice, we will use $H_m$ to denote any such line defect with magnetic charge $m$ and trivial electric charge (in the sense that it is left invariant  by the action of ${\cal D}^{(1)}_{p/m}$, see Section \ref{sec:gauss}).

\subsubsection*{Axion String Worldsheet}

Consider an axion string worldsheet $S_w$ with winding symmetry charge $w$ supported on a closed 2-manifold $\Sigma^{(2)}$.
Near the string, the axion field $\theta$ becomes singular and we have (see \eqref{thetawind})
\begin{equation}
    d\left(d\theta\right) = 2\pi w \delta(\Sigma^{(2)}) \,,
\end{equation}
where $\delta(\Sigma^{(2)})$ is the delta function 2-form localized on $\Sigma^{(2)}$.
However, such a singular configuration of $\theta$ is not gauge invariant.
To see this, we first integrate by parts the axion-photon coupling term to write it as
\begin{equation}
    \frac{i}{8\pi^2} \oint_{X^{(4)}} A \wedge d\theta \wedge F
\end{equation}
where $X^{(4)}$ is the 4-dimensional closed spacetime manifold.
Under the $U(1)$ gauge transformation $A \rightarrow A + d\lambda$, the axion-photon coupling term transforms as
\begin{align} \label{eq:inflow_string}
\begin{split}
    \delta \left(\frac{i}{8\pi^2} \oint_{X^{(4)}} A \wedge d\theta \wedge F \right) &= \frac{i}{8\pi^2} \oint_{X^{(4)}} d\lambda \wedge d\theta \wedge F 
= -\frac{iw}{4\pi} \oint_{\Sigma^{(2)}} \lambda F \,.
\end{split}
\end{align}
(Here we assume the absence of monopoles.)
The anomalous variation \eqref{eq:inflow_string} shows that there is an anomaly inflow from the 3+1d bulk to the axion string worldsheet.
Through the descent procedure, the worldsheet anomaly is characterized by the anomaly polynomial 4-form,
\begin{equation} \label{eq:poly_string}
    \mathcal{I}^{\text{string}}_4 = -\frac{w}{2(2\pi)^2} F \wedge F \,.
\end{equation}
The same result can also be derived by dualizing the axion field, see, for example, \cite[Appendix~B]{Heidenreich:2021yda}.

To have a well-defined, gauge-invariant 2-dimensional extended defect, we therefore need to dress the worldsheet with additional charged degrees of freedom coupled to the bulk eletromagnetic gauge field, which carry the anomaly that is opposite to \eqref{eq:poly_string}. 
For instance, we can have $w$ flavors of 1+1d left-moving complex Weyl fermions $\psi_i$ of gauge charge 1, where $i = 1, \cdots w$.
In this case, the wolrdsheet action of axion string   includes
\begin{equation} \label{eq:worldsheet_weyl}
     \int \left[D\bar{\psi}_i D\psi_i \right]_{\Sigma^{(2)}} \exp \left[i \oint_{\Sigma^{(2)}} \sqrt{h}d^2 x \, \sum_{i=1}^{w} \bar{\psi}_i \left(
        \slashed{\partial} - i\slashed{A}        
    \right) \psi_i  \right] \,,
\end{equation}
where $\sqrt{h}d^2 x$ is the worldsheet volume form and $A$ is the bulk electromagnetic gauge field. 

Again, the choice of the worldsheet degrees of freedom is not unique, and \eqref{eq:worldsheet_weyl} is only one possible choice.
Independent of any particular choice, $S_w$ will denote any such surface defect with winding symmetry charge $w$.

\subsection{Junctions Involving Monopoles and Strings}\label{sec:junction}

\begin{figure}[!t]
    \centering
    \includegraphics[width=0.8\textwidth]{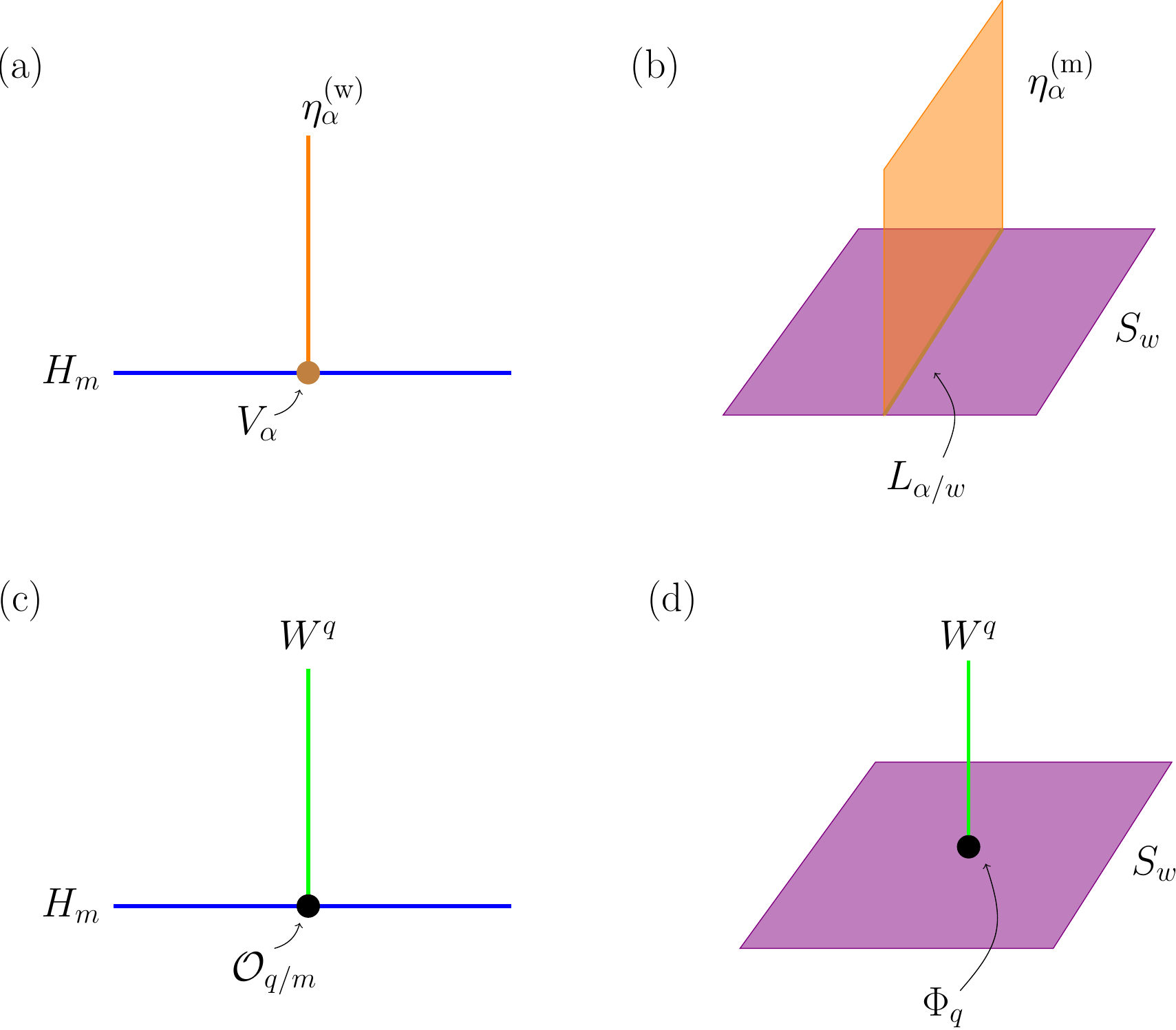}
    \caption{Junctions between extended operators in the axion-Maxwell theory. (a) A winding 2-form symmetry line ending on an 't Hooft line. (b) A magnetic 1-form symmetry surface ending on an axion string worldsheet. (c) A Wilson line ending on an 't Hooft line with $q=0$ mod $m$. (d) A Wilson line ending on an axion string worldsheet with  $q=0$ mod $w$.
    The junctions (a) and (b) are topological, whereas the junctions (c) and (d) are non-topological.}
    \label{Fig:junctions}
\end{figure}

Below we discuss four different junctions involving either an 't Hooft line or an axion string worldsheet.
Two of them are topological junctions, and the other two are not.
They are summarized in Figure \ref{Fig:junctions}.

\subsubsection*{Topological Junction between $\eta^\text{(w)}_{\alpha}$ and $H_m$} 

The winding symmetry line defect $\eta^\text{(w)}_\alpha$ can topologically end on an 't Hooft line $H_m$ for any values of $\alpha$ and $m$, as shown in Figure \ref{Fig:junctions}(a).
For rational value of $\alpha$, such a topological junction is required to exist from the action of the non-invertible 1-form symmetry defect $\mathcal{D}^{(1)}_{p/N}$ on $H_m$, which was discussed in Section \ref{sec:action}.

More generally, this topological junction can be understood as follows.
Recall that the 't Hooft line $H_m$ supports quantum mechanical degrees of freedom to cancel the anomaly  \eqref{eq:inflow_monopole}.
To be concrete, we will choose to dress $H_m$ with the particle on a circle quantum mechanics as in \eqref{eq:particle_on_circle}, but the conclusion will be the same for any abstract worldline quantum mechanics which properly cancels the anomaly.

The  quantum mechanics has a global $U(1)^{(0)}$ symmetry, whose charge operator is given by
\begin{equation}
    Q = i l_\sigma \star d\sigma \,,
\end{equation}
with integer eigenvalues.
When $\theta$ is a constant, the charge is conserved, that is, $dQ=0$, due to the equation of motion.
The $U(1)^{(0)}$ symmetry is generated by the topological operator which we will denote as 
\ie\label{Valpha}
V_\alpha \equiv \exp (i\alpha Q)
\fe
 where $\alpha \in [0,2\pi)$ labels a $U(1)^{(0)}$ group element.

When $\theta$ is dynamical or is not a constant, we have
\begin{equation} \label{eq:anomalous_Q}
    dQ = d\left( i l_\sigma \star d\sigma \right)
    = \frac{d\theta}{2\pi} \,.
\end{equation}
This anomalous conservation of the $U(1)^{(0)}$ symmetry charge in the presence of varying $\theta$ is a general consequence of the anomaly in the space of coupling constant.
By  Stokes' theorem, \eqref{eq:anomalous_Q} implies that the winding symmetry line $\eta^\text{(w)}_\alpha= \exp(i\alpha \int \frac{d\theta}{2\pi})$ from the bulk can terminate topologically at  the point operator \eqref{Valpha}  
 on $H_m$.
The existence of such a topological junction also implies that the 't Hooft line can freely absorb the winding symmetry operator.

\subsubsection*{Topological Junction between $\eta^\text{(m)}_{\alpha}$ and $S_w$}

The magnetic symmetry surface defect $\eta^\text{(m)}_{ \alpha}$ can end topologically  on an axion string worldsheet $S_w$ for any values of $\alpha$ and $w$ as shown in Figure \ref{Fig:junctions}(b).
For  rational values of $\alpha$, this  is implied by the action of the non-invertible 1-form symmetry defect $\mathcal{D}^{(1)}_{p/N}$ on $S_w$ discussed in Section \ref{sec:action}.

Recall that on the axion string worldsheet $S_w$, there is a 1+1d QFT which cancels the anomaly inflow from the bulk given in \eqref{eq:poly_string}.
For concreteness, we will choose $w$ copies of left-moving Weyl fermions as the 1+1 worldsheet QFT on $S_w$, as in \eqref{eq:worldsheet_weyl}.
However, the  conclusion holds independent of the choice of the worldsheet QFT as long as it  cancels the anomaly inflow.

Before coupling to the bulk axion-Maxwell theory, the 1+1d theory of $w$ left-moving Weyl fermions has a global $U(1)^{(0)}$ chiral symmetry.
Denote the corresponding current as $J_{\text{chiral}}^{(1)}$, which is conserved, $d\star J_{\text{chiral}}^{(1)} = 0$.
The $U(1)^{(0)}$ symmetry is generated by the topological line operator
\ie
L_{\alpha'} \equiv \exp (i\alpha' \oint \star J_{\text{chiral}}^{(1)})
\fe
 with $\alpha' \in [0,2\pi)$.

On the axion string worldsheet $S_w$, this $U(1)^{(0)}$ symmetry   is coupled to the bulk electromagnetic gauge field $A$.
The anomaly carried by the Weyl fermions implies the anomalous conservation equation for the current $J_{\text{chiral}}^{(1)}$,\footnote{The partition function of $w$ left-moving Weyl fermions satisfies
\begin{equation*}
    Z[A+d\lambda] = Z[A] \exp \left( \frac{iw}{4\pi} \int \lambda F \right) \,.
\end{equation*}
If one defines the current for the $U(1)^{(0)}$ symmetry as $\tilde{J}^{(1)}_{\text{chiral}}[A] \equiv i \frac{\delta \text{log} Z[A]}{\delta A}$, 
then the anomalous variation of the partition function implies $d\star \tilde{J}^{(1)}_{\text{chiral}} = \frac{w}{4\pi} F$ (see, for instance, \cite{Bilal:2008qx}).
However, the current $\tilde{J}^{(1)}_{\text{chiral}}$ is not gauge-invariant.
Instead, it transforms as $ \star \tilde{J}^{(1)}_{\text{chiral}}[A+d\lambda] = \star \tilde{J}^{(1)}_{\text{chiral}}[A] - \frac{w}{4\pi} d\lambda$,
which directly follows from the definition of $\tilde{J}^{(1)}_{\text{chiral}}$ and the anomalous variation of the partition function.
We define the gauge-invariant current as $\star J_{\text{chiral}}^{(1)} \equiv \star \tilde{J}^{(1)}_{\text{chiral}} + \frac{w}{4\pi} A$. 
This gauge-invariant current then satisfies \eqref{eq:anomalous_Weyl}.
}
\begin{equation} \label{eq:anomalous_Weyl}
    d\star J_{\text{chiral}}^{(1)} = \frac{w}{2\pi} F \,.
\end{equation}
Using  Stokes' theorem, we deduce from \eqref{eq:anomalous_Weyl} that the magnetic symmetry surface defect $\eta^\text{(m)}_{\alpha} = \exp(i\alpha \int \frac{F}{2\pi} )$ from the bulk can end topologically along  the line operator 
 $L_{\alpha /w} = \exp\left(i\frac{\alpha}{w} \oint \star J_{\text{chiral}}^{(1)}\right)$ on  $S_w$.  
The existence of this 1-dimensional topological junction implies that the axion string worldsheet can freely absorb the magnetic symmetry defect.

We note that when $|w|>1$, there is a $\mathbb{Z}_{|w|}$ subgroup of the $U(1)^{(0)}$ symmetry of the worldsheet QFT which is free of anomaly.
Correspondingly, there are topological line operators living on the axion string worldsheet $S_w$,
\begin{equation} \label{eq:Zw}
    \hat{L}_{2\pi n /w} \equiv \exp\left[ i \oint
    \left(
        \frac{2\pi n}{w} \star J_{\text{chiral}}^{(1)} - nA
    \right)
    \right] \,,
\end{equation}
for $n= 0, 1, \cdots, |w|-1$ mod $|w|$.
This is the topological line operator on $S_w$ that appears in \eqref{eq:inv_action_string}.\footnote{In \eqref{eq:inv_action_string}, we have $w=Nk$ for some integer $k$, and $\hat{L}_{2\pi p /N}$ is obtained by setting $n=pk$ in \eqref{eq:Zw}.}

\subsubsection*{Junction between $W^q$ and $H_m$} 

Consider the configuration of a Wilson line $W^q$ ending on an 't Hooft line $H_m$ at a 0-dimensional junction.
For such a configuration to be gauge-invariant, one needs to insert an operator at the junction which has charge $q$ under the gauge group.
This may or may not be possible for arbitrary values of $q$ and $m$, depending on the choice of the quantum mechanical degrees of freedom that we put on $H_m$.
We claim that for any choice of the worldline quantum mechanics, as long as it properly cancels the anomaly inflow \eqref{eq:inflow_monopole}, the Wilson lines with charge $q = 0$ mod $m$ can always end on an 't Hooft line $H_m$.

For instance, if we choose the particle on a circle quantum mechanics in \eqref{eq:particle_on_circle} as the worldline quantum mechanics, then the operator 
\ie
\mathcal{O}_k \equiv \exp(i k \sigma)
\fe
 on the 't Hooft line, with $k$ being an arbitrary integer, has charge $km$ under the gauge group, and thus the Wilson line with charge $q=km$ can end on this operator to form a gauge-invariant 0-dimensional junction with $H_m$.
This is shown in Figure \ref{Fig:junctions}(c).

More abstractly, the claim is that any quantum mechanics with an anomaly opposite to \eqref{eq:inflow_monopole} should have an operator with charge $m$ under its $U(1)^{(0)}$ symmetry.
This can be easily proven as follows. 
Assume that the charge of  every operator  is a multiple of some integer $m'>0$. 
This implies  that the $\mathbb{Z}_{m'}$ subgroup of the $U(1)^{(0)}$ symmetry does not act faithfully in the quantum mechanics.
We can then couple the faithfully acting symmetry $U(1)^{(0)}/\mathbb{Z}_{m'}$ to a (properly quantized) background $U(1)$ gauge field $A' $.
The background gauge field $A$ for the original $U(1)^{(0)}$ and $A'$ are related by $A' = m' A$.
However, this then implies that the anomaly polynomial is
\begin{equation}
    \frac{m}{(2\pi)^2} d\theta \wedge dA = \frac{m/m'}{(2\pi)^2} d\theta \wedge  dA' \,,
\end{equation}
which is properly quantized  only if $m = 0$ mod $m'$. 
This proves that the Wilson line $W^q$ with $q=0$ mod $m$ can always end on $H_m$.

Independent of the choice of the specific worldline quantum mechanics on $H_m$, we will denote this non-topological junction between the Wilson line with charge $q=km$ and the 't Hooft line $H_m$ as ${\cal O}_k$.

\subsubsection*{Junction between $W^q$ and $S_w$}

Similarly, we claim that the Wilson line $W^q$   can end on an axion string worldsheet $S_w$ if $q = 0$ mod $w$, at a 0-dimensional junction with $S_w$. 
This junction exists independent of the choice of the worldsheet QFT as long as it has a $U(1)^{(0)}$ global symmetry with an anomaly that is opposite to \eqref{eq:poly_string}.\footnote{Depending on the choice of the worldsheet QFT on $S_w$, there will in general be more allowed values of $q$ for which $W^q$ is allowed to end on $S_w$.
For instance, if we choose the $w$ copies of left-moving Weyl fermions as the worldsheet QFT as in \eqref{eq:worldsheet_weyl}, then Wilson lines of any charge can end on $S_w$ as we have charge 1 Weyl fermions on the worldsheet.
However, this is not necessarily the case for an arbitrary choice of a worldsheet QFT that cancels the anomaly inflow \eqref{eq:poly_string}.} 
That is, the claim is that any such 1+1d QFT has a local operator with charge $w$.

The proof is similar to the 't Hooft line case.
Let $w'$ be the minimal charge under the $U(1)^{(0)}$ symmetry of the worldsheet QFT.
Then, turning on the background $U(1)$ gauge field $A'$ for the faithfully acting symmetry group $U(1)^{(0)}/\mathbb{Z}_{w'}$, we find that the anomaly polynomial for $A'$ is given by
\begin{equation}
    \frac{w}{2(2\pi)^2} dA \wedge dA = \frac{w/w'^2}{2(2\pi)^2} dA' \wedge dA' \,.
\end{equation}
The coefficient of the anomaly polynomial $w/w'^2$ must be an integer, which implies $w=0$ mod $w'$.
Thus, there are  local operators with charge $w$ under the $U(1)^{(0)}$ symmetry.
This proves that the Wilson line with charge $q=0$ mod $w$ can end on $S_w$.

 In general, we will denote any local operator on $S_w$ with charge $q$ under the $U(1)^{(0)}$ symmetry as $\Phi_q$, which gives a non-topological junction where the bulk Wilson line $W^q$ can end. 
The allowed values of $q$  depend on $w$ and also on the choice of the worldsheet QFT, but as we explained above, $\Phi_q$ with $q=0$ mod $w$  always exist.
See Figure \ref{Fig:junctions}(d).

\subsection{Crossing Relations}\label{sec:crossing}

We now derive two crossing relations involving the junctions that we discussed above,  and show that they are consistent with the action of the non-invertible 1-form symmetry ${\cal D}^{(1)}_{p/N}$ discussed in Section \ref{sec:action}. 
These are shown in Figures \ref{Fig:crossing1} and \ref{Fig:crossing2}. 

\subsubsection*{Crossing on $H_m$}

\begin{figure}
    \centering
    \includegraphics[width=\textwidth]{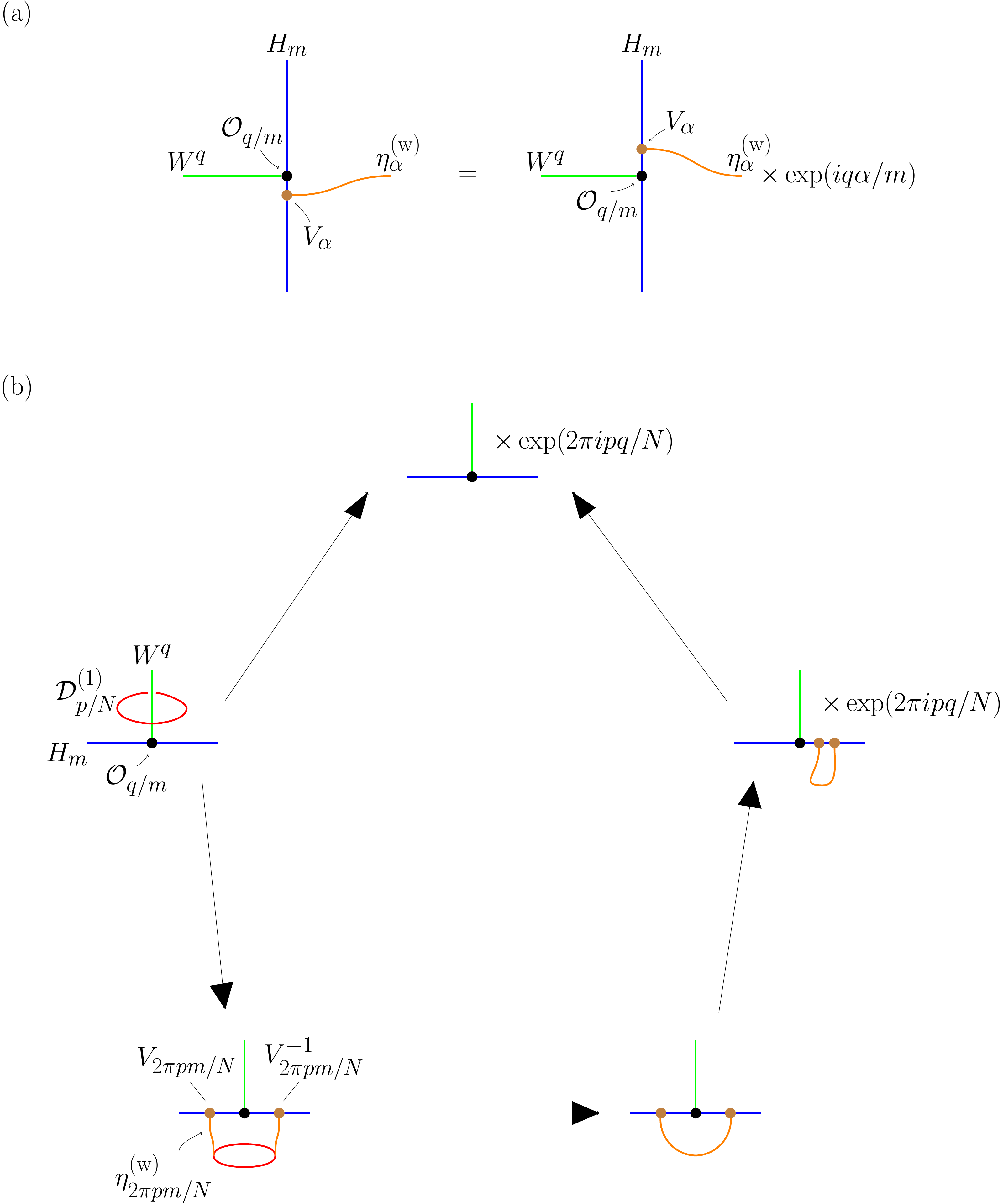}
    \caption{(a) Local crossing relation of junctions on an 't Hooft line $H_m$. Here $q$ is an integer multiple of $m$. (b) The nontrivial crossing relation is necessary for the action of the non-invertible 1-form symmetry to be consistent. In the leftmost picture, the $\mathcal{D}^{(1)}_{p/N}$ defect shown in red is supported on a 2-sphere which links with the Wilson line $W^q$. We choose the Euler counterterm to set the expectation value of $\mathcal{D}^{(1)}_{p/N}$ on a 2-sphere to be 1.}
    \label{Fig:crossing1}
\end{figure}

\begin{figure}
    \centering
    \includegraphics[width=\textwidth]{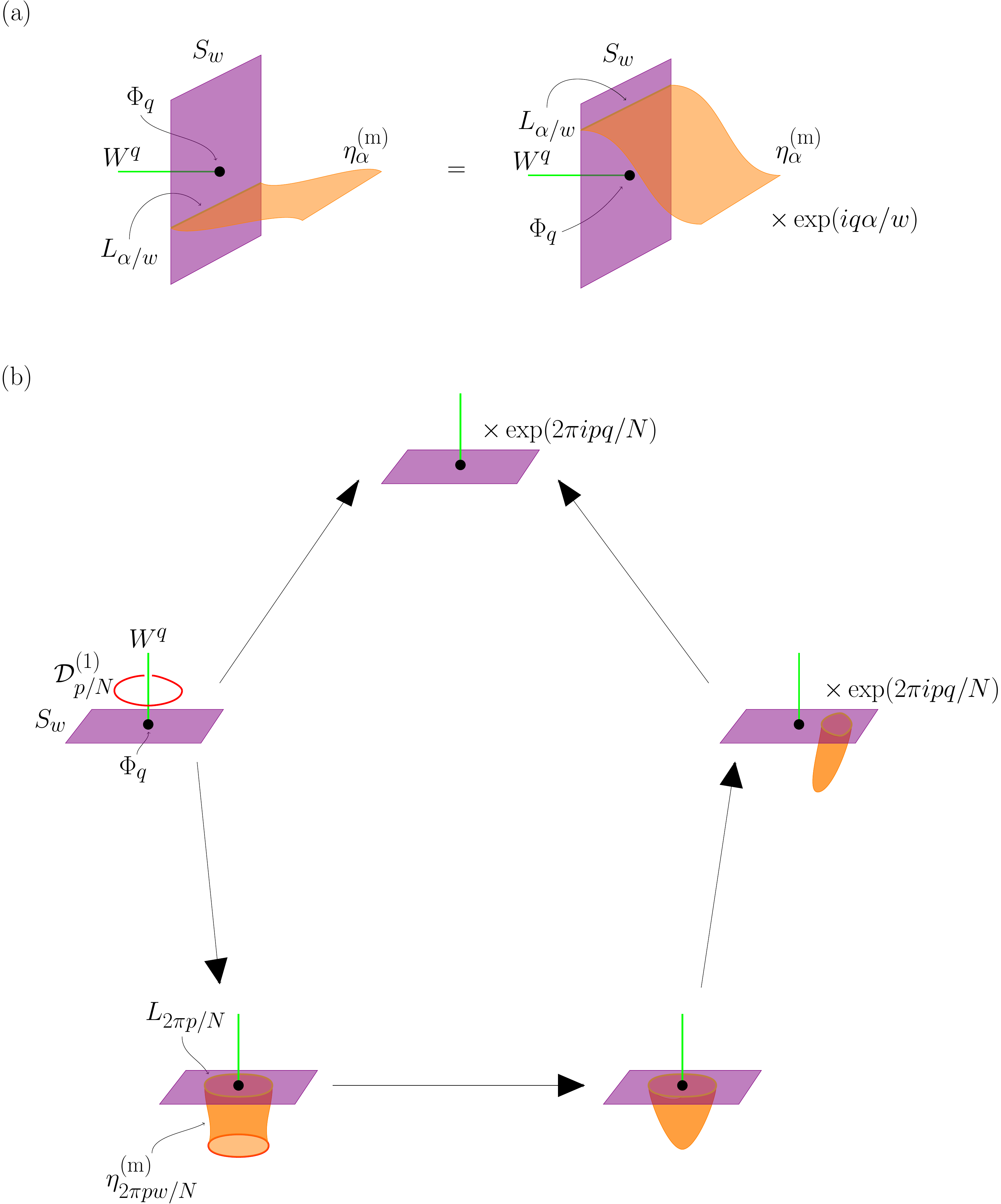}
    \caption{(a) Local crossing relation of junctions on an axion string worldsheet $S_w$. Here $q$ is an integer multiple of $w$. (b) The nontrivial crossing relation is necessary for the action of the non-invertible 1-form symmetry to be consistent. Again, we choose the Euler counterterm to set the expectation value of $\mathcal{D}^{(1)}_{p/N}$ on a 2-sphere to be 1.}
    \label{Fig:crossing2}
\end{figure}

There are two kinds of junctions on $H_m$: the bulk $\eta^\text{(w)}_\alpha$ line can end topologically on $V_\alpha$, and the bulk Wilson line $W^q$ can end on ${\cal O}_{q/m}$ with $q=mk$ and $k$ some integer. 
Since ${\cal O}_k$ carries charge $k$ under the $U(1)^{(0)}$ symmetry generated by $V_\alpha$, we have the following crossing relation:
\begin{equation} \label{eq:crossing1}
    V_\alpha \mathcal{O}_k = e^{ik\alpha} \mathcal{O}_k V_\alpha \,.
\end{equation}
This is depicted in Figure \ref{Fig:crossing1}(a). 
Recall that ${\cal O}_k$ is the (non-topological) junction between the Wilson line $W^q$ and $H_m$ with $q=mk$.

For rational $\alpha$, the crossing relation obeys a consistency condition with the action of the non-invertible 1-form symmetry ${\cal D}^{(1)}_{p/N}$. This is shown in Figure \ref{Fig:crossing1}(b).

\subsubsection*{Crossing on $S_w$}

There are two kinds of junctions on $S_w$: the bulk $\eta^\text{(m)}_\alpha$ can end topologically on $L_{\alpha'}$ with $\alpha' = \alpha/w$, and the bulk Wilson line $W^q$ can end on $\Phi_q$. 
Since $\Phi_q$ carries charge $q$ under the $U(1)^{(0)}$ symmetry generated by $L_{\alpha'}$, they obey the following crossing relation
\begin{equation} \label{eq:crossing2}
    L_{\alpha'} \Phi_q = e^{iq\alpha'} \Phi_q L_{\alpha'} \,.
\end{equation}
This is shown in Figure \ref{Fig:crossing2}(a).

For rational $\alpha$, the crossing relation obeys a consistency condition with the action of the non-invertible 1-form symmetry ${\cal D}^{(1)}_{p/N}$. This is shown in Figure \ref{Fig:crossing2}(b).

\begin{figure}[!t]
    \centering
    \includegraphics[width=.8\textwidth]{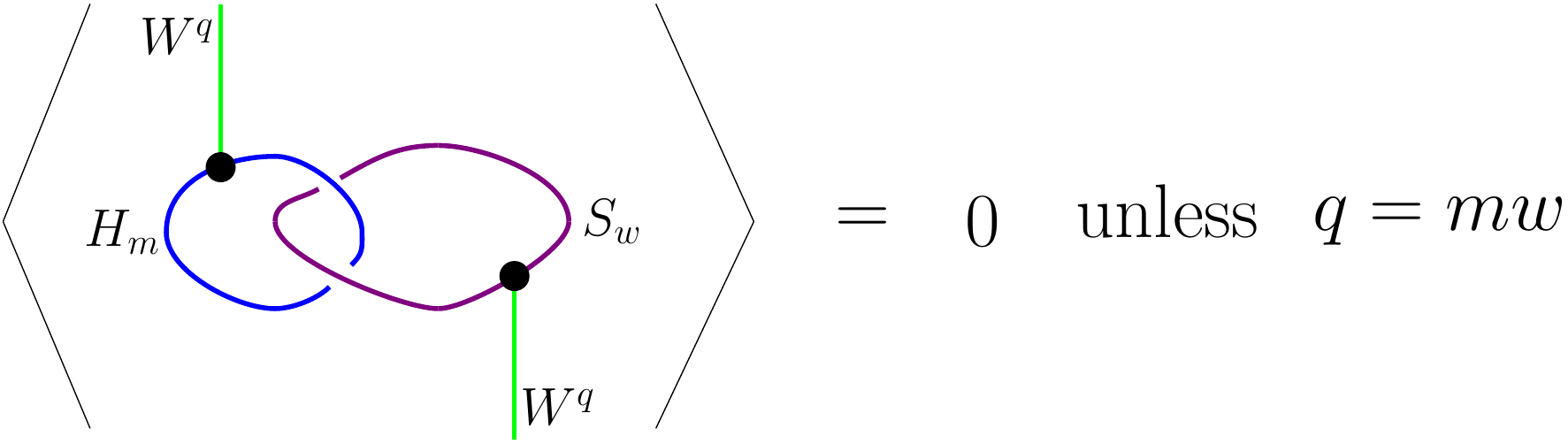}
    \caption{Selection rule from the non-invertible symmetry. The 't Hooft line $H_m$ is supported on a closed loop and the axion string worldsheet $S_w$ is supported on a closed 2-dimensional surface. The two are linked in the 4d  spacetime. The selection rule states that any correlation function, independent of other operator insertions, including this   configuration in a local region  must vanish unless $q=mw$.}
    \label{Fig:selection1}
\end{figure}

\subsection{Selection Rules and  the Witten Effect}\label{sec:selection}

First let us recall the Witten effect \cite{Witten:1979ey} in the ordinary free Maxwell theory without axions.
It states that a magnetic monopole carries an electric charge  proportional to the $\theta$ angle.
In particular, when $\theta \rightarrow \theta+2\pi$, the bulk Maxwell theory does not change,  but there is a nontrivial spectral flow among the line operators, with an 't Hooft line  mapped to a dyonic line.

Now, let us promote the constant $\theta$ angle to a dynamical axion field that couples to a $U(1)$ gauge field.
As discussed in Section \ref{sec:gauss}, there is no conserved and gauge-invariant electric charge in axion-Maxwell theory. 
Then what  does the Witten effect mean?

We claim that the Witten effect should be interpreted as  an exact selection rule involving the 't Hooft line $H_m$, axion string worldsheet $S_w$, and Wilson lines $W^q$ as shown in Figure \ref{Fig:selection1}. 
The selection rule states that any correlation function containing this local configuration vanishes unless 
\ie
q=mw\,.
\fe 
This holds true independent of how the Wilson lines are extended outside this local configuration, and also independent of the other operator insertions.

\begin{figure}[!t]
    \centering
    \includegraphics[width=\textwidth]{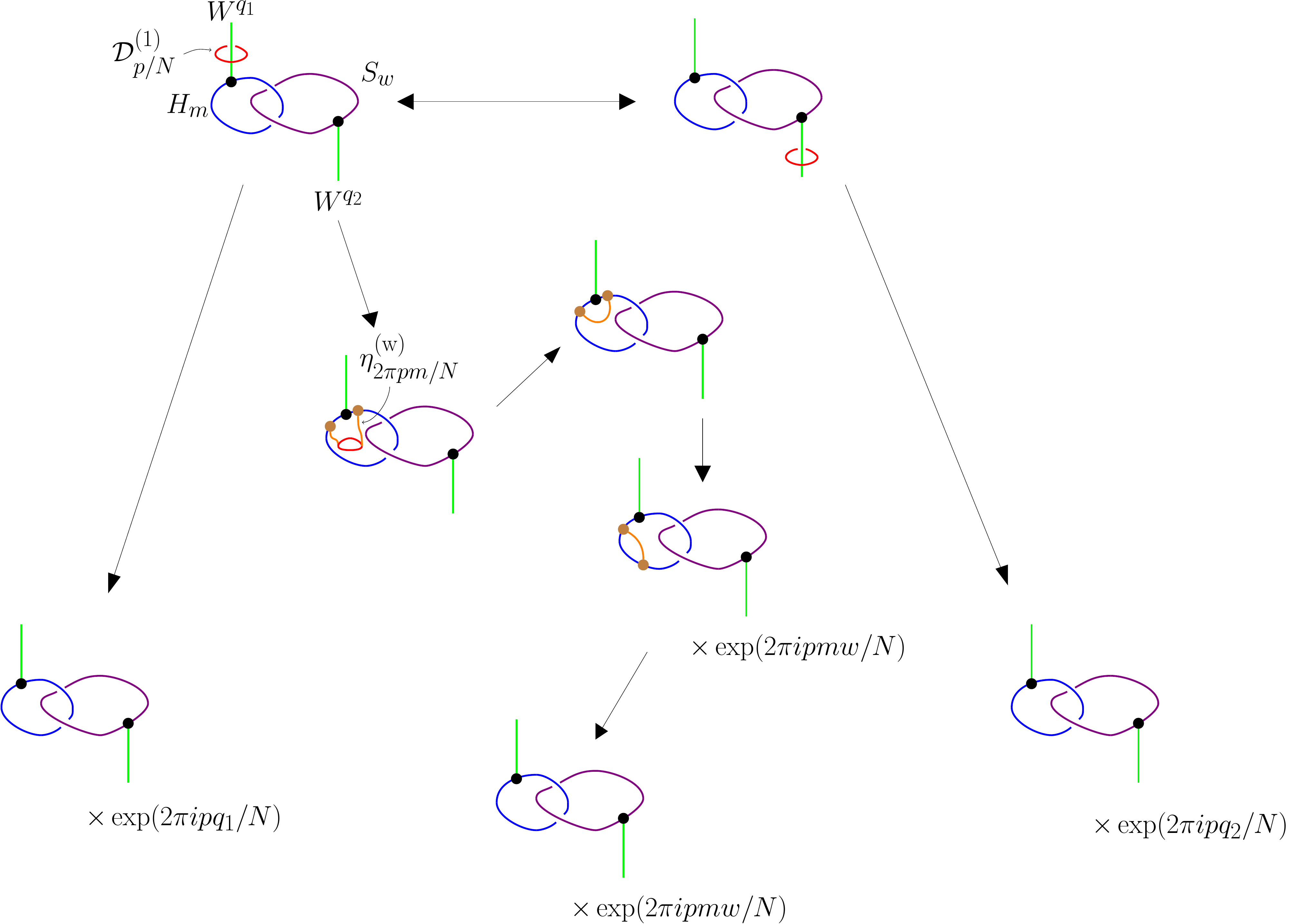}
    \caption{Derivation of the selection rule in Figure \ref{Fig:selection1}. We link the configuration of the 't Hooft line $H_m$, axion string $S_w$, Wilson lines $W^{q_i}$ with  a non-invertible 1-form symmetry defect $\mathcal{D}^{(1)}_{p/N}$. There are various ways to shrink the $\mathcal{D}^{(1)}_{p/N}$ defect, and comparing them implies that any correlation function involving this local configuration vanishes unless $q_1 = q_2 = mw$ mod $N$. By using ${\cal D}_{p/N}^{(1)}$ with different coprime pairs $p,N$, we further demand $q_1 = q_2 = mw$ as integers. We choose the Euler counterterm on $\mathcal{D}^{(1)}_{p/N}$ so that its expectation value on a 2-sphere is 1.}
    \label{Fig:selection2}
\end{figure}

This selection rule follows from the non-invertible 1-form symmetry. 
We demonstrate the proof in  Figure \ref{Fig:selection2}. 
The various phase factors are only consistent if $q\equiv q_1=q_2=mw$. 
One can also derive the selection rule using either one of the crossing relations that we discussed previously.

\begin{figure}[!t]
    \centering
    \includegraphics[width=.7\textwidth]{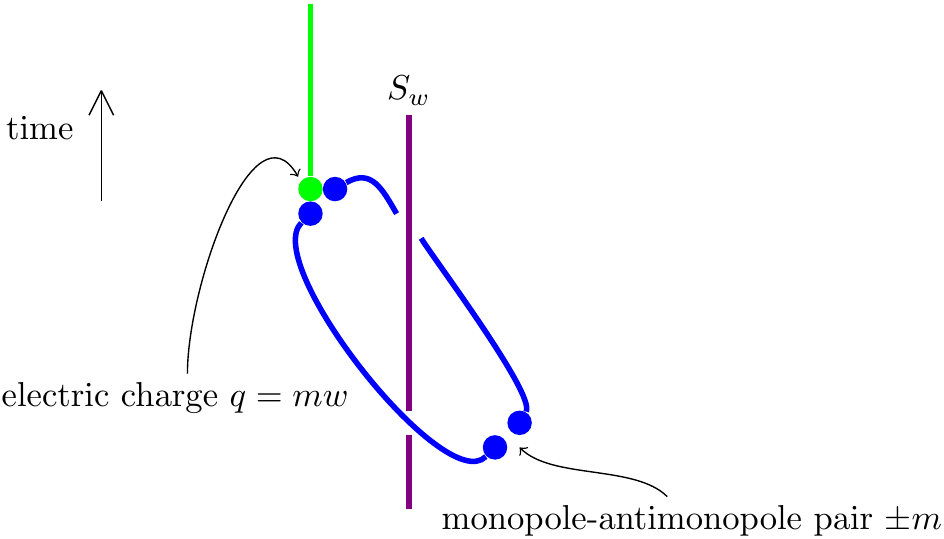}
    \caption{A real time process constrained by the selection rule. The axion string worldsheet $S_w$ extends  along the time direction, and it forms a closed loop along a spatial direction that is suppressed in  the figure. We create a monopole-antimonopole pair of magnetic charges $\pm m$, let them travel around the axion string worldsheet, and then pair annihilate them. After the pair annihiliation, there is an electrically charged particle with charge $q=mw$ created.}
    \label{Fig:realtime}
\end{figure}

 We can translate the Euclidean selection rule of Figure \ref{Fig:selection1} into a real time process in Lorentzian signature in Figure \ref{Fig:realtime}. 
First, near an axion string $S_w$, we pair create a monopole-antimonopole pair of magnetic charge $\pm m$.
Next, we  bring the monopole and anti-monopole  around the axion string clockwise and couterclockwise, respectively, and then pair annihilate them. 
The selection rule implies that such a process is forbidden, unless during the process there is an electrically charged particle of charge $q=mw$ coming out, as dictated by the Witten effect.\footnote{Unlike in the Euclidean configuration, here the axion string extends in the time direction indefinitely, and the worldsheet therefore is noncompact. In this case, the selection rule does not require another Wilson line attached to the axion string worldsheet.}
The net effect is that we gain a particle of charge $q=mw$ out of the vacuum. 
This is consistent with the non-invertible Gauss law in Section \ref{sec:gauss}, where we found that, in the presence of  a charge $m$ monopole, the electric charge is conserved modulo $m$. 

Related processes have been discussed \cite{Kogan:1993yw,Fukuda:2020imw}, which are sometimes called ``charge teleportation." 
Here we provide a new interpretation of these phenomena in terms of a non-invertible global symmetry and its selection rules.

\section{Non-Invertible Generalizations of Higher Groups} \label{sec:higher_structure}

In this section, we discuss the topological junctions of the symmetry defects.\footnote{This is to be contrasted with the junctions discussed in Section \ref{sec:junction} which always involve at least one non-topological defect (either the monopole or the axion string).} 
In particular, we show that symmetries of different form degrees mix in a way reminiscent of a higher-group symmetry for invertible symmetries. For higher-group symmetries, when the symmetry defects intersect, their junctions emit symmetry defects of higher form degrees (see, for example, \cite{Benini:2018reh} for the case of a 2-group).
In \cite{Hidaka:2020iaz,Hidaka:2020izy,Brennan:2020ehu}, it was pointed out that there exists a higher-group symmetry in the axion-Maxwell theory when the axion-photon coupling is greater than 1, i.e., $K>1$ (see the review in Appendix \ref{app:higher_group}). In this section, we show that the mixing of symmetries of different form degrees exist even at $K=1$, except that the symmetries of interest  are now non-invertible.

\subsection{Junctions between Non-Invertible Defects}

\subsubsection*{Junction between $\mathcal{D}^{(1)}_{p/N}$ and $\mathcal{D}^{(1)}_{p'/N'}$}

Two non-invertible 1-form symmetry defects $\mathcal{D}^{(1)}_{p/N}$, $\mathcal{D}^{(1)}_{p'/N'}$ can intersect topologically at a 0-dimensional junction from which an invertible winding 2-form symmetry line defect $\eta^\text{(w)}_{-2\pi p p'/NN'}$ is emitted as shown in Figure \ref{Fig:highergroup}(a). The emission of the winding symmetry line is related to the topological property of the junction. Suppose we move $\mathcal{D}^{(1)}_{p/N}$ from $\Sigma^{(2)}$ to $\widetilde{\Sigma}^{(2)}$ as depicted in Figure \ref{Fig:highergroup}(b). This sweeps out a 3-dimensional volume $ N^{(3)}$ bounded by $\Sigma^{(2)}$ and $\widetilde{\Sigma}^{(2)}$, and implements a transformation
\ie
A\rightarrow A+\frac{2\pi p}{N} \, \delta(N^{(3)})\,,
\fe
where $\delta(N^{(3)})$ is the delta function 1-form localized on $N^{(3)}$.
Locally, let $y$ be a normal coordinate such that $N^{(3)}$ is at $y=0$, then $\delta(N^{(3)}) = \delta(y)dy$.
Different choices of $N^{(3)}$'s bounded by the same $\Sigma^{(2)}$ and $\widetilde{\Sigma}^{(2)}$ are related by gauge transformations. 
From the worldsheet action \eqref{eq:electric_worldvolume} of $\mathcal{D}^{(1)}_{p'/N'}(\Sigma'^{(2)})$, we see that this transformation generates a line defect
\ie\label{eq:line_defect}
\eta^\text{(w)}_{-2\pi p p'/NN'}(M^{(1)})=\exp\left(\frac{i p}{N}\int_{M^{(1)}} d\phi\right)=\exp\left(-\frac{i pp'}{NN'}\int_{M^{(1)}} d\theta\right)
\fe
where $M^{(1)}=N^{(3)}\cap\Sigma'^{(2)}$ is a line interval on $\Sigma'^{(2)}$ that connects the initial and final intersections. In the second equality, we used the equation of motion of $c$ to relate $d\phi=-p'd\theta/N'$. The line defect \eqref{eq:line_defect} moves the endpoint of the winding symmetry line $\eta^\text{(w)}_{-2\pi p p'/NN'}$ from the initial junction to the finial junction. Therefore, such a junction between two non-invertible 1-form symmetry defects is topological.

\subsubsection*{Junction between $\mathcal{D}^{(0)}_{p/N}$ and $\mathcal{D}^{(1)}_{p'/N'}$}

The non-invertible 0-form symmetry defect $\mathcal{D}^{(0)}_{p/N}(\Sigma^{(3)})$ and the non-invertible 1-form symmetry defect $\mathcal{D}^{(1)}_{p'/N'}(\Sigma'^{(2)})$ intersect topologically at a 1-dimensional junction from which an invertible magnetic 1-form symmetry surface defect $\eta^\text{(m)}_{-2\pi p p'/NN'}$ is emitted. 
The emission of the magnetic symmetry surface  is related to the topological property of the junction.
Suppose we move $\mathcal{D}^{(0)}_{p/N}$ from $\Sigma^{(3)}$ to $\widetilde\Sigma^{(3)}$. It implements a transformation 
\ie
\theta\rightarrow\theta+\frac{2\pi p}{N}
\fe
 in the region bounded by $\Sigma^{(3)}$ and $\widetilde\Sigma^{(3)}$. 
From the worldsheet action \eqref{eq:electric_worldvolume} of   $\mathcal{D}^{(1)}_{p'/N'}(\Sigma'^{(2)})$, we see that the transformation generates a surface defect 
\ie\label{eq:surface_defect}
\eta^\text{(m)}_{-2\pi p p'/NN'}(M^{(2)})=\exp\left(\frac{i pp'}{N}\int_{M^{(2)}} dc\right)=\exp\left(-\frac{i pp'}{NN'}\int_{M^{(2)}} dA\right)~,
\fe
where $M^{(2)}$ is the region on $\Sigma'^{(2)}$ bounded by $\Sigma'^{(2)}\cap \Sigma^{(3)}$ and $\Sigma'^{(2)}\cap \widetilde\Sigma^{(3)}$. 
We have used the equation of motion of $\phi$ to relate $dc=-dA/N'$ in the second equality. The surface defect \eqref{eq:surface_defect} moves the end of the magnetic symmetry surface  $\eta^\text{(m)}_{-2\pi p p'/NN'}$ from the initial 1-dimensional junction to the final one. Hence, such a junction between $\mathcal{D}^{(0)}_{p/N}$ and $\mathcal{D}^{(1)}_{p'/N'}$ is topological.

\begin{figure}[!t]
    \centering
    \includegraphics[width=0.9\textwidth]{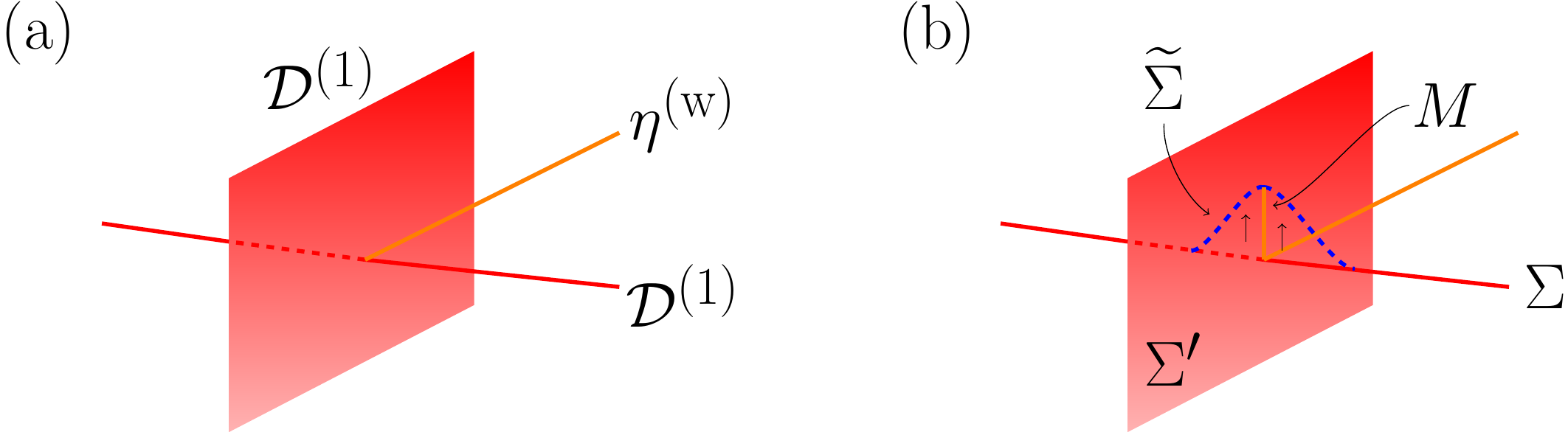}
    \caption{
        (a) The 0-dimensional junction between ${\cal D}^{(1)}_{p/N}(\Sigma^{(2)})$ and ${\cal D}^{(1)}_{p'/N'}(\Sigma'^{(2)})$ emits a winding symmetry line $\eta^\text{(w)}_{-{2\pi pp'/NN'}}$. This is reminiscent of the junction between the invertible defects in a higher-group symmetry. For the ${\cal D}^{(1)}_{p/N}$ on $\Sigma^{(2)}$, we suppress one of its dimensions and draw it as a line in the figure. (b) A topological deformation of the junction. We omit various super/subscripts in the figure but they can be found in this caption.}
    \label{Fig:highergroup}
\end{figure}

This junction and deformation are similar to the ones in Figure \ref{Fig:highergroup}, with ${\cal D}^{(0)}_{p/N}$ and ${\cal D}^{(1)}_{p'/N'}$ shown as lines and surfaces there, respectively. Note that two out of the three dimensions of ${\cal D}^{(0)}_{p/N}$ are suppressed in that figure.

The emission of  a lower dimensional topological defects at the above two kinds of  junctions is reminiscent of the higher-group symmetry. 
It would be interesting to understand the precise mathematical structure of this non-invertible generalization of the higher-group symmetry.

For completeness, below we discuss other junctions between topological defects.

\subsection{Junctions between Invertible and Non-Invertible Defects}

\begin{figure}[!t]
    \centering
    \includegraphics[width=0.9\textwidth]{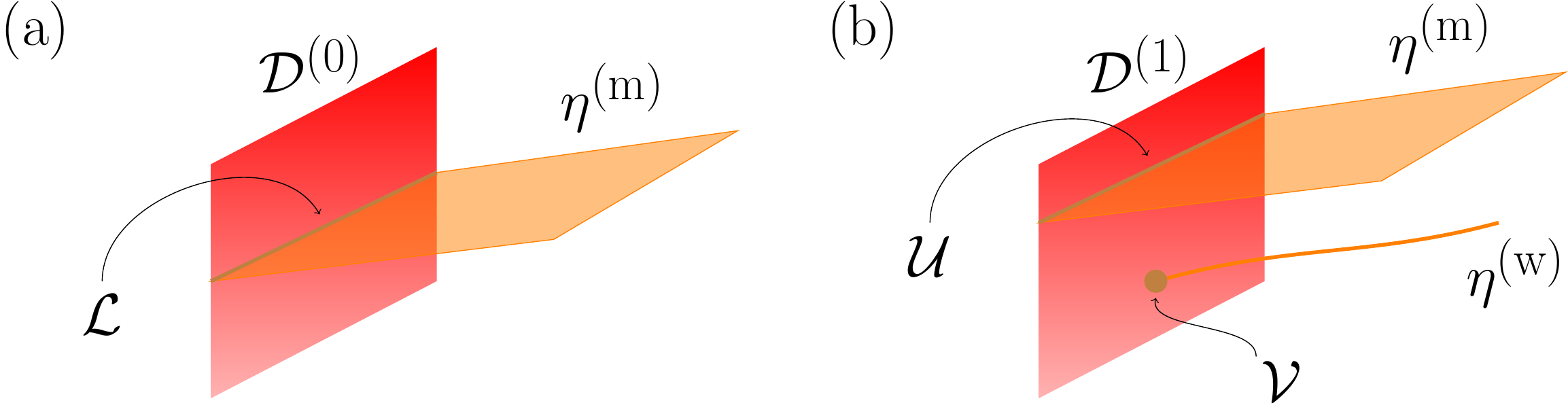}
    \caption{
        (a) The 1-dimensional topological junction ${\cal L}^{-p^{-1}\ell}$ between the non-invertible 0-form symmetry defect ${\cal D}^{(0)}_{p/N}$ and the magnetic symmetry surface $\eta^\text{(m)}_{2\pi \ell/N}$. Here $\mathcal{L}$ denotes the Wilson line in the 2+1d minimal TQFT ${\cal A}^{N,p}$ living on  ${\cal D}^{(0)}_{p/N}$.  (b) Top: 1-dimensional topological junction ${\cal U}^\ell$ between the non-invertible 1-form symmetry defect ${\cal D}^{(1)}_{p/N}$ and $\eta^\text{(m)}_{2\pi \ell/N}$. Bottom: The 0-dimensional topological junction ${\cal V}^{p^{-1}\ell}$ between ${\cal D}^{(1)}_{p/N}$ and the winding symmetry line $\eta^\text{(w)}_{2\pi\ell/N}$. Here $\mathcal{U}$ and $\mathcal{V}$ are the $\mathbb{Z}_N$ topological line and point operators of the 1+1d $\mathbb{Z}_N$ gauge theory living on ${\cal D}^{(1)}_{p/N}$.
        We omit various super/subscripts and powers in the figure but they can be found in this caption.}
    \label{Fig:topjunction}
\end{figure}

\subsubsection*{Junction between $\eta^\text{(m)}_{2\pi \ell/N}$ and $\mathcal{D}^{(0)}_{p/N}$} 
The non-invertible 0-form symmetry defect $\mathcal{D}^{(0)}_{p/N}$ and the magnetic  symmetry surface $\eta^\text{(m)}_{2\pi \ell/N} = \exp(\frac{i\ell}{N} \oint F)$ (with $\ell = 0 ,\cdots, N-1$ mod $N$) meet at a 1-dimensional topological junction as shown in Figure \ref{Fig:topjunction}(a).
The existence of such a junction follows from the half gauging construction of $\mathcal{D}^{(0)}_{p/N}$ \cite{Choi:2021kmx,Choi:2022zal,Choi:2022jqy}.

To understand this junction in more detail, recall that the worldvolume action of $\mathcal{D}^{(0)}_{p/N}$ in \eqref{eq:0form} supports the 2+1 minimal $\mathbb{Z}_N$ TQFT $\mathcal{A}^{N,p}$.
The minimal TQFT $\mathcal{A}^{N,p}$ has $N$ topological lines, which we denote as $\mathcal{L}^s$ for $s = 0, \cdots, N-1$ mod $N$, generating a $\mathbb{Z}_N^{(1)}$ 1-form symmetry.
The lines themselves are charged under the $\mathbb{Z}_N^{(1)}$ symmetry.
In particualr, the line $\mathcal{L}^s$ has charge $-ps$ mod $N$ under the $\mathbb{Z}_N^{(1)}$ symmetry \cite{Hsin:2018vcg}.

This means that when we couple the minimal TQFT $\mathcal{A}^{N,p}[B]$ to a background 2-form gauge field $B$ for the $\mathbb{Z}_N^{(1)}$ symmetry, the line operator $\mathcal{L}^s$ must be attached to the boundary of the surface $\exp(-ips \int B)$, for it to remain gauge-invariant.\footnote{Our normalization convention here is such that the $\mathbb{Z}_N$ 2-form gauge field $B$ satisfies $\oint B \in \frac{2\pi}{N}\mathbb{Z}$ on 2-cycles. Similar conventions apply to $B^{(1)}$ and $B^{(2)}$ below.}
In the definition of the defect $\mathcal{D}^{(0)}_{p/N}$, we couple the minimal TQFT $\mathcal{A}^{N,p}$ to the bulk electromagnetic gauge field by setting 
\ie
B = {1\over N}F\,.
\fe
Therefore, we learn that the magnetic 1-form symmetry surface defect $\eta^\text{(m)}_{2\pi \ell/N} $ ends on the topological line $\mathcal{L}^{-p^{-1}\ell}$. 
This  defines the 1-dimensional topological junction between $\mathcal{D}^{(0)}_{p/N}$ and $\eta^\text{(m)}_{2\pi \ell/N}$.
Recall that $\text{gcd}(p,N)=1$, so $-p^{-1} \ell$ is a well-defined integer mod $N$.

\subsubsection*{Junction between $\eta^{\text{(w)/(m)}}_{2\pi \ell/N}$ and $\mathcal{D}^{(1)}_{p/N}$} 
The winding symmetry lines $\eta^\text{(w)}_{2\pi \ell/N} = \exp(\frac{i\ell}{N} \oint d\theta)$ and the magnetic  symmetry surfaces $\eta^\text{(m)}_{2\pi \ell/N} = \exp(\frac{i\ell}{N} \oint F)$ meet with the non-invertible 1-form symmetry defect $\mathcal{D}^{(1)}_{p/N}$ at a topological 0- and  1-dimensional junction, respectively.
This is shown in \ref{Fig:topjunction}(b).
Similar to before, this can be understood from the 1+1d $\mathbb{Z}_N$ gauge theory on the worldsheet  of $\mathcal{D}^{(1)}_{p/N}$ in \eqref{eq:electric_worldvolume}. 

The 1+1d $\mathbb{Z}_N$ gauge theory is characterized by topological point operators $\mathcal{V}^s$ and  line operators $\mathcal{U}^s$, with $s= 0, \cdots, N-1$ mod $N$, see Appendix \ref{app:2dZN}.
These topological operators generate a $\mathbb{Z}_N^{(1)} \times \mathbb{Z}_N^{(0)}$ symmetry.
The topological point operators are charged under the $\mathbb{Z}_N^{(0)}$ 0-form symmetry.
Specifically, the $\mathbb{Z}_N^{(0)}$ charge of $\mathcal{V}^s$ is $s$ mod $N$.
Similarly, the topological line operators are charged under the $\mathbb{Z}_N^{(1)}$ 1-form symmetry, and $\mathcal{U}^s$ carries  charge $s$ mod $N$ under $\mathbb{Z}_N^{(1)}$. 

Therefore, when we turn on the background 1-form and 2-form gauge fields $B^{(1)}$ and $B^{(2)}$ for the $\mathbb{Z}_N^{(0)} \times \mathbb{Z}_N^{(1)}$ symmetry, the topological point operator $\mathcal{V}^s$ now lives on the boundary of the line $\exp(is\int B^{(1)})$, whereas the topological line operator $\mathcal{U}^s$ lives on the boundary of the surface $\exp(is\int B^{(2)})$. 
The non-invertible 1-form symmetry defect $\mathcal{D}^{(1)}_{p/N}$ is obtained by coupling the 1+1d $\mathbb{Z}_N$ gauge theory to the bulk by setting 
\ie
B^{(1)} = \frac{p}{N} d\theta\,,~~~~B^{(2)} = \frac{1}{N}F\,.
\fe
Thus, the winding symmetry line $\eta^\text{(w)}_{2\pi \ell/N} $ from the bulk can end on   $\mathcal{V}^{p^{-1}\ell}$. 
 Similarly, the magnetic symmetry surface $\eta^\text{(m)}_{2\pi \ell/N}  $ from the bulk can end on  $\mathcal{U}^{\ell}$.
These give the topological junctions in Figure \ref{Fig:topjunction}(b).

\section{Applications}\label{sec:application}

\subsection{Constraints on Symmetry Breaking Scales} \label{sec:scales}

The generalized global symmetries discussed so far are typically emergent in a renormalization group flow to the axion-Maxwell theory in the low energy limit. 
Not all symmetries are on the same footing: some of them are subordinate to others in the sense that they cannot exist if the others are broken. 
This structure leads to constraints on the energy scales where these symmetries are broken as we go up in energy. 
Such constraints have been derived for the higher-group symmetry in the $K>1$ theory in \cite{Brennan:2020ehu}. Below, we will focus on the $K=1$ case and show that the non-invertible global symmetry lead to similar universal constraints.

We define the following $E$'s to be the approximate symmetry breaking scales of the corresponding generalized global symmetries:
\ie
&E_\text{shift}:~\text{non-invertible 0-form symmetry ${\cal D}^{(0)}_{p/N}$}\,,\\
&E_\text{electric}:~\text{non-invertible 1-form symmetry ${\cal D}^{(1)}_{p/N}$}\,,\\
&E_\text{magnetic}:~\text{magnetic 1-form symmetry $U(1)^{(1)}_\text{magnetic}$}\,,\\
&E_\text{winding}:~\text{winding 2-form symmetry $U(1)^{(2)}_\text{winding}$}\,.
\fe
That is, the symmetry becomes emergent below the corresponding energy scale $E$.

The constraints on the symmetry breaking scales can be understood from the non-invertible fusion rules. 
In \eqref{DDdagger0}, two non-invertible 0-form symmetry defects fuse to a condensation defect.  
Next, using \eqref{condensation1form}, we see that wrapping the condensation defect around an $S^2\times S^1$ yields a magnetic 1-form symmetry defect on the $S^2$.
Therefore the condensation defect in turn cannot exist without the magnetic 1-form symmetry.\footnote{More generally, it is possible to break a higher-form symmetry while preserving its condensation defect. Here is one such example in 2+1d. The charge conjugation symmetry in the 2+1d $U(1)_4$ Chern-Simons theory is a condensation defect of the $\mathbb{Z}_2^{(1)}$ 1-form symmetry generated by the fermion line \cite{Roumpedakis:2022aik}. Consider a renormalization group flow from a  $U(1)$ gauge theory with a charge 1, massive scalar field and a bare Chern-Simons level 4 in the UV, to $U(1)_4$ in the IR. The charge conjugation symmetry is preserved along the flow, but the 1-form symmetry is only emergent in the IR. For the condensation defects discussed in this paper, however, the explicit expressions \eqref{condensation1form} and \eqref{condensation2form} make it clear that an appropriately wrapped condensation defect yields the underlying higher-form symmetry defect, and therefore they cannot exist independently.} 
Following this chain of reasoning, it follows that the non-invertible 0-form symmetry cannot exist without the magnetic 1-form symmetry.\footnote{This is analogous to the non-invertible symmetry in the 1+1d Ising CFT, where the Kramers-Wannier duality line $\cal D$ obeys ${\cal D}\times{\cal D}=1+\eta$ with $\eta$ being the $\mathbb{Z}_2$ line. Hence, the Kramers-Wannier line $\cal D$ cannot exist on its own without the $\mathbb{Z}_2$ symmetry. } 
Equivalently, since the non-invertible 0-form symmetry is realized from half gauging the magnetic 1-form symmetry \cite{Choi:2022jqy}, it cannot exist on its own without the latter. 
Hence we derive the inequality
\ie\label{eq:inq1}
E_{\text{shift}}\lesssim E_{\text{magnetic}}~.
\fe
The inequality is not strict because the symmetry breaking scale is only approximately defined.

Similarly, the fusion rule \eqref{DDdagger1} implies that the non-invertible 1-form symmetry cannot exist without the condensation defects in \eqref{condensation1}. 
Using \eqref{condensation1form} and \eqref{condensation2form}, these condensation defects in turn rely on the existence of the magnetic 1-form  and  winding 2-form symmetries.  
Equivalently, since the non-invertible 1-form symmetry is realized from half higher gauging the magnetic and winding symmetries (see Section \ref{sec:gauging}), it cannot exist on its own without the latter two invertible symmetries. 
Hence we obtain the inequality
\ie\label{eq:inq2}
E_{\text{electric}}\lesssim \text{min}\{E_{\text{magnetic}},E_{\text{winding}}\}~.
\fe

We comment that these inequalities are consistent with the topological junctions in Section \ref{sec:higher_structure}. 
The intersection between the non-invertible 0- and 1-form symmetry defects emits a magnetic 1-form symmetry defect. 
Therefore, such a topological junction cannot exist without the magnetic symmetry, implying that $\text{min}\{E_{\text{shift}},E_{\text{electric}}\}\lesssim E_{\text{magnetic}}$. Indeed, this inequality is implied by \eqref{eq:inq1}. 
Similarly, the intersection of two non-invertible 1-form symmetry defects emits a winding symmetry defect. 
Since this topological junction cannot exist without the winding symmetry, we have $E_{\text{electric}}\lesssim E_{\text{winding}}$. Indeed, this is implied by \eqref{eq:inq2}.

We now give some physical interpretations to these inequalities.  
We start with the inequality \eqref{eq:inq1} involving the non-invertible shift symmetry. 
Since the magnetic 1-form symmetry can be broken by the dynamical magnetic monopoles, the symmetry breaking scale $E_{\text{magnetic}}$ is naturally associated with the mass $m_{\text{magnetic}}$ of the lightest magnetic monopoles, i.e., $E_{\text{magnetic}}\approx m_{\text{magnetic}}$.\footnote{We assume that the various generalized global symmetries can only be broken by their canonically charged objects. For example, we assume a $q$-form global symmetry is broken by a dynamical $(q-1)$-dimensional charged object (whose worldvolume is a $q$-dimensional manifold in spacetime). For a 0-form symmetry, we assume it's explicitly broken by a symmetry violating term in the Lagrangian. We do not explore more general possibilities here.}  
On the other hand, $E_\text{shift}$ is associated with the scale of the axion potential term. 
Our inequality \eqref{eq:inq1} is then consistent with the calculation in \cite{Fan:2021ntg}. 
The authors of that paper show that virtual monopoles running in the loops generate a potential for the axion. 
Therefore, as we go up in energy, an axion potential is generated before we reach the energy scale of a dynamical monopole, i.e., $E_\text{shift}\lesssim m_\text{magnetic}$. 
See also \cite{Cordova:2022ieu} for related discussions.

Next, we consider the inequality \eqref{eq:inq2} involving the non-invertible 1-form symmetry. The latter is explicitly broken when there are dynamical electrically charged particles. 
Therefore, the symmetry breaking scale $E_{\text{electric}}$ is naturally associated to the mass $m_{\text{electric}}$ of the lightest electrically charged particles, i.e., $E_{\text{electric}}\approx m_{\text{electric}}$. 
As for the winding 2-form symmetry, it is broken by the dynamical axion strings. Therefore, we expect the 2-form symmetry is broken at a scale $E_{\text{winding}}$ no larger than the scale set by the axion string tension $\sqrt{T}$ (see \cite{Brennan:2020ehu} for a concrete example)
\ie
E_{\text{winding}}\lesssim \sqrt{T}~.
\fe
The inequality \eqref{eq:inq2} now translates into an inequality of the mass scales associated to these dynamical objects
\ie\label{eq:inequality_mass}
m_{\text{electric}}\lesssim\text{min}\{m_{\text{magnetic}},\sqrt{T}\}~.
\fe
It means that there exists electrically charged particles that are (approximately) lighter than the lightest magnetic monopoles and the mass scale set by the axion string tension.

This hierarchy of mass scales can be understood more directly by the excitations of the magnetic monopoles and the axion strings. Because of the anomaly inflow, the magnetic monopole worldline contains a nontrivial quantum mechanics \eqref{eq:particle_on_circle}. The excitations of the magnetic monopoles are dyons that carry electric charges under the $U(1)$ gauge group \cite{Jackiw:1975ep}.  Hence, the mass $m_{\text{electric}}$ of the lightest electrically charged particles cannot be significantly larger than the mass of these dyons. 
The latter is of order of  $m_{\text{magnetic}}$  since the energy gap of these excitations are small compared to the mass of the magnetic monopoles, therefore $m_\text{electric}\lesssim m_\text{monopole}$. Similarly, because of anomaly inflow, the axion string worldsheet supports a nontrivial 1+1d degrees of freedom \eqref{eq:worldsheet_weyl}, whose excitations carry electric charges under the $U(1)$ gauge group. Hence, we have $m_{\text{electric}}\lesssim \sqrt{T}$.\footnote{We thank M.\ Reece for discussions on this point.}
 
\subsection{Weak Gravity Conjecture Mixing}

Next, we discuss applications of non-invertible symmetries to the Weak Gravity Conjecture Mixing.

The Weak Gravity Conjecture (WGC) states that in a consistent theory of quantum gravity, there must exist some particles whose electric charges are greater than their mass in Planck units \cite{Arkani-Hamed:2006emk}. 
In 3+1d, this means 
\ie
m_{\text{electric}}\lesssim e M_\text{Pl}~,
\fe
where we are schematic about the order one coefficients. Below, we will focus on 3+1d. The WGC also applies to magnetic monopoles with the electric charges $e$ replaced by the magnetic charges $e_m\approx 1/e$
\ie
m_{\text{magnetic}}\lesssim e_m M_\text{Pl}\approx \frac{M_\text{Pl}}{e}~.
\fe
The WGC has also been generalized to $p$-form gauge symmetries. The conjecture states that there exists a $(p-1)$-brane whose tension $T_p$ obeys
\ie
T_p\lesssim e_pM_\text{Pl}~,
\fe
where $e_p$ is the gauge coupling of the $p$-form gauge field. Viewing the axion as a $0$-form gauge field, it is natural to generalize the WGC further to the axion WGC, which states that there must exist an instanton with action $S_{\text{inst}}$ satisfying 
\ie
S_{\text{inst}}\lesssim \frac{M_\text{Pl}}{f}~.
\fe
See, for example, \cite{Harlow:2022gzl} for a recent review of these  WGC's.

In the axion-Maxwell theory, there is a mixing between different WGC's. More specifically, the axion WGC together with the WGC for strings implies the standard WGC in the axion-Maxwell theory \cite{Heidenreich:2021yda,Kaya:2022edp}. When $K>1$, this mixing can be argued using the higher-group symmetry \cite{Kaya:2022edp}. But the argument is not applicable to the $K=1$ theory because of the absence of the higher-group symmetry. Below, we will use the non-invertible symmetries to argue that the mixing of WGC continues to hold in the $K=1$ theory.

Starting from the axion-Maxwell theory, we can dualize the axion field to a 2-form gauge field with gauge coupling $e_2\approx f$. The WGC for strings states that there exists an axion string whose tension $T$ satisfies
\ie\label{eq:inequality_T}
T\lesssim f M_\text{Pl}~.
\fe
In the case of an abelian gauge group, the instanton arises from monopoles running in the loop, whose action is of order \cite{Fan:2021ntg}
\ie
S_{\text{inst}}\approx \frac{1}{e^2}~.
\fe
Together with the axion WGC, it implies that
\ie\label{eq:inequality_f}
f\lesssim e^2 M_\text{Pl}~.
\fe
Combining the inequalities \eqref{eq:inequality_T}, \eqref{eq:inequality_f} and the inequalities \eqref{eq:inequality_mass} deduced from the non-invertible symmetries, we have
\ie
m_{\text{electric}}\lesssim \sqrt{T}\lesssim \sqrt{fM_{\text{Pl}}}\lesssim eM_\text{Pl}~.
\fe
This recovers the standard WGC for particles.
We conclude that using the non-invertible 1-form symmetry, the WGC's for axions and strings together imply the WGC for particles even at minimal axion-photon coupling $K=1$.

\subsection{Completeness Hypothesis}

There are two pieces of lore about symmetries in quantum gravity: (1) there is no global symmetry \cite{Misner:1957mt,Banks:2010zn,Banks:1988yz,Harlow:2018tng}, and (2) when there is a gauge symmetry, the spectrum of gauge charges has to be complete -- the Completeness Hypothesis \cite{Polchinski:2003bq,Banks:2010zn,Harlow:2018tng}. 
The two statements are not unrelated. 
For example, for an ordinary $U(1)$ gauge theory, the absence of the electric 1-form global symmetry is equivalent to complete spectrum of $U(1)$ gauge charges. 
The complete spectrum means that all the Wilson lines can terminate on the fields for the electrically charged particles. 
The endable Wilson lines then imply that it cannot link topologically with the electric 1-form global symmetry defect, and the latter therefore has to be broken. 
See \cite{Rudelius:2020orz} for more details on this argument in terms of the topological defects. 

However, this equivalence breaks down if one considers more general gauge theories, such as  non-abelian finite group gauge theory  \cite{Harlow:2018tng}. 
This tension was then resolved by extending the no global symmetry conjecture  to include not only the invertible symmetries, but also the non-invertible global symmetries \cite{Rudelius:2020orz,Heidenreich:2021xpr} (see also \cite{McNamara:2021cuo,Arias-Tamargo:2022nlf}). 
It follows that the absence of invertible \textit{and} non-invertible global symmetries is equivalent to the completeness hypothesis in diverse setups. 
This more generalized notion of global symmetries consolidate different conjectures in quantum gravity and provide a coherent picture for symmetries in field theory and gravity.

Here we illustrate the connection between these two statements for axions in quantum gravity.\footnote{We thank 
I.\ Valenzuela   for insightful discussions on this point.}
Consider the axion-Maxwell theory \eqref{Lagrangian} as a low-energy sector of a full-fledged quantum gravity theory, such as string theory. 
In the absence of additional matter fields, even though the spectrum of electrically charged particles is not complete (i.e., all the Wilson lines are not endable), there is no associated invertible electric 1-form global symmetry. 
This signals the breakdown of the equivalence between the no global symmetry conjecture and the completeness hypothesis. 
This question was raised in  \cite{Heidenreich:2021xpr}, and the authors argued that the presence of the magnetic 1-form global symmetry is related to the incompleteness of the spectrum.  
However, the magnetic symmetry does not act on the Wilson lines, so it is not entirely clear how the two statements are tied together.

This is where the non-invertible 1-form global symmetry  \eqref{eq:electric_worldvolume} comes to rescue. 
Even though there is no invertible 1-form global symmetry in the axion-Maxwell theory, there is a non-invertible one, which acts on the Wilson lines by topological linking.  
To break this non-invertible global symmetry in quantum gravity, we need to include a complete set of electrically charged particles into the spectrum by the same argument in \cite{Rudelius:2020orz,Heidenreich:2021xpr}. 
Hence the completeness of the gauge spectrum is equivalent to the absence of the invertible and non-invertible 1-form global symmetries.  
The non-invertible 1-form global symmetry gives a purpose in life for the electrically charged particles when there is an axion.

As emphasized before, the existence of the non-invertible 1-form global symmetry is tied to its invertible parents: the magnetic 1-form  and the winding 2-form symmetries.  
Indeed, the non-invertible 1-form symmetry can also be broken   by including dynamical magnetic monopoles or axion strings.  
As discussed in Section \ref{sec:selection_rules}, because of anomaly inflow, these dynamical objects necessarily carry electrically charged states that generate a complete spectrum of gauge charges.
This clarifies the observation in \cite{Heidenreich:2021xpr} on the relation between the magnetic 1-form symmetry  and the completeness of the gauge spectrum.

We conclude that the  no \textit{generalized} global symmetry conjecture is equivalent to the completeness hypothesis.

\section{Summary and Outlook}\label{sec:summary}

\begin{table}[h!]
\begin{align*}
\left.\begin{array}{|c|c|c|c|c|}
\hline &\text{Non-inv.}&\text{Non-inv.}&&\\
  &~\text{0-form sym.}~&~\text{1-form sym.}~& ~U(1)_\text{magnetic}^{(1)} ~& ~U(1)_\text{winding}^{(2)}~\\
 &&&&\\
& {\cal D}^{(0)}_{p\over N} & {\cal D}^{(1)}_{p\over N} &\eta^\text{(m)}_\alpha& \eta^\text{(w)}_\alpha\\
&&&&\\
 \hline \text{Axion field}&&&&\\
 e^{i\theta} &~~e^{2\pi i p\over N}\, e^{i\theta}& - & - & -  \\
 &&&&\\
 \hline~ \text{Wilson line} &  &  &  & \\
W & - & e^{2\pi i p\over N}\, W& - & - \\
&&&&\\
 \hline~ \text{'t Hooft line} &&  &  &  \\
H &0&0& ~~  e^{i\alpha}\, H ~~&-\\
 &&&&\\
 \hline~ ~\text{Axion string surface}~~ &&&& \\
S& - & 0 & - & ~~e^{i\alpha} \,S~~ \\
&&&&\\
 \hline 
 \end{array}\right.
 \end{align*}
 \caption{Action of the generalized global symmetries (non-invertible 0- and 1-form symmetries, magnetic 1-form symmetry, winding 2-form symmetry) on the charged objects (axion field, Wilson line, 't Hooft line, and axion string worldsheet) with minimal charges. The symbol $-$ means that the global symmetry acts trivially on that charged object.  For the diagonal entries, the symmetry defects act on the charged objects by  canonical linking in spacetime. The action of the non-invertible 1-form symmetry on $H$ and $S$ are shown in Figure \ref{Fig:action2}. The non-invertible 0-form symmetry acts on $H$ by wrapping ${\cal D}^{(0)}_{p\over N}$ around a $S^2\times S^1$ with the $S^2$ linked with $H$ and the $S^1$ extended along $H$. The 0 entries mean that the symmetries annihilate the charged object, which is the hallmark of the non-invertible symmetries.}
 \label{table}
\end{table}

In this paper we explore generalized global symmetries of axion-Maxwell theory at the minimal level $K=1$, in which case the previously discovered higher group \cite{Hidaka:2020iaz,Hidaka:2020izy,Brennan:2020ehu} trivializes. 
In addition to the invertible $U(1)^{(1)}_\text{magnetic}$ winding 1-form symmetry and the $U(1)^{(2)}_\text{winding}$ 2-form symmetry, we find  non-invertible 0- and 1-form global symmetries. 
We summarize these symmetries and their charged objects in Table \ref{table}. 

The  non-invertible 0- and 1-form symmetries are constructed by coupling a TQFT to the naive shift and center symmetry operators, respectively (see Section \ref{sec:noninv}). 
More rigorously, they can be realized by half higher gauging a higher form global symmetry. 
Specifically, the non-invertible 0-form symmetry is realized via half 0-gauging  $\mathbb{Z}_N^{(1)}\subset U(1)^{(1)}_\text{magnetic}$, while the non-invertible 1-form symmetry is realized by half 1-gauging $\mathbb{Z}_N^{(1)}\times \mathbb{Z}_N^{(2)}\subset U(1)^{(1)}_\text{magnetic}\times U(1)^{(2)}_\text{winding}$ (see Section \ref{sec:gauging}).

Because of  \eqref{Je}, it is well known that there is no gauge-invariant, conserved, and quantized electric charge in axion-Maxwell theory. 
In particular, the Page charge $Q_\text{Page} = \oint_{\Sigma^{(2)}} (\star J^{(2)}_\text{electric} -{1\over 4\pi^2} \theta dA)$ is conserved, quantized, but not gauge-invariant. 
Instead, we define a new operator ${\cal D}^{(1)}_{p/N}$ with properties itemized below:
\begin{itemize}
\item It can be placed on any closed 2-manifold and is gauge-invariant.
\item It is topological, and in particular, conserved under time evolution. 
\item It does not obey a group multiplication law.
\item It leads to the non-invertible Gauss law: it measures invertibly the ordinary electric charge of  a Wilson line, but annihilates the minimal 't Hooft line (see Section \ref{sec:gauss}). 
\end{itemize}
Our non-invertible 1-form symmetry ${\cal D}^{(1)}_{\alpha}$ can be loosely viewed as a gauge-invariant fix of $``\exp(i\alpha Q_\text{Page})"$, with $\alpha=2\pi p/N $. 
However, one should not equate the two since ${\cal D}^{(1)}_{p/N}$ has a kernel -- it is  non-invertible.

Looking forward, we discuss interesting targets for future work:
\begin{itemize}
\item The full categorical structure of the generalized global symmetries in axion-Maxwell contains many other junctions and crossing relations. In this paper we have only explored the tip of the iceberg. Furthermore, at non-minimal axion-photon coupling $K>1$, the non-invertible symmetries mix in an intricate way with the higher-group symmetries. It would be interesting to understand these categorical symmetries more completely.
\item Can we interpret the massless   photon as the Goldstone boson for the non-invertible  1-form global symmetry? See \cite{GarciaEtxebarria:2022jky} for discussions on the Goldstone theorem for non-invertible 0-form symmetries. 
\item Anomalous conservation equations similar to  \eqref{Je} are ubiquitous in string and M-theory. For example, the Chern-Weil symmetries are of this type \cite{Heidenreich:2020pkc}. It would be interesting to generalize the discussion in this paper and in \cite{Apruzzi:2021nmk,Apruzzi:2022rei,GarciaEtxebarria:2022vzq,Heckman:2022muc,Antinucci:2022vyk} to understand the emergent non-invertible global symmetries in the low-energy limit of string/M-theory, and to incorporate  earlier works such as \cite{Diaconescu:2003bm,Moore:2004jv} into this framework.  
\end{itemize}

\section*{Acknowledgements}

We are grateful to  I.\ Bah, T.\ D.\ Brennan,  C.\ Cordova, B.\  Heidenreich, P.-S.\ Hsin, Z.\ Komargodski,  J.\ Maldacena, G.\ W.\ Moore, K.\ Ohmori, S.\ Pufu, M.\ Reece and S.\ Seifnashri for useful discussions. 
We thank T.\ D.\ Brennan, J.\ Kaidi, K.\ Ohmori, 
I.\ Valenzuela, and Y.\ Zheng  for comments on a draft.
HTL is supported in part by a Croucher fellowship from the Croucher Foundation, the Packard Foundation and the Center for Theoretical Physics at MIT. 
The work of SHS was supported in part by NSF grant PHY-2210182. 
We thank the Simons Collaboration on Global Categorical Symmetries  for its hospitality during a conference and a school. 
SHS thanks Harvard University for its hospitality during the course of this work. 
The authors of this paper were ordered alphabetically.

\appendix
 
\section{Higher Groups of the Axion-Maxwell Theory with $K>1$}\label{app:higher_group}

In this appendix, we review the higher-group symmetry in the axion-Maxwell theory when $K>1$ following \cite{Hidaka:2020iaz,Hidaka:2020izy,Brennan:2020ehu} and then rephrase it using symmetry operators. We consider only the invertible symmetries. They include:\
\begin{itemize}
	\item
	A $\mathbb{Z}_K^{(0)}$ zero-form shift symmetry generated by 
	\ie\label{eq:U0}
	\hat U_{2\pi p/K}^{(0)}(\Sigma^{(3)})=\exp\left[\frac{2\pi i p}{K}\oint_{\Sigma^{(3)}}\left(\star J^{(1)}_\text{shift}-\frac{K}{8\pi^2}A\wedge dA\right)\right]
	\fe
	where $J^{(1)}_\text{shift}$ is defined in \eqref{eq:currents}. The background gauge field for this symmetry is a $\mathbb{Z}_K$ 1-form gauge field. We represent it by a $U(1)$ 1-form gauge field $C^{(1)}$ obeying a constraint $K C^{(1)}=d\Gamma^{(0)}$.
	\item
	A $\mathbb{Z}_K^{(1)}$ electric 1-form symmetry generated by
	\ie\label{eq:U1}
	\hat U_{2\pi p/K}^{(1)}(\Sigma^{(2)})=\exp\left[\frac{2\pi i p}{K}\oint_{\Sigma^{(2)}}\left(\star J^{(2)}_\text{electric}-\frac{K}{4\pi^2}\theta dA\right)\right]
	\fe
	where $J^{(2)}_\text{electric}$ is defined in \eqref{eq:currents}. The background gauge field for this symmetry is a $\mathbb{Z}_K$ 2-form background gauge field. We represent it by a $U(1)$ 2-form gauge field $C_e^{(2)}$ obeying a constraint $K C_e^{(2)}=d\Gamma_e^{(1)}$.
	\item
	A $U(1)^{(1)}_\text{magnetic}$ magnetic 1-form symmetry generated by the symmetry operator
	\ie
	\eta^\text{(m)}_\alpha (\Sigma^{(2)}) \equiv \exp \left(i\alpha \oint_{\Sigma^{(2)}} \frac{F}{2\pi}  \right)~.
	\fe
	The background gauge field for this symmetry is a $U(1)$ 2-form background gauge field, which we denote by $C_m^{(2)}$.
	\item
	A $U(1)^{(2)}_\text{winding}$ winding 2-form symmetry generated by the symmetry operator
	\ie
	\eta^\text{(w)}_\alpha(\Sigma^{(1)}) \equiv \exp\left( i \alpha \oint_{\Sigma^{(1)}} {d\theta\over2\pi}\right).
	\fe
	The background gauge field for this symmetry is a $U(1)$ 3-form background gauge field, which we denote by $C^{(3)}$.
\end{itemize}
Turning on all the background gauge fields modifies the Lagrangian \eqref{Lagrangian} to
\ie\label{eq:Lagrangian_coupling}
&{f^2\over 2} \left(d\theta-C^{(1)}\right) \wedge \star \left(d\theta-C^{(1)}\right) + {1\over 2e^2} \left(F-C_e^{(2)}\right)\wedge \star\left(F-C_e^{(2)}\right) 
\\
&- {i K \over 8\pi^2 } \left[\theta F\wedge F-A\wedge F\wedge C^{(1)}-2\theta F\wedge C_e^{(2)}\right]
+\frac{i}{2\pi}A\wedge dC_m^{(2)}+\frac{i}{2\pi}\theta dC^{(3)}\,.
\fe
The gauge symmetry is
\ie
&C^{(1)}\rightarrow C^{(1)}+d\gamma^{(0)}~,\qquad\qquad\ \theta\rightarrow\theta+\gamma^{(0)}
\\ 
&C_e^{(2)}\rightarrow C_e^{(2)}+d\gamma^{(1)}_e~,\qquad\qquad A\rightarrow A+\gamma_e^{(1)}
\\
&C^{(2)}_m\rightarrow C^{(2)}_m+d\gamma^{(1)}_m-\frac{K}{2\pi}\left(\gamma^{(1)}_e\wedge d\gamma^{(0)}+\gamma^{(0)} C_e^{(2)}+\gamma^{(1)}_e\wedge C^{(1)}\right)~,
\\
&C^{(3)}\rightarrow C^{(3)}+d\gamma^{(2)}-\frac{K}{2\pi}\left(\gamma^{(1)}_e\wedge C_e^{(2)}+\frac{1}{2}\gamma^{(1)}_e\wedge d\gamma^{(1)}_e\right)~.
\fe
$C^{(2)}_m$ transforms under the gauge symmetry of $\gamma^{(0)}$ and $\gamma^{(1)}_e$, and $C^{(3)}$ transforms under the gauge symmetry of $\gamma^{(1)}_2$. This is the signature of a higher-group symmetry.
The gauge invariant field strength is
\ie\label{eq:field_strength}
&G^{(3)}_m=dC^{(2)}_m+\frac{K}{2\pi}C^{(2)}_e\wedge C^{(1)}~,
\\
&G^{(4)}=dC^{(3)}+\frac{K}{4\pi}C^{(2)}_e\wedge C^{(2)}_e~.
\fe
There is an 't Hooft anomaly which can be canceled by a 4+1d invertible field theory described by the Euclidean Lagrangian
\ie\label{eq:anomaly}
-\frac{i}{2\pi}C^{(1)}\wedge dC^{(3)}-\frac{i}{2\pi}C^{(2)}_e\wedge dC^{(2)}_m-\frac{iK}{4\pi^2}C^{(1)}\wedge C^{(2)}_e\wedge C^{(2)}_e~.
\fe
It is easy to check the anomaly cancellation by lifting the Lagrangian \eqref{eq:Lagrangian_coupling} from the boundary to the bulk. The second line of \eqref{eq:Lagrangian_coupling} combines with \eqref{eq:anomaly} into a gauge invariant term in 4+1d
\ie
&-\frac{iK}{8\pi^2}(d\theta-C^{(1)})\wedge (F-C^{(2)}_e)\wedge (F-C^{(2)}_e)
\\
&+\frac{i}{2\pi}(F-C^{(2)}_e)\wedge \left(dC^{(2)}_m+\frac{K}{2\pi}C^{(2)}_e\wedge C^{(1)}\right)+\frac{i}{2\pi}(d\theta-C^{(1)})\wedge\left(dC^{(3)}+\frac{K}{4\pi}C^{(2)}_e\wedge C^{(2)}_e\right)~.
\fe

We now rephrase the higher-group symmetry using the symmetry operators. Recall that turning on a flat background with $G_m^{(3)}=G^{(4)}=0$ is equivalent to inserting the symmetry operators into the partition function. The first equation of \eqref{eq:field_strength} implies that the one-dimensional intersection of $\hat U_{2\pi /K}^{(0)}$ and $\hat U_{2\pi /K}^{(1)}$ emits $\eta^\text{(m)}_{-2\pi/K}$. 
This is the signature of a higher-group symmetry (see \cite{Benini:2018reh} for the case of a 2-group symmetry).
The emission of $\eta^\text{(m)}_{-2\pi/K}$ is crucial for the junction to be topological. 
Consider a deformation of $\hat U_{2\pi /K}^{(0)}$ from $\Sigma^{(3)}$ to $\widetilde{\Sigma}^{(3)}$. This implements a transformation 
\ie
\theta\rightarrow \theta+\frac{2\pi}{K}
\fe
 in the region between $\Sigma^{(3)}$ and $\tilde{\Sigma}^{(3)}$. 
 Because of the worldvolume action \eqref{eq:U1} of $\hat U_{2\pi /K}^{(1)}(\Sigma^{(2)})$, this transformation generates an operator $\exp\left(-\frac{2\pi i }{K}\int_{ M^{(2)}} \frac{F}{2\pi}\right)$ on $\Sigma^{(2)}$ where $M^{(2)}$ is a region on $\Sigma^{(2)}$ bounded by $\Sigma^{(2)}\cap \Sigma^{(3)}$ and $\Sigma^{(2)}\cap \widetilde\Sigma^{(3)}$. This operator moves the boundary of $\eta^\text{(m)}_{-2\pi/K}$ from $\Sigma^{(2)}\cap \Sigma^{(3)}$ to $\Sigma^{(2)}\cap \widetilde\Sigma^{(3)}$ and therefore preserves the topological property of the intersection.

This junction and deformation are similar to the one in Figure \ref{Fig:highergroup}, with $\hat U^{(0)}_{2\pi/K}(\Sigma^{(3)})$ and $\hat{U}^{(1)}_{2\pi/K}(\Sigma^{(2)})$ shown as lines and surfaces  there, respectively. Note that two out of the three dimensions of $\hat U^{(0)}_{2\pi/K}(\Sigma^{(3)})$ are suppressed in that figure.
 
 Similarly, the second equation of \eqref{eq:field_strength} implies that the zero-dimensional intersection of two $\hat U_{2\pi /K}^{(1)}$  emits $\eta^\text{(w)}_{-2\pi/K}(M^{(1)})$ similar to Figure \ref{Fig:highergroup}, and the emission of $\eta^\text{(w)}_{-2\pi/K}(M^{(1)})$ preserves the topological property of the intersection.

\section{1+1d $\mathbb{Z}_N$ Gauge Theory} \label{app:2dZN}

In this Appendix, we will show that the 1+1d $\mathbb{Z}_N$ gauge theory can be realized on the boundary of a 2+1d invertible field theory, and derive \eqref{eq:ZN}.

The action for the 1+1d $\mathbb{Z}_N$ gauge theory is given by \cite{Maldacena:2001ss,Banks:2010zn,Kapustin:2014gua,Gaiotto:2014kfa}
\begin{equation} \label{eq:2dZN}
    S_{\text{1+1d}} = -\frac{iN}{2\pi} \int_{\Sigma^{(2)}} \phi dc \,,
\end{equation}
where $\phi \sim \phi+2\pi$ is a periodic scalar and $c$ is a $U(1)$ gauge field.
The sign of the action \eqref{eq:2dZN} is conventional since it can be absorbed by a field redifintion $\phi \rightarrow -\phi$.
Upon integrating out $\phi$, $c$ becomes a $\mathbb{Z}_N$ 1-form gauge field.

The theory has a $\mathbb{Z}_N^{(0)} \times \mathbb{Z}_N^{(1)}$ global symmetry, generated by the topological operators
\begin{equation}
    \mathcal{U}(\gamma) = e^{i\oint_\gamma c} \,, \quad \mathcal{V}(P) = e^{i\phi(P) } \,.
\end{equation}
Here, $\gamma \subset \Sigma^{(2)}$ is a closed curve and $P \in \Sigma^{(2)}$ is a point, and we have $\mathcal{U}^N = \mathcal{V}^N = 1$.
The point operator $\mathcal{V}$ carries charge 1 under the $\mathbb{Z}_N^{(0)}$ 0-form symmetry generated by the line operator $\mathcal{U}$. 
Similarly,  $\mathcal{U}$ carries charge 1 under the $\mathbb{Z}_N^{(1)}$ 1-form symmetry generated by $\mathcal{V}$.
When quantized on a circle, these two operators generate a clock-and-shift algebra,
\begin{equation} \label{eq:clockshift}
    \mathcal{U}\mathcal{V} = e^{\frac{2\pi i}{N}} \mathcal{V}\mathcal{U} \,.
\end{equation}

The fact that the 0-form symmetry generator carries a nonzero 1-form symmetry charge and vice versa implies that there is a  mixed 't Hooft anomaly between the two symmetries.
To see this, we can couple \eqref{eq:2dZN} to the background gauge fields $B^{(1)}$ and $B^{(2)}$ for the $\mathbb{Z}_N^{(0)}$ and $\mathbb{Z}_N^{(1)}$ symmetries, respectively. 
The background gauge fields are normalized such that $\oint B^{(1)}$ and $\oint B^{(2)}$ are valued in $\frac{2\pi}{N} \mathbb{Z}$ on 1-cycles and 2-cycles, respectively.
The action becomes
\begin{equation} \label{eq:2dZNB}
    S_{\text{1+1d}}[B^{(1)},B^{(2)}] = -\frac{iN}{2\pi}\int_{\Sigma^{(2)}} \left( \phi dc + cB^{(1)} + \phi B^{(2)} 
    \right)   \,.
\end{equation}
The partition function of the 1+1d $\mathbb{Z}_N$ gauge theory in the presence of both background gauge fields $B^{(1)}$ and $B^{(2)}$ is not invariant under the background gauge transformations due to the mixed 't Hooft anomaly.
The inflow action, that is, the classical action for the corresponding 2+1d $\mathbb{Z}_N^{(0)} \times \mathbb{Z}_N^{(1)}$ symmetry protected topological (SPT) phase for this mixed 't Hooft anomaly is
\begin{equation} \label{eq:3dSPT}
    -\frac{iN}{2\pi} \int_{\Sigma^{(3)}}  B^{(1)} B^{(2)} \,,
\end{equation}
where $\Sigma^{(3)}$ is a 3-manifold with the boundary $\partial \Sigma^{(3)} = \Sigma^{(2)}$.

The combined system
\begin{equation} \label{eq:combined}
    S_{\text{1+1d}}[B^{(1)},B^{(2)}]
     - \frac{iN}{2\pi}\int_{\Sigma^{(3)}} B^{(1)} B^{(2)}
\end{equation}
is invariant under the gauge transformations of both $B^{(1)}$ and $B^{(2)}$.

Now, we claim that \eqref{eq:combined} can be realized as a 2+1d twisted gauge theory of dynamical $\mathbb{Z}_N$ 1-form and 2-form gauge fields, with a suitable choice of the boundary condition.
To see this, consider a 2+1d gauge theory given by the action
\begin{equation} \label{eq:3daction}
    S_{\text{2+1d}}[B^{(1)},B^{(2)}] = -\frac{iN}{2\pi} \int_{\Sigma^{(3)}} \left(
        u^{(1)} dv^{(1)} + u^{(2)} d v^{(0)} + u^{(1)} u^{(2)}
        + u^{(1)} B^{(2)} - u^{(2)} B^{(1)}
    \right) \,.
\end{equation}
Here, $u^{(1)}$ and $u^{(2)}$ are $U(1)$ 1-form and 2-form gauge fields, respectively.
$v^{(1)}$ is a $U(1)$ 1-form gauge field, and $v^{(0)} \sim v^{(0)} + 2\pi$ is a periodic scalar.
Upon integrating out $v^{(0)}$ and $v^{(1)}$, $u^{(1)}$ and $u^{(2)}$ become discrete $\mathbb{Z}_N$ gauge fields, and the $u^{(1)} u^{(2)}$ term corresponds to a  twist (or equivalently, a discrete torsion).
The gauge transformations of dynamical gauge fields are given by
\begin{align}
\begin{split}
    u^{(1)} &\rightarrow u^{(1)} + d\Lambda^{(0)}\,, \\
    u^{(2)} &\rightarrow u^{(2)} + d\Lambda^{(1)}\,, \\
    v^{(0)} &\rightarrow v^{(0)} - \Lambda^{(0)} \,, \\
    v^{(1)} &\rightarrow v^{(1)} + d\lambda^{(0)} - \Lambda^{(1)} \,.
\end{split}
\end{align}

For simplicity, first consider the case where $\partial \Sigma^{(3)} = \Sigma^{(2)} = \emptyset$, that is, $\Sigma^{(3)}$ is a closed manifold without boundary.
In this case, it is easy to see that integrating out all the dynamical fields $v^{(0)}$, $v^{(1)}$, $u^{(1)}$ and $u^{(2)}$ in \eqref{eq:3daction} leaves behind the classical action for the SPT \eqref{eq:3dSPT}.
Thus, the 2+1d gauge theory \eqref{eq:3daction} is an invertible theory and there is no nontrivial operator in the theory, in the absence of the boundary.

When $\partial \Sigma^{(3)} = \Sigma^{(2)} \neq \emptyset$, we impose the Dirichlet boundary condition for $u^{(1)}$ and $u^{(2)}$,
\begin{equation} \label{eq:Dirichlet}
    u^{(1)}|=0 \,, \quad u^{(2)}|=0 \,,
\end{equation}
where the notation $|$ means the restriction of a field to the boundary.
In this case, the boundary operators
\begin{equation} \label{eq:boundary_op}
    \mathcal{U} \equiv \exp \left( i \oint v^{(1)}| \right) \,, \quad \mathcal{V} \equiv \exp \left( i v^{(0)}| \right) \,,
\end{equation}
become gauge-invariant as the Dirichlet boundary condition \eqref{eq:Dirichlet} sets the gauge parameters $\Lambda^{(0)}$ and $\Lambda^{(1)}$ to be zero at the boundary.

The boundary operators \eqref{eq:boundary_op} have nontrivial correlation functions, and in particular, they generate the clock-and-shift algebra \eqref{eq:clockshift} of the 1+1d $\mathbb{Z}_N$ gauge theory.
One way to show this is to identify the $\mathcal{U} \equiv \exp \left( i \oint v^{(1)}| \right)$ and $\mathcal{V} \equiv \exp \left( i v^{(0)}| \right)$ operators on the boundary as trivial surface and line operators $\left( i \int u^{(2)} \right)$ and $\left( i \int u^{(1)} \right)$ in the bulk ending on the boundary, respectively, which is justified due to the equations of motion in the bulk.
When we commute the $\mathcal{U}$ and $\mathcal{V}$ operators on the boundary, the intersection number between the corrsponding trivial surface and line operators in the bulk change by 1.
To such an intersection point, a phase factor of $\exp (2\pi i/N)$ is assigned, due to the discrete torsion.
This corresponds to the clock-and-shift algebra \eqref{eq:clockshift} generated by $\mathcal{U}$ and $\mathcal{V}$.

We see that on the boundary of the 2+1d invertible theory \eqref{eq:3daction}, with the boundary condition \eqref{eq:Dirichlet}, lives a 1+1d $\mathbb{Z}_N$ gauge theory.
In particular, we can identify
\begin{equation}
    v^{(1)}| = c \,, \quad v^{(0)}| = \phi \,,
\end{equation}
where $c$ and $\phi$ are the fields of the 1+1d $\mathbb{Z}_N$ gauge theory \eqref{eq:2dZN}.
Inside the bulk we simply have the classical SPT \eqref{eq:3dSPT}.
Therefore, we have
\begin{align}   \label{eq:3d_to_2d}
\begin{split}
    &\int [Dv^{(0)}Dv^{(1)}Du^{(1)}Du^{(2)}]_{\Sigma^{(3)},u^{(1)}|=0,u^{(2)}|=0} \exp \left( -S_{\text{2+1d}}[B^{(1)},B^{(2)}] \right) \\
    =&\exp\left(
        \frac{iN}{2\pi}\int_{\Sigma^{(3)}} B^{(1)} B^{(2)}
    \right)
    \times \int [D\phi Dc]_{\Sigma^{(2)}} \exp \left( -S_{\text{1+1d}}[B^{(1)},B^{(2)}]\right) \,.
\end{split}
\end{align}
On the lefthand side of \eqref{eq:3d_to_2d}, we can integrate out $v^{(0)}$ and $v^{(1)}$.
This makes $u^{(1)}$ and $u^{(2)}$ to become $\mathbb{Z}_N$ gauge fields,
\begin{equation}
    u^{(1)} \rightarrow \frac{2\pi}{N} b^{(1)} \,, \quad
    u^{(2)} \rightarrow \frac{2\pi}{N} b^{(2)} \,.
\end{equation}
The remaining path integral becomes a summation over $b^{(1)} \in H^1_\partial$ and $b^{(2)} \in H^2_\partial$, divided by the volume of the gauge group $|H^1_\partial|$.
By multiplying $\exp\left(-
    \frac{iN}{2\pi}\int_{\Sigma^{(3)}} B^{(1)} B^{(2)}
\right)$ on both sides, we obtain
\begin{align}
	\begin{split}
		{\frac{1}{|H^1_\partial|}}\sum_{\substack{b^{(1)}\in H^1_\partial \\  b^{(2)}\in H^2_\partial}} &\,\exp \left[
		\frac{2\pi i}{N} \int_{\Sigma^{(3)}} \left(b^{(1)} - \frac{N}{2\pi}B^{(1)} \right) \cup \left(b^{(2)}+\frac{N}{2\pi}B^{(2)} \right)
		\right] \\
		= 
		\int [D\phi\,D c]_{\Sigma^{(2)}}
		&\,\exp \left[ \oint_{\Sigma^{(2)}}\left(
		\frac{iN}{2\pi} \phi dc + \frac{iN}{2\pi} c B^{(1)}
		+ \frac{iN}{2\pi} \phi B^{(2)}
		\right)
		\right]
	\end{split}
\end{align}
Setting $B^{(1)} = pd\theta/N$ and $B^{(2)} = F/N$ gives us \eqref{eq:ZN} as desired.

\bibliographystyle{JHEP}
\bibliography{ref}

\providecommand{\href}[2]{#2}\begingroup\raggedright\begin{thebibliography}{100}

\bibitem{Koide:2021zxj}
M.~Koide, Y.~Nagoya, and S.~Yamaguchi, {\it {Non-invertible topological defects
  in 4-dimensional $\mathbb {Z}_2$ pure lattice gauge theory}},  {\em PTEP}
  {\bf 2022} (2022), no.~1 013B03, [\href{http://arxiv.org/abs/2109.05992}{{\tt
  arXiv:2109.05992}}].

\bibitem{Choi:2021kmx}
Y.~Choi, C.~Cordova, P.-S. Hsin, H.~T. Lam, and S.-H. Shao, {\it {Noninvertible
  duality defects in 3+1 dimensions}},  {\em Phys. Rev. D} {\bf 105} (2022),
  no.~12 125016, [\href{http://arxiv.org/abs/2111.01139}{{\tt
  arXiv:2111.01139}}].

\bibitem{Kaidi:2021xfk}
J.~Kaidi, K.~Ohmori, and Y.~Zheng, {\it {Kramers-Wannier-like Duality Defects
  in (3+1)D Gauge Theories}},  {\em Phys. Rev. Lett.} {\bf 128} (2022), no.~11
  111601, [\href{http://arxiv.org/abs/2111.01141}{{\tt arXiv:2111.01141}}].

\bibitem{Cordova:2022rer}
C.~Cordova, K.~Ohmori, and T.~Rudelius, {\it {Generalized Symmetry Breaking
  Scales and Weak Gravity Conjectures}},
  \href{http://arxiv.org/abs/2202.05866}{{\tt arXiv:2202.05866}}.

\bibitem{Benini:2022hzx}
F.~Benini, C.~Copetti, and L.~Di~Pietro, {\it {Factorization and global
  symmetries in holography}},  \href{http://arxiv.org/abs/2203.09537}{{\tt
  arXiv:2203.09537}}.

\bibitem{Roumpedakis:2022aik}
K.~Roumpedakis, S.~Seifnashri, and S.-H. Shao, {\it {Higher Gauging and
  Non-invertible Condensation Defects}},
  \href{http://arxiv.org/abs/2204.02407}{{\tt arXiv:2204.02407}}.

\bibitem{Bhardwaj:2022yxj}
L.~Bhardwaj, L.~Bottini, S.~Schafer-Nameki, and A.~Tiwari, {\it {Non-Invertible
  Higher-Categorical Symmetries}},  \href{http://arxiv.org/abs/2204.06564}{{\tt
  arXiv:2204.06564}}.

\bibitem{Arias-Tamargo:2022nlf}
G.~Arias-Tamargo and D.~Rodriguez-Gomez, {\it {Non-Invertible Symmetries from
  Discrete Gauging and Completeness of the Spectrum}},
  \href{http://arxiv.org/abs/2204.07523}{{\tt arXiv:2204.07523}}.

\bibitem{Hayashi:2022fkw}
Y.~Hayashi and Y.~Tanizaki, {\it {Non-invertible self-duality defects of
  Cardy-Rabinovici model and mixed gravitational anomaly}},  {\em JHEP} {\bf
  08} (2022) 036, [\href{http://arxiv.org/abs/2204.07440}{{\tt
  arXiv:2204.07440}}].

\bibitem{Choi:2022zal}
Y.~Choi, C.~Cordova, P.-S. Hsin, H.~T. Lam, and S.-H. Shao, {\it
  {Non-invertible Condensation, Duality, and Triality Defects in 3+1
  Dimensions}},  \href{http://arxiv.org/abs/2204.09025}{{\tt
  arXiv:2204.09025}}.

\bibitem{Kaidi:2022uux}
J.~Kaidi, G.~Zafrir, and Y.~Zheng, {\it {Non-invertible symmetries of $
  \mathcal{N} $ = 4 SYM and twisted compactification}},  {\em JHEP} {\bf 08}
  (2022) 053, [\href{http://arxiv.org/abs/2205.01104}{{\tt arXiv:2205.01104}}].

\bibitem{Choi:2022jqy}
Y.~Choi, H.~T. Lam, and S.-H. Shao, {\it {Noninvertible Global Symmetries in
  the Standard Model}},  {\em Phys. Rev. Lett.} {\bf 129} (2022), no.~16
  161601, [\href{http://arxiv.org/abs/2205.05086}{{\tt arXiv:2205.05086}}].

\bibitem{Cordova:2022ieu}
C.~Cordova and K.~Ohmori, {\it {Non-Invertible Chiral Symmetry and Exponential
  Hierarchies}},  \href{http://arxiv.org/abs/2205.06243}{{\tt
  arXiv:2205.06243}}.

\bibitem{Antinucci:2022eat}
A.~Antinucci, G.~Galati, and G.~Rizi, {\it {On Continuous 2-Category Symmetries
  and Yang-Mills Theory}},  \href{http://arxiv.org/abs/2206.05646}{{\tt
  arXiv:2206.05646}}.

\bibitem{Bashmakov:2022jtl}
V.~Bashmakov, M.~Del~Zotto, and A.~Hasan, {\it {On the 6d Origin of
  Non-invertible Symmetries in 4d}},
  \href{http://arxiv.org/abs/2206.07073}{{\tt arXiv:2206.07073}}.

\bibitem{Damia:2022rxw}
J.~A. Damia, R.~Argurio, and L.~Tizzano, {\it {Continuous Generalized
  Symmetries in Three Dimensions}},
  \href{http://arxiv.org/abs/2206.14093}{{\tt arXiv:2206.14093}}.

\bibitem{Damia:2022bcd}
J.~A. Damia, R.~Argurio, and E.~Garcia-Valdecasas, {\it {Non-Invertible Defects
  in 5d, Boundaries and Holography}},
  \href{http://arxiv.org/abs/2207.02831}{{\tt arXiv:2207.02831}}.

\bibitem{Moradi:2022lqp}
H.~Moradi, S.~F. Moosavian, and A.~Tiwari, {\it {Topological Holography:
  Towards a Unification of Landau and Beyond-Landau Physics}},
  \href{http://arxiv.org/abs/2207.10712}{{\tt arXiv:2207.10712}}.

\bibitem{Choi:2022rfe}
Y.~Choi, H.~T. Lam, and S.-H. Shao, {\it {Non-invertible Time-reversal
  Symmetry}},  \href{http://arxiv.org/abs/2208.04331}{{\tt arXiv:2208.04331}}.

\bibitem{Bhardwaj:2022lsg}
L.~Bhardwaj, S.~Schafer-Nameki, and J.~Wu, {\it {Universal Non-Invertible
  Symmetries}},  \href{http://arxiv.org/abs/2208.05973}{{\tt
  arXiv:2208.05973}}.

\bibitem{Bartsch:2022mpm}
T.~Bartsch, M.~Bullimore, A.~E.~V. Ferrari, and J.~Pearson, {\it
  {Non-invertible Symmetries and Higher Representation Theory I}},
  \href{http://arxiv.org/abs/2208.05993}{{\tt arXiv:2208.05993}}.

\bibitem{Lin:2022xod}
L.~Lin, D.~G. Robbins, and E.~Sharpe, {\it {Decomposition, condensation
  defects, and fusion}},  {\em Fortsch. Phys.} {\bf 70} (2022) 2200130,
  [\href{http://arxiv.org/abs/2208.05982}{{\tt arXiv:2208.05982}}].

\bibitem{GarciaEtxebarria:2022vzq}
I.~Garc\'\i{}a~Etxebarria, {\it {Branes and Non-Invertible Symmetries}},
  \href{http://arxiv.org/abs/2208.07508}{{\tt arXiv:2208.07508}}.

\bibitem{Apruzzi:2022rei}
F.~Apruzzi, I.~Bah, F.~Bonetti, and S.~Schafer-Nameki, {\it {Non-Invertible
  Symmetries from Holography and Branes}},
  \href{http://arxiv.org/abs/2208.07373}{{\tt arXiv:2208.07373}}.

\bibitem{Heckman:2022muc}
J.~J. Heckman, M.~H\"ubner, E.~Torres, and H.~Y. Zhang, {\it {The Branes Behind
  Generalized Symmetry Operators}},
  \href{http://arxiv.org/abs/2209.03343}{{\tt arXiv:2209.03343}}.

\bibitem{Freed:2022qnc}
D.~S. Freed, G.~W. Moore, and C.~Teleman, {\it {Topological symmetry in quantum
  field theory}},  \href{http://arxiv.org/abs/2209.07471}{{\tt
  arXiv:2209.07471}}.

\bibitem{Niro:2022ctq}
P.~Niro, K.~Roumpedakis, and O.~Sela, {\it {Exploring Non-Invertible Symmetries
  in Free Theories}},  \href{http://arxiv.org/abs/2209.11166}{{\tt
  arXiv:2209.11166}}.

\bibitem{Kaidi:2022cpf}
J.~Kaidi, K.~Ohmori, and Y.~Zheng, {\it {Symmetry TFTs for Non-Invertible
  Defects}},  \href{http://arxiv.org/abs/2209.11062}{{\tt arXiv:2209.11062}}.

\bibitem{Mekareeya:2022spm}
N.~Mekareeya and M.~Sacchi, {\it {Mixed Anomalies, Two-groups, Non-Invertible
  Symmetries, and 3d Superconformal Indices}},
  \href{http://arxiv.org/abs/2210.02466}{{\tt arXiv:2210.02466}}.

\bibitem{Antinucci:2022vyk}
A.~Antinucci, F.~Benini, C.~Copetti, G.~Galati, and G.~Rizi, {\it {The
  holography of non-invertible self-duality symmetries}},
  \href{http://arxiv.org/abs/2210.09146}{{\tt arXiv:2210.09146}}.

\bibitem{Chen:2022cyw}
S.~Chen and Y.~Tanizaki, {\it {Solitonic symmetry beyond homotopy:
  invertibility from bordism and non-invertibility from TQFT}},
  \href{http://arxiv.org/abs/2210.13780}{{\tt arXiv:2210.13780}}.

\bibitem{Bashmakov:2022uek}
V.~Bashmakov, M.~Del~Zotto, A.~Hasan, and J.~Kaidi, {\it {Non-invertible
  Symmetries of Class $\mathcal{S}$ Theories}},
  \href{http://arxiv.org/abs/2211.05138}{{\tt arXiv:2211.05138}}.

\bibitem{Karasik:2022kkq}
A.~Karasik, {\it {On anomalies and gauging of U(1) non-invertible symmetries in
  4d QED}},  \href{http://arxiv.org/abs/2211.05802}{{\tt arXiv:2211.05802}}.

\bibitem{Cordova:2022fhg}
C.~Cordova, S.~Hong, S.~Koren, and K.~Ohmori, {\it {Neutrino Masses from
  Generalized Symmetry Breaking}},  \href{http://arxiv.org/abs/2211.07639}{{\tt
  arXiv:2211.07639}}.

\bibitem{Decoppet:2022dnz}
T.~D. D\'ecoppet and M.~Yu, {\it {Gauging Noninvertible Defects: A
  2-Categorical Perspective}},  \href{http://arxiv.org/abs/2211.08436}{{\tt
  arXiv:2211.08436}}.

\bibitem{GarciaEtxebarria:2022jky}
I.~Garc\'\i{}a~Etxebarria and N.~Iqbal, {\it {A Goldstone theorem for
  continuous non-invertible symmetries}},
  \href{http://arxiv.org/abs/2211.09570}{{\tt arXiv:2211.09570}}.

\bibitem{Rudelius:2020orz}
T.~Rudelius and S.-H. Shao, {\it {Topological Operators and Completeness of
  Spectrum in Discrete Gauge Theories}},  {\em JHEP} {\bf 12} (2020) 172,
  [\href{http://arxiv.org/abs/2006.10052}{{\tt arXiv:2006.10052}}].

\bibitem{Heidenreich:2021xpr}
B.~Heidenreich, J.~McNamara, M.~Montero, M.~Reece, T.~Rudelius, and
  I.~Valenzuela, {\it {Non-invertible global symmetries and completeness of the
  spectrum}},  {\em JHEP} {\bf 09} (2021) 203,
  [\href{http://arxiv.org/abs/2104.07036}{{\tt arXiv:2104.07036}}].

\bibitem{Nguyen:2021yld}
M.~Nguyen, Y.~Tanizaki, and M.~\"Unsal, {\it {Semi-Abelian gauge theories,
  non-invertible symmetries, and string tensions beyond $N$-ality}},  {\em
  JHEP} {\bf 03} (2021) 238, [\href{http://arxiv.org/abs/2101.02227}{{\tt
  arXiv:2101.02227}}].

\bibitem{Kaidi:2021gbs}
J.~Kaidi, Z.~Komargodski, K.~Ohmori, S.~Seifnashri, and S.-H. Shao, {\it
  {Higher central charges and topological boundaries in 2+1-dimensional
  TQFTs}},  {\em SciPost Phys.} {\bf 13} (2022), no.~3 067,
  [\href{http://arxiv.org/abs/2107.13091}{{\tt arXiv:2107.13091}}].

\bibitem{Wang:2021vki}
J.~Wang and Y.-Z. You, {\it {Gauge Enhanced Quantum Criticality Between Grand
  Unifications: Categorical Higher Symmetry Retraction}},
  \href{http://arxiv.org/abs/2111.10369}{{\tt arXiv:2111.10369}}.

\bibitem{McGreevy:2022oyu}
J.~McGreevy, {\it {Generalized Symmetries in Condensed Matter}},
  \href{http://arxiv.org/abs/2204.03045}{{\tt arXiv:2204.03045}}.

\bibitem{Cordova:2022ruw}
C.~Cordova, T.~T. Dumitrescu, K.~Intriligator, and S.-H. Shao, {\it {Snowmass
  White Paper: Generalized Symmetries in Quantum Field Theory and Beyond}},  in
  {\em {2022 Snowmass Summer Study}}, 5, 2022.
\newblock \href{http://arxiv.org/abs/2205.09545}{{\tt arXiv:2205.09545}}.

\bibitem{Gaiotto:2014kfa}
D.~Gaiotto, A.~Kapustin, N.~Seiberg, and B.~Willett, {\it {Generalized Global
  Symmetries}},  {\em JHEP} {\bf 02} (2015) 172,
  [\href{http://arxiv.org/abs/1412.5148}{{\tt arXiv:1412.5148}}].

\bibitem{Verlinde:1988sn}
E.~P. Verlinde, {\it {Fusion Rules and Modular Transformations in 2D Conformal
  Field Theory}},  {\em Nucl. Phys. B} {\bf 300} (1988) 360--376.

\bibitem{Petkova:2000ip}
V.~B. Petkova and J.~B. Zuber, {\it {Generalized twisted partition functions}},
   {\em Phys. Lett. B} {\bf 504} (2001) 157--164,
  [\href{http://arxiv.org/abs/hep-th/0011021}{{\tt hep-th/0011021}}].

\bibitem{Fuchs:2002cm}
J.~Fuchs, I.~Runkel, and C.~Schweigert, {\it {TFT construction of RCFT
  correlators 1. Partition functions}},  {\em Nucl. Phys. B} {\bf 646} (2002)
  353--497, [\href{http://arxiv.org/abs/hep-th/0204148}{{\tt hep-th/0204148}}].

\bibitem{Frohlich:2004ef}
J.~Frohlich, J.~Fuchs, I.~Runkel, and C.~Schweigert, {\it {Kramers-Wannier
  duality from conformal defects}},  {\em Phys. Rev. Lett.} {\bf 93} (2004)
  070601, [\href{http://arxiv.org/abs/cond-mat/0404051}{{\tt
  cond-mat/0404051}}].

\bibitem{Frohlich:2006ch}
J.~Frohlich, J.~Fuchs, I.~Runkel, and C.~Schweigert, {\it {Duality and defects
  in rational conformal field theory}},  {\em Nucl. Phys.} {\bf B763} (2007)
  354--430, [\href{http://arxiv.org/abs/hep-th/0607247}{{\tt hep-th/0607247}}].

\bibitem{Feiguin:2006ydp}
A.~Feiguin, S.~Trebst, A.~W.~W. Ludwig, M.~Troyer, A.~Kitaev, Z.~Wang, and
  M.~H. Freedman, {\it {Interacting anyons in topological quantum liquids: The
  golden chain}},  {\em Phys. Rev. Lett.} {\bf 98} (2007), no.~16 160409,
  [\href{http://arxiv.org/abs/cond-mat/0612341}{{\tt cond-mat/0612341}}].

\bibitem{Frohlich:2009gb}
J.~Frohlich, J.~Fuchs, I.~Runkel, and C.~Schweigert, {\it {Defect lines,
  dualities, and generalised orbifolds}},  in {\em {Proceedings, 16th
  International Congress on Mathematical Physics (ICMP09): Prague, Czech
  Republic, August 3-8, 2009}}, 2009.
\newblock \href{http://arxiv.org/abs/0909.5013}{{\tt arXiv:0909.5013}}.

\bibitem{Carqueville:2012dk}
N.~Carqueville and I.~Runkel, {\it {Orbifold completion of defect
  bicategories}},  {\em Quantum Topol.} {\bf 7} (2016) 203,
  [\href{http://arxiv.org/abs/1210.6363}{{\tt arXiv:1210.6363}}].

\bibitem{Aasen:2016dop}
D.~Aasen, R.~S.~K. Mong, and P.~Fendley, {\it {Topological Defects on the
  Lattice I: The Ising model}},  {\em J. Phys. A} {\bf 49} (2016), no.~35
  354001, [\href{http://arxiv.org/abs/1601.07185}{{\tt arXiv:1601.07185}}].

\bibitem{Bhardwaj:2017xup}
L.~Bhardwaj and Y.~Tachikawa, {\it {On finite symmetries and their gauging in
  two dimensions}},  {\em JHEP} {\bf 03} (2018) 189,
  [\href{http://arxiv.org/abs/1704.02330}{{\tt arXiv:1704.02330}}].

\bibitem{Tachikawa:2017gyf}
Y.~Tachikawa, {\it {On gauging finite subgroups}},  {\em SciPost Phys.} {\bf 8}
  (2020), no.~1 015, [\href{http://arxiv.org/abs/1712.09542}{{\tt
  arXiv:1712.09542}}].

\bibitem{Chang:2018iay}
C.-M. Chang, Y.-H. Lin, S.-H. Shao, Y.~Wang, and X.~Yin, {\it {Topological
  Defect Lines and Renormalization Group Flows in Two Dimensions}},  {\em JHEP}
  {\bf 01} (2019) 026, [\href{http://arxiv.org/abs/1802.04445}{{\tt
  arXiv:1802.04445}}].

\bibitem{Ji:2019ugf}
W.~Ji, S.-H. Shao, and X.-G. Wen, {\it {Topological Transition on the Conformal
  Manifold}},  {\em Phys. Rev. Res.} {\bf 2} (2020), no.~3 033317,
  [\href{http://arxiv.org/abs/1909.01425}{{\tt arXiv:1909.01425}}].

\bibitem{Lin:2019hks}
Y.-H. Lin and S.-H. Shao, {\it {Duality Defect of the Monster CFT}},  {\em J.
  Phys. A} {\bf 54} (2021), no.~6 065201,
  [\href{http://arxiv.org/abs/1911.00042}{{\tt arXiv:1911.00042}}].

\bibitem{Thorngren:2019iar}
R.~Thorngren and Y.~Wang, {\it {Fusion Category Symmetry I: Anomaly In-Flow and
  Gapped Phases}},  \href{http://arxiv.org/abs/1912.02817}{{\tt
  arXiv:1912.02817}}.

\bibitem{Gaiotto:2020iye}
D.~Gaiotto and J.~Kulp, {\it {Orbifold groupoids}},  {\em JHEP} {\bf 02} (2021)
  132, [\href{http://arxiv.org/abs/2008.05960}{{\tt arXiv:2008.05960}}].

\bibitem{Komargodski:2020mxz}
Z.~Komargodski, K.~Ohmori, K.~Roumpedakis, and S.~Seifnashri, {\it {Symmetries
  and strings of adjoint QCD$_{2}$}},  {\em JHEP} {\bf 03} (2021) 103,
  [\href{http://arxiv.org/abs/2008.07567}{{\tt arXiv:2008.07567}}].

\bibitem{Aasen:2020jwb}
D.~Aasen, P.~Fendley, and R.~S.~K. Mong, {\it {Topological Defects on the
  Lattice: Dualities and Degeneracies}},
  \href{http://arxiv.org/abs/2008.08598}{{\tt arXiv:2008.08598}}.

\bibitem{Chang:2020imq}
C.-M. Chang and Y.-H. Lin, {\it {Lorentzian dynamics and factorization beyond
  rationality}},  {\em JHEP} {\bf 10} (2021) 125,
  [\href{http://arxiv.org/abs/2012.01429}{{\tt arXiv:2012.01429}}].

\bibitem{Nguyen:2021naa}
M.~Nguyen, Y.~Tanizaki, and M.~\"Unsal, {\it {Noninvertible 1-form symmetry and
  Casimir scaling in 2D Yang-Mills theory}},  {\em Phys. Rev. D} {\bf 104}
  (2021), no.~6 065003, [\href{http://arxiv.org/abs/2104.01824}{{\tt
  arXiv:2104.01824}}].

\bibitem{Thorngren:2021yso}
R.~Thorngren and Y.~Wang, {\it {Fusion Category Symmetry II: Categoriosities at
  $c$ = 1 and Beyond}},  \href{http://arxiv.org/abs/2106.12577}{{\tt
  arXiv:2106.12577}}.

\bibitem{Sharpe:2021srf}
E.~Sharpe, {\it {Topological operators, noninvertible symmetries and
  decomposition}},  \href{http://arxiv.org/abs/2108.13423}{{\tt
  arXiv:2108.13423}}.

\bibitem{Huang:2021zvu}
T.-C. Huang, Y.-H. Lin, and S.~Seifnashri, {\it {Construction of
  two-dimensional topological field theories with non-invertible symmetries}},
  {\em JHEP} {\bf 12} (2021) 028, [\href{http://arxiv.org/abs/2110.02958}{{\tt
  arXiv:2110.02958}}].

\bibitem{Huang:2021nvb}
T.-C. Huang, Y.-H. Lin, K.~Ohmori, Y.~Tachikawa, and M.~Tezuka, {\it {Numerical
  Evidence for a Haagerup Conformal Field Theory}},  {\em Phys. Rev. Lett.}
  {\bf 128} (2022), no.~23 231603, [\href{http://arxiv.org/abs/2110.03008}{{\tt
  arXiv:2110.03008}}].

\bibitem{Vanhove:2021zop}
R.~Vanhove, L.~Lootens, M.~Van~Damme, R.~Wolf, T.~J. Osborne, J.~Haegeman, and
  F.~Verstraete, {\it {Critical Lattice Model for a Haagerup Conformal Field
  Theory}},  {\em Phys. Rev. Lett.} {\bf 128} (2022), no.~23 231602,
  [\href{http://arxiv.org/abs/2110.03532}{{\tt arXiv:2110.03532}}].

\bibitem{Burbano:2021loy}
I.~M. Burbano, J.~Kulp, and J.~Neuser, {\it {Duality defects in E$_{8}$}},
  {\em JHEP} {\bf 10} (2022) 186, [\href{http://arxiv.org/abs/2112.14323}{{\tt
  arXiv:2112.14323}}].

\bibitem{Inamura:2022lun}
K.~Inamura, {\it {Fermionization of fusion category symmetries in 1+1
  dimensions}},  \href{http://arxiv.org/abs/2206.13159}{{\tt
  arXiv:2206.13159}}.

\bibitem{Chang:2022hud}
C.-M. Chang, J.~Chen, and F.~Xu, {\it {Topological Defect Lines in Two
  Dimensional Fermionic CFTs}},  \href{http://arxiv.org/abs/2208.02757}{{\tt
  arXiv:2208.02757}}.

\bibitem{Lin:2022dhv}
Y.-H. Lin, M.~Okada, S.~Seifnashri, and Y.~Tachikawa, {\it {Asymptotic density
  of states in 2d CFTs with non-invertible symmetries}},
  \href{http://arxiv.org/abs/2208.05495}{{\tt arXiv:2208.05495}}.

\bibitem{Robbins:2022wlr}
D.~Robbins, E.~Sharpe, and T.~Vandermeulen, {\it {Decomposition,
  Trivially-Acting Symmetries, and Topological Operators}},
  \href{http://arxiv.org/abs/2211.14332}{{\tt arXiv:2211.14332}}.

\bibitem{Hidaka:2020iaz}
Y.~Hidaka, M.~Nitta, and R.~Yokokura, {\it {Higher-form symmetries and 3-group
  in axion electrodynamics}},  {\em Phys. Lett. B} {\bf 808} (2020) 135672,
  [\href{http://arxiv.org/abs/2006.12532}{{\tt arXiv:2006.12532}}].

\bibitem{Hidaka:2020izy}
Y.~Hidaka, M.~Nitta, and R.~Yokokura, {\it {Global 3-group symmetry and 't
  Hooft anomalies in axion electrodynamics}},  {\em JHEP} {\bf 01} (2021) 173,
  [\href{http://arxiv.org/abs/2009.14368}{{\tt arXiv:2009.14368}}].

\bibitem{Brennan:2020ehu}
T.~D. Brennan and C.~Cordova, {\it {Axions, higher-groups, and emergent
  symmetry}},  {\em JHEP} {\bf 02} (2022) 145,
  [\href{http://arxiv.org/abs/2011.09600}{{\tt arXiv:2011.09600}}].

\bibitem{Nakajima:2022feg}
T.~Nakajima, T.~Sakai, and R.~Yokokura, {\it {Higher-group structure in
  $2n$-dimensional axion-electrodynamics}},
  \href{http://arxiv.org/abs/2211.13861}{{\tt arXiv:2211.13861}}.

\bibitem{Cardy:1989ir}
J.~L. Cardy, {\it {Boundary Conditions, Fusion Rules and the Verlinde
  Formula}},  {\em Nucl. Phys. B} {\bf 324} (1989) 581--596.

\bibitem{Page:1983mke}
D.~N. Page, {\it {Classical Stability of Round and Squashed Seven Spheres in
  Eleven-dimensional Supergravity}},  {\em Phys. Rev. D} {\bf 28} (1983) 2976.

\bibitem{Marolf:2000cb}
D.~Marolf, {\it {Chern-Simons terms and the three notions of charge}},  in {\em
  {International Conference on Quantization, Gauge Theory, and Strings:
  Conference Dedicated to the Memory of Professor Efim Fradkin}}, pp.~312--320,
  6, 2000.
\newblock \href{http://arxiv.org/abs/hep-th/0006117}{{\tt hep-th/0006117}}.

\bibitem{Witten:1979ey}
E.~Witten, {\it {Dyons of Charge $e \theta/2 \pi$}},  {\em Phys. Lett. B} {\bf
  86} (1979) 283--287.

\bibitem{Gorantla:2022eem}
P.~Gorantla, H.~T. Lam, N.~Seiberg, and S.-H. Shao, {\it {Global dipole
  symmetry, compact Lifshitz theory, tensor gauge theory, and fractons}},  {\em
  Phys. Rev. B} {\bf 106} (2022), no.~4 045112,
  [\href{http://arxiv.org/abs/2201.10589}{{\tt arXiv:2201.10589}}].

\bibitem{Cordova:2018cvg}
C.~C\'{o}rdova, T.~T. Dumitrescu, and K.~Intriligator, {\it {Exploring 2-Group
  Global Symmetries}},  {\em JHEP} {\bf 02} (2019) 184,
  [\href{http://arxiv.org/abs/1802.04790}{{\tt arXiv:1802.04790}}].

\bibitem{Yokokura:2022alv}
R.~Yokokura, {\it {Non-invertible symmetries in axion electrodynamics}},
  \href{http://arxiv.org/abs/2212.05001}{{\tt arXiv:2212.05001}}.

\bibitem{Seiberg:2018ntt}
N.~Seiberg, Y.~Tachikawa, and K.~Yonekura, {\it {Anomalies of Duality Groups
  and Extended Conformal Manifolds}},  {\em PTEP} {\bf 2018} (2018), no.~7
  073B04, [\href{http://arxiv.org/abs/1803.07366}{{\tt arXiv:1803.07366}}].

\bibitem{Cordova:2019uob}
C.~C\'ordova, D.~S. Freed, H.~T. Lam, and N.~Seiberg, {\it {Anomalies in the
  Space of Coupling Constants and Their Dynamical Applications II}},  {\em
  SciPost Phys.} {\bf 8} (2020), no.~1 002,
  [\href{http://arxiv.org/abs/1905.13361}{{\tt arXiv:1905.13361}}].

\bibitem{Hsin:2018vcg}
P.-S. Hsin, H.~T. Lam, and N.~Seiberg, {\it {Comments on One-Form Global
  Symmetries and Their Gauging in 3d and 4d}},  {\em SciPost Phys.} {\bf 6}
  (2019), no.~3 039, [\href{http://arxiv.org/abs/1812.04716}{{\tt
  arXiv:1812.04716}}].

\bibitem{Putrov:2022pua}
P.~Putrov, {\it {$\mathbb{Q}/\mathbb{Z}$ symmetry}},
  \href{http://arxiv.org/abs/2208.12071}{{\tt arXiv:2208.12071}}.

\bibitem{Fischler:1983sc}
W.~Fischler and J.~Preskill, {\it {DYON - AXION DYNAMICS}},  {\em Phys. Lett.
  B} {\bf 125} (1983) 165--170.

\bibitem{Diaconescu:2003bm}
E.~Diaconescu, G.~W. Moore, and D.~S. Freed, {\it {The M theory three form and
  E(8) gauge theory}},  \href{http://arxiv.org/abs/hep-th/0312069}{{\tt
  hep-th/0312069}}.

\bibitem{Moore:2004jv}
G.~W. Moore, {\it {Anomalies, Gauss laws, and Page charges in M-theory}},  {\em
  Comptes Rendus Physique} {\bf 6} (2005) 251--259,
  [\href{http://arxiv.org/abs/hep-th/0409158}{{\tt hep-th/0409158}}].

\bibitem{Kong:2013aya}
L.~Kong, {\it {Anyon condensation and tensor categories}},  {\em Nucl. Phys. B}
  {\bf 886} (2014) 436--482, [\href{http://arxiv.org/abs/1307.8244}{{\tt
  arXiv:1307.8244}}].

\bibitem{Kong:2014qka}
L.~Kong and X.-G. Wen, {\it {Braided fusion categories, gravitational
  anomalies, and the mathematical framework for topological orders in any
  dimensions}},  \href{http://arxiv.org/abs/1405.5858}{{\tt arXiv:1405.5858}}.

\bibitem{Else:2017yqj}
D.~V. Else and C.~Nayak, {\it {Cheshire charge in (3+1)-dimensional topological
  phases}},  {\em Phys. Rev. B} {\bf 96} (2017), no.~4 045136,
  [\href{http://arxiv.org/abs/1702.02148}{{\tt arXiv:1702.02148}}].

\bibitem{Gaiotto:2019xmp}
D.~Gaiotto and T.~Johnson-Freyd, {\it {Condensations in higher categories}},
  \href{http://arxiv.org/abs/1905.09566}{{\tt arXiv:1905.09566}}.

\bibitem{Chen:2014wse}
X.~Chen, F.~J. Burnell, A.~Vishwanath, and L.~Fidkowski, {\it {Anomalous
  Symmetry Fractionalization and Surface Topological Order}},  {\em Phys. Rev.
  X} {\bf 5} (2015), no.~4 041013, [\href{http://arxiv.org/abs/1403.6491}{{\tt
  arXiv:1403.6491}}].

\bibitem{Barkeshli:2014cna}
M.~Barkeshli, P.~Bonderson, M.~Cheng, and Z.~Wang, {\it {Symmetry
  Fractionalization, Defects, and Gauging of Topological Phases}},  {\em Phys.
  Rev. B} {\bf 100} (2019), no.~11 115147,
  [\href{http://arxiv.org/abs/1410.4540}{{\tt arXiv:1410.4540}}].

\bibitem{Benini:2018reh}
F.~Benini, C.~C\'{o}rdova, and P.-S. Hsin, {\it {On 2-Group Global Symmetries
  and their Anomalies}},  {\em JHEP} {\bf 03} (2019) 118,
  [\href{http://arxiv.org/abs/1803.09336}{{\tt arXiv:1803.09336}}].

\bibitem{Delmastro:2022pfo}
D.~Delmastro, J.~Gomis, P.-S. Hsin, and Z.~Komargodski, {\it {Anomalies and
  Symmetry Fractionalization}},  \href{http://arxiv.org/abs/2206.15118}{{\tt
  arXiv:2206.15118}}.

\bibitem{Brennan:2022tyl}
T.~D. Brennan, C.~Cordova, and T.~T. Dumitrescu, {\it {Line Defect Quantum
  Numbers \& Anomalies}},  \href{http://arxiv.org/abs/2206.15401}{{\tt
  arXiv:2206.15401}}.

\bibitem{Vafa:1989ih}
C.~Vafa, {\it {Quantum Symmetries of String Vacua}},  {\em Mod. Phys. Lett. A}
  {\bf 4} (1989) 1615.

\bibitem{Bachas:2000ik}
C.~Bachas, M.~R. Douglas, and C.~Schweigert, {\it {Flux stabilization of
  D-branes}},  {\em JHEP} {\bf 05} (2000) 048,
  [\href{http://arxiv.org/abs/hep-th/0003037}{{\tt hep-th/0003037}}].

\bibitem{Taylor:2000za}
W.~Taylor, {\it {D2-branes in B fields}},  {\em JHEP} {\bf 07} (2000) 039,
  [\href{http://arxiv.org/abs/hep-th/0004141}{{\tt hep-th/0004141}}].

\bibitem{Callan:1984sa}
C.~G. Callan, Jr. and J.~A. Harvey, {\it {Anomalies and Fermion Zero Modes on
  Strings and Domain Walls}},  {\em Nucl. Phys. B} {\bf 250} (1985) 427--436.

\bibitem{Naculich:1987ci}
S.~G. Naculich, {\it {Axionic Strings: Covariant Anomalies and Bosonization of
  Chiral Zero Modes}},  {\em Nucl. Phys. B} {\bf 296} (1988) 837--867.

\bibitem{Fukuda:2020imw}
H.~Fukuda and K.~Yonekura, {\it {Witten effect, anomaly inflow, and charge
  teleportation}},  {\em JHEP} {\bf 01} (2021) 119,
  [\href{http://arxiv.org/abs/2010.02221}{{\tt arXiv:2010.02221}}].

\bibitem{Fan:2021ntg}
J.~Fan, K.~Fraser, M.~Reece, and J.~Stout, {\it {Axion Mass from Magnetic
  Monopole Loops}},  {\em Phys. Rev. Lett.} {\bf 127} (2021), no.~13 131602,
  [\href{http://arxiv.org/abs/2105.09950}{{\tt arXiv:2105.09950}}].

\bibitem{Heidenreich:2021yda}
B.~Heidenreich, M.~Reece, and T.~Rudelius, {\it {The Weak Gravity Conjecture
  and axion strings}},  {\em JHEP} {\bf 11} (2021) 004,
  [\href{http://arxiv.org/abs/2108.11383}{{\tt arXiv:2108.11383}}].

\bibitem{Cordova:2019jnf}
C.~C\'ordova, D.~S. Freed, H.~T. Lam, and N.~Seiberg, {\it {Anomalies in the
  Space of Coupling Constants and Their Dynamical Applications I}},  {\em
  SciPost Phys.} {\bf 8} (2020), no.~1 001,
  [\href{http://arxiv.org/abs/1905.09315}{{\tt arXiv:1905.09315}}].

\bibitem{Gaiotto:2017yup}
D.~Gaiotto, A.~Kapustin, Z.~Komargodski, and N.~Seiberg, {\it {Theta, Time
  Reversal, and Temperature}},  {\em JHEP} {\bf 05} (2017) 091,
  [\href{http://arxiv.org/abs/1703.00501}{{\tt arXiv:1703.00501}}].

\bibitem{Kikuchi:2018gfo}
Y.~Kikuchi, {\em {'t Hooft anomaly, global inconsistency, and some of their
  applications}}.
\newblock PhD thesis, Kyoto U., 2018.

\bibitem{Jackiw:1975ep}
R.~Jackiw, {\it {Charge and Mass Spectrum of Quantum Solitons}},  {\em Conf.
  Proc. C} {\bf 750926} (1975) 377--401.

\bibitem{Bilal:2008qx}
A.~Bilal, {\it {Lectures on Anomalies}},
  \href{http://arxiv.org/abs/0802.0634}{{\tt arXiv:0802.0634}}.

\bibitem{Kogan:1993yw}
I.~I. Kogan, {\it {Axions, monopoles and cosmic strings}},  in {\em
  {International Workshop on Supersymmetry and Unification of Fundamental
  Interactions (SUSY 93)}}, 5, 1993.
\newblock \href{http://arxiv.org/abs/hep-ph/9305307}{{\tt hep-ph/9305307}}.

\bibitem{Arkani-Hamed:2006emk}
N.~Arkani-Hamed, L.~Motl, A.~Nicolis, and C.~Vafa, {\it {The String landscape,
  black holes and gravity as the weakest force}},  {\em JHEP} {\bf 06} (2007)
  060, [\href{http://arxiv.org/abs/hep-th/0601001}{{\tt hep-th/0601001}}].

\bibitem{Harlow:2022gzl}
D.~Harlow, B.~Heidenreich, M.~Reece, and T.~Rudelius, {\it {The Weak Gravity
  Conjecture: A Review}},  \href{http://arxiv.org/abs/2201.08380}{{\tt
  arXiv:2201.08380}}.

\bibitem{Kaya:2022edp}
S.~Kaya and T.~Rudelius, {\it {Higher-Group Symmetries and Weak Gravity
  Conjecture Mixing}},  \href{http://arxiv.org/abs/2202.04655}{{\tt
  arXiv:2202.04655}}.

\bibitem{Misner:1957mt}
C.~W. Misner and J.~A. Wheeler, {\it {Classical physics as geometry:
  Gravitation, electromagnetism, unquantized charge, and mass as properties of
  curved empty space}},  {\em Annals Phys.} {\bf 2} (1957) 525--603.

\bibitem{Banks:2010zn}
T.~Banks and N.~Seiberg, {\it {Symmetries and Strings in Field Theory and
  Gravity}},  {\em Phys. Rev. D} {\bf 83} (2011) 084019,
  [\href{http://arxiv.org/abs/1011.5120}{{\tt arXiv:1011.5120}}].

\bibitem{Banks:1988yz}
T.~Banks and L.~J. Dixon, {\it {Constraints on String Vacua with Space-Time
  Supersymmetry}},  {\em Nucl. Phys. B} {\bf 307} (1988) 93--108.

\bibitem{Harlow:2018tng}
D.~Harlow and H.~Ooguri, {\it {Symmetries in quantum field theory and quantum
  gravity}},  {\em Commun. Math. Phys.} {\bf 383} (2021), no.~3 1669--1804,
  [\href{http://arxiv.org/abs/1810.05338}{{\tt arXiv:1810.05338}}].

\bibitem{Polchinski:2003bq}
J.~Polchinski, {\it {Monopoles, duality, and string theory}},  {\em Int. J.
  Mod. Phys. A} {\bf 19S1} (2004) 145--156,
  [\href{http://arxiv.org/abs/hep-th/0304042}{{\tt hep-th/0304042}}].

\bibitem{McNamara:2021cuo}
J.~McNamara, {\it {Gravitational Solitons and Completeness}},
  \href{http://arxiv.org/abs/2108.02228}{{\tt arXiv:2108.02228}}.

\bibitem{Heidenreich:2020pkc}
B.~Heidenreich, J.~McNamara, M.~Montero, M.~Reece, T.~Rudelius, and
  I.~Valenzuela, {\it {Chern-Weil global symmetries and how quantum gravity
  avoids them}},  {\em JHEP} {\bf 11} (2021) 053,
  [\href{http://arxiv.org/abs/2012.00009}{{\tt arXiv:2012.00009}}].

\bibitem{Apruzzi:2021nmk}
F.~Apruzzi, F.~Bonetti, I.~G. Etxebarria, S.~S. Hosseini, and
  S.~Schafer-Nameki, {\it {Symmetry TFTs from String Theory}},
  \href{http://arxiv.org/abs/2112.02092}{{\tt arXiv:2112.02092}}.

\bibitem{Maldacena:2001ss}
J.~M. Maldacena, G.~W. Moore, and N.~Seiberg, {\it {D-brane charges in
  five-brane backgrounds}},  {\em JHEP} {\bf 10} (2001) 005,
  [\href{http://arxiv.org/abs/hep-th/0108152}{{\tt hep-th/0108152}}].

\bibitem{Kapustin:2014gua}
A.~Kapustin and N.~Seiberg, {\it {Coupling a QFT to a TQFT and Duality}},  {\em
  JHEP} {\bf 04} (2014) 001, [\href{http://arxiv.org/abs/1401.0740}{{\tt
  arXiv:1401.0740}}].

\end{thebibliography}\endgroup

\end{document}